\documentclass[english,aps,prl,twocolumn,amsmath,amssymb,showpacs,superscriptaddress,notitlepage,nofootinbib]{revtex4-2}

\usepackage[T1]{fontenc}
\usepackage{color}
\definecolor{page_backgroundcolor}{rgb}{1, 1, 1}
\pagecolor{page_backgroundcolor}
\usepackage{amsmath}
\usepackage{graphicx}
\usepackage{microtype}
\usepackage{bm}
\usepackage{makecell}
\usepackage[normalem]{ulem}
\usepackage{enumitem}
\usepackage[marginal]{footmisc}
\usepackage{braket}

\makeatletter
\usepackage{amssymb}
\usepackage{amsmath}
\usepackage{graphicx}
\usepackage[caption=false]{subfig}
\usepackage{lipsum}
\usepackage{mathrsfs}

\usepackage[svgnames]{xcolor}

\usepackage[colorlinks=true,linkcolor=RoyalBlue,anchorcolor=red,citecolor=RoyalBlue,urlcolor=RoyalBlue]{hyperref}

\newcommand{\dd}{\mathrm{d}}

\makeatother

\usepackage{babel}

\newcommand{\ii}{\hspace{1pt}\mathrm{i}\hspace{1pt}}

\begin{document}
\global\long\def\figurename{Fig.}%

\title{Quantum Many-Body Lattice $\mathcal{C}$-$\mathcal{R}$-$\mathcal{T}$ Symmetry: \\
Fractionalization, Anomaly, and Symmetric Mass Generation}

\author{Yang-Yang Li}
\email{yang-yang.li@stonybrook.edu}
\affiliation{Department of Physics and Astronomy, State University of New York at Stony Brook, NY 11794, USA}

\author{Juven Wang}
\email{jw@lims.ac.uk}
\homepage{http://sns.ias.edu/~juven/}
\affiliation{London Institute for Mathematical Sciences, Royal Institution, W1S 4BS, UK}
\affiliation{Center of Mathematical Sciences and Applications, Harvard University, MA 02138, USA}

\author{Yi-Zhuang You}
\email{yzyou@physics.ucsd.edu}
\affiliation{Department of Physics, University of California San Diego, CA 92093, USA}

\begin{abstract}
Charge conjugation ($\mathcal{C}$), mirror reflection ($\mathcal{R}$), and time reversal ($\mathcal{T}$) symmetries, along with internal symmetries, are essential for massless Majorana and Dirac fermions. These symmetries are sufficient to rule out potential fermion bilinear mass terms, thereby establishing a gapless free fermion fixed point phase, pivotal for symmetric mass generation (SMG) transition. In this work, we systematically study the anomaly of $\mathcal{C}$-$\mathcal{R}$-$\mathcal{T}$-internal symmetry in all spacetime dimensions by analyzing the projective representation (i.e. the fractionalization) of the $\mathcal{C}$-$\mathcal{R}$-$\mathcal{T}$-internal symmetry group in the quantum many-body Hilbert space on the lattice. By discovering the fermion-flavor-number-dependent $\mathcal{C}$-$\mathcal{R}$-$\mathcal{T}$-internal symmetry’s anomaly structure, we demonstrate an alternative way to derive the minimal flavor number for SMG, which shows consistency with known results from K\"ahler-Dirac fermion or cobordism classification. 
Our findings reveal that, in general spatial dimensions, either 8 copies of staggered Majorana fermions or 4 copies of staggered Dirac fermions admit SMG. By directly searching for 4-fermion interactions that form commuting stabilizers respecting all symmetry constraints, we can prove the explicit SMG gapping retained a unique ground state in the codespace. Furthermore, we establish the correspondence between the symmetry operators of staggered fermions and free fermions, which is instrumental in facilitating the analysis of symmetry fractionalization at the field theory level. Previous research has shown that in 1+1d, the IR chiral symmetry $\mathbb{Z}_2^{\chi}$ is an emanant symmetry from the UV translational $\mathbb{Z}_L^T$, their symmetry groups are not the subgroups of each other, but their 't Hooft anomalies matches, and the UV symmetry in the low-energy limit gives the IR counterpart. In this work, we show that in higher dimensions, all IR internal symmetries are emanant symmetries from UV translations in different directions, and some IR $\mathcal{C}$, $\mathcal{R}$, or $\mathcal{T}$ symmetries can emanate from their UV versions combined with translation.
\end{abstract}

\maketitle

\tableofcontents

\section{Introduction}

Charge conjugation ($\mathcal{C}$), mirror reflection ($\mathcal{R}$), and time reversal ($\mathcal{T}$) are fundamental symmetries in physics~\cite{PhysRev.82.914,Luders:1954zz,10.1063/1.3060063,Pauli:1957voo,LUDERS19571,streater2000pct}. Recent studies have shown that the $\mathcal{C}$-$\mathcal{R}$-$\mathcal{T}$ symmetry $G=\mathbb{Z}_2^\mathcal{C}\times\mathbb{Z}_2^\mathcal{R}\times\mathbb{Z}_2^\mathcal{T}$ does not acts on fermions like a direct product group structure. Instead, it is extended by the fermion parity symmetry $\mathbb{Z}_2^F$ to form a non-Abelian Pauli group in ($1+1$) dimensions ($1+1$d) \cite{2022PhRvD.106j5009W}, or by larger internal symmetries to yield a more involved group structure $\tilde{G}$ in higher dimensions \cite{2023arXiv231217126W,singleparticle}. Consequently, the fermion field does not transform as a linear representation of the original symmetry group $G$, but rather as a projective representation of $G$ or a linear representation of the extended group $\tilde{G}$. This phenomenon, known as symmetry fractionalization~\cite{2022PhRvD.106j5009W}, is a unique feature of symmetry actions on fermion fields. 

From another perspective, fractionalized symmetry groups exhibit quantum anomalies~\cite{Hooft1980,Adler19692426,bell1969pcac,1976PhRvL..37....8T,Blaer19811364,Kapustin2015,Bhardwaj2017,Kapustin2017,Thorngren2020,10.21468/SciPostPhys.7.1.007,PhysRevResearch.2.033317,Lin_2021,PhysRevLett.126.195701,Delmastro2021,Grigoletto2023,10.21468/SciPostPhys.15.5.216,2023ScPP...15...51C,10.21468/SciPostPhys.16.3.064,2024arXiv240912220C}, where the partition function acquires a complex phase, $Z \rightarrow e^{\ii\alpha}Z$, upon insertion of symmetry defects. This non-trivial symmetry response prevents the system from deforming into a symmetrically gapped and featureless state, thereby preserving gapless fermionic excitations at low energy in many common cases.

Considerable evidence indicates that $\mathcal{C}$-$\mathcal{R}$-$\mathcal{T}$ and internal symmetries are sufficient to exclude all possible fermion bilinear mass terms~\cite{2023arXiv231217126W,singleparticle} from the Hamiltonian, thereby obstructing the gap formation (or mass generation) for fermions in the non-interacting limit. Therefore, the conventional fermion mass generation mechanism, known as the Higgs mechanism, must rely on spontaneous symmetry breaking (SSB) of at least some of these protecting symmetries. However, a symmetric mass (or excitation gap in the fermion spectrum) can still be generated via multi-fermion interaction or condensation, without breaking the $\mathcal{C}$-$\mathcal{R}$-$\mathcal{T}$-internal symmetry, provided that the fermion flavor number satisfies certain conditions to cancel the anomaly associated with these symmetries. This alternative process is known as symmetric mass generation (SMG)~\cite{2010PhRvB..81m4509F,2011PhRvB..83g5103F,2023PhRvB.107a4311W,2015PhRvB..91k5121S,PhysRevD.91.065035,Catterall2016,2022JHEP...07..001T,2022Symm...14.1475W,PhysRevX.8.011026,PhysRevB.107.195133,PhysRevLett.128.185301,2021arXiv210315865X,PhysRevB.97.125112,PhysRevLett.132.156503}.

The study of SMG originally emerges from the classification problem of interacting fermionic symmetry-protected topological (iFSPT) phases~\cite{2012arXiv1201.2648G,2015arXiv150101313C,2015PhRvB..92l5104M,2015JHEP...12..052K,2016arXiv160406527F,doi:10.1142/S0217751X16450445,RevModPhys.89.041004,2017arXiv170310937W,2017JHEP...10..080K,10.1093/ptep/pty051,2018arXiv181100536W,2019JHEP...05..007G,2019arXiv190607199T,2019PhRvB.100w5141L,Guo2020,2021ChPhL..38l7101O,2021arXiv210910911A,PhysRevB.105.235143,2022PhRvD.106a4513H,2023PhRvB.107p5126M,2023PhRvX..13c1005Z,PhysRevLett.129.275301}, where it reduces the $\mathbb{Z}$-classified free fermionic symmetry-protected topological (fFSPT) phases to $\mathbb{Z}_N$-classified iFSPT phases, and the required fermion flavor (copy) number $N \sim 2^{d/2}$ to enable SMG generally grows exponentially with the spacetime dimension $d$ asymptotically. Since then, SMG has found broader and richer applications in high-energy physics and plays a significant role in the lattice regularization of chiral gauge theories~\cite{PhysRevLett.128.185301,Wen_2013,2014arXiv1402.4151Y,2015PhRvB..91l5147Y,BenTov2015,2017arXiv170604648D,2018arXiv180911171W,PhysRevD.99.111501,10.1093/ptep/ptz055,2021PhRvX..11a1063R,PhysRevD.104.094504}.

It was recently noticed that for K\"ahler-Dirac fermion---one of the lattice realizations of the Dirac fermion, the classification is simply $\mathbb{Z}_2$ in all spacetime dimensions~\cite{PhysRevD.104.094504,2022PhRvD.106a4509C,2023PhRvD.107a4501C,PhysRevD.104.014503,2023PhRvB.108k5139G}. There is an elegant geometric interpretation for SMG in terms of Euler characteristic $\chi$~\cite{2023PhRvD.107a4501C}. The SMG flavor number patterns for Majorana fermion, Dirac fermion, K\"ahler-Dirac fermion and staggered fermions are listed in Tab.~\ref{Tab: class_all}.

\begin{table}[h]
\centering
\caption{The required fermion flavor number $N$ for SMG to occur. The flavor numbers are counted in terms of Majorana root states, Dirac root states, staggered Majorana fermions, staggered Dirac fermions, and K\"ahler-Dirac fermions. $d+1$ is the spacetime dimension.}
  \begin{tabular}{c|cccccccc}
  \hline
    $d+1$   & 1   & 2     & 3    & 4 & 5 & 6 & 7 & 8      \\ \hline
    Majorana         & 8 & 8 & 16 & 16 & 16 & 16 & 32 & 64 \\
    Dirac         & 4 & 4 & 8 & 8 & 16 & 16 & 32 & 32 \\
    \hline
    staggered Majorana    & 8 & 8 & 8 & 8 & 8 & 8 & 8 & 8 \\
    staggered Dirac         & 4 & 4 & 4 & 4 & 4 & 4 & 4 & 4  \\
    K\"ahler-Dirac     & 2 & 2 & 2 & 2 & 2 & 2 & 2 & 2 \\
\hline
    \end{tabular}\label{Tab: class_all}
\end{table}

In this work, we employ the staggered fermion model~\cite{PhysRevD.11.395, Becher1982, Smit_2002, 2007PhRvD..75e4502F, 2013PhRvD..87e4505B, 2006slft.confE..22S, 2008arXiv0812.3110G, 2006PhRvD..74a4511H} to scrutinize the anomaly structure within the quantum many-body Hilbert space on a lattice and to classify SMG. Our approach begins with a concise review of the $\mathcal{R}$-$\mathcal{T}$ and internal symmetries for a free Majorana fermion within the single-particle framework, as detailed in Sec.~\ref{Sec:free_Maj}. Subsequently, we concentrate on the $0+1$d scenario in Sec.~\ref{Sec: 0+1d}, delineating the projectively realized symmetry group, or symmetry fractionalization, akin to its single-particle counterpart. In this simple case, we observe that the fractionalized symmetry group encompasses a non-Abelian $\mathbb{D}_8$ structure or Kramers' doubling for 2 and 4 copies of the Majorana system, posing an impediment to achieving a unique ground state for SMG. For 8 copies of the Majorana system, we successfully reproduce the original invariant symmetry group and identify an explicit gapping interaction through direct searching. Building on this foundational analysis, we extend our investigation to the $1+1$d Majorana chain in Sec.~\ref{Sec:1+1d_M}. From Sec.~\ref{Sec:1+1d_M_1} to Sec.~\ref{Sec. 1+1d lattice}, we initiate our analysis with a finite length $L$ (subsequently setting $L\to\infty$) and articulate the lattice realization of the single-particle $\mathcal{R}$-$\mathcal{T}$-internal symmetry group. In Sec.~\ref{Sec.1+1d_1copy_e} to Sec.~\ref{Sec:4copy_o}, we promote the operators to the quantum many-body Hilbert space and scrutinize the symmetry fractionalization patterns with 1, 2, and 4 copies. The anomalies within the fractionalized symmetry group preclude the realization of SMG. Ultimately, in Sec.~\ref{Sec.8copy_D}, we faithfully reproduce the original symmetry group and uncover explicit on-site gapping interactions. To extend our methodology to $2+1$d, $3+1$d, and higher dimensions, as presented in Secs.~\ref{Sec:2+1d_M}-\ref{Sec:SMG_M_higher}, we introduce the staggered fermion model as a lattice realization of the Majorana fermion. We ascertain that an invariant subgroup, isomorphic to the invariant group of the $(d-1)+1$ (spacetime) dimensional system, is present across all spacetime dimensions $d+1$. Moreover, by setting $L_i$ to be odd, the anomalies within the subgroup align with those of the invariant group of the $(d-1)+1$ dimensional system, facilitating the application of dimensional reduction for the classification of SMG. By applying this dimensional reduction method, we identify the $0+1$d $\mathbb{Z}_8$ anomaly for time-reversal symmetry across all dimensions. We demonstrate that every 8 copies of the staggered Majorana fermion can induce a symmetric mass through a specified on-site gapping interaction. The correspondence between lattice and free symmetry operators, as well as the classification of SMG, are tabulated in Sec.~\ref{Sec:dim_reduc} and Sec.~\ref{Sec:class}, respectively.

For Dirac fermions, we offer a parallel review of the $\mathcal{C}$-$\mathcal{R}$-$\mathcal{T}$ and internal symmetries for a free Dirac fermion within the single-particle framework, as outlined in Sec.~\ref{Sec:free_Dirac}. However, in Sec.~\ref{Sec.0+1d_D}, we determine that the $U(1)$ symmetry leads to a $\mathbb{Z}$ classification, obscuring the SMG mechanism. A more effective approach to probing SMG involves breaking the full $U(1)$ symmetry to descrete $\mathbb{Z}_4^F$ symmetry. Echoing the process for the Majorana case, we arrive at a $\mathbb{Z}_4$ SMG classification. In Sec.~\ref{Sec:1+1d_D}, we scrutinize the fractionalized symmetry group on a $1+1$d Dirac chain and obtain a $\mathbb{Z}_4$ classification for SMG. In Sec.~\ref{Sec:SMG_D_higher}, we reiterate the dimensional reduction approach to assess the SMG classification within the staggered fermion model for higher dimensions. The correspondence between lattice and free symmetry operators, along with the classification of SMG, are delineated in Sec.~\ref{Sec.dimred_D} and Sec.~\ref{Sec.class_D}, respectively.

\section{Symmetric Mass Generation for Majorana Fermion \label{sec:AlterBHZ}}

\subsection{Free Majorana Fermion Model \label{Sec:free_Maj}}

Before starting a lattice version of Majorana fermion theory, let's first see what the invariant group looks like for the free Majorana system. In order to write down a free massless Majorana Hamiltonian in $d+1$ dimensional spacetime, we need to find $d$ real Hermitian matrices $\alpha_i$ that anticommute with each other. By coupling these real matrices with momentum (i.e., $\mathrm{i}\partial_i$), we can write the Hamiltonian for free Majorana fermion as:
\begin{equation}
    H=\frac{1}{2}\int \dd^d x \chi^T \left(\sum_{i=1}^d \alpha_i \mathrm{i}\partial_i\right)\chi,\label{Eq:freebasis}
\end{equation}
where $\alpha_i$ is chosen in the real Clifford algebra $\mathcal{C}\ell (d,0)$ (see Appendix.~\ref{Ap. bastrans} for the explicit representation), and the corresponding vector space $\chi$ is a Majorana spinor with dim$_{\mathbb{R}}\chi_{\mathcal{C}\ell (d,0)}$ flavors. 

The orthogonal transformation of the Majorana spinor preserving the Clifford algebra of $\alpha_i$ defines the internal symmetry of the Majorana system. We can assign $\mathcal{R}$-$\mathcal{T}$ symmetry to the Hamiltonian. However, charge conjugation $\mathcal{C}$ is not defined for Majorana fermion since it's charge neutral. The reflection operator on the first direction $\mathcal{R}_1$ acts as

\begin{equation}
    \mathcal{R}_1\partial_i\mathcal{R}_1^{-1}=\begin{cases}
        -\partial_i & i=1,\\
        \partial_i & i\neq 1,
    \end{cases}
\end{equation}
and

\begin{equation}
    \mathcal{R}_1\chi\mathcal{R}_1^{-1}=M_{\mathcal{R}_1} \chi,
\end{equation}
where $M_{\mathcal{R}_1}\in O(\text{dim}_{\mathbb{R}}\mathcal{C}\ell(d,0))$ is defined as

\begin{equation}
    M_{\mathcal{R}_1}^\mathsf{T}\alpha_i M_{\mathcal{R}_1}=\begin{cases}
        -\alpha_i & i=1,\\
        \alpha_i & i\neq 1,
    \end{cases}
\end{equation}
and time-reversal operator $\mathcal{T}$ acts as

\begin{equation}
    \mathcal{T}\chi\mathcal{T}^{-1}=\mathcal{K}M_{\mathcal{T}}\chi,\quad \mathcal{T}\mathrm{i}\mathcal{T}^{-1}=-\mathrm{i},
\end{equation}
where $\mathcal{K}$ is complex conjugation acting on the components of matrices, and $M_{\mathcal{T}}\in O(\text{dim}_{\mathbb{R}}\mathcal{C}\ell(d,0))$ is defined as

\begin{equation}
    M_{\mathcal{T}}^\mathsf{T}\alpha_i M_{\mathcal{T}}=-\alpha_i,\quad \forall i=1,...,d.
\end{equation}

By combining the internal symmetry group and $\mathcal{R}$-$\mathcal{T}$ symmetry group, we can provide the invariant group $G_{\mathcal{C}\text{-}\mathcal{R}\text{-}\mathcal{T}\text{-internal}}$, listed in Tab.~\ref{tab: CPT-M} (we will still call the full invariant symmetry as $\mathcal{C}$-$\mathcal{R}$-$\mathcal{T}$-internal symmetry just to be general, although $\mathcal{C}$ is not defined form Majorana fermions). A systematic classification of the invariant group for free Majorana fermion in all spacetime dimensions are elaborated in Ref.~\cite{singleparticle}.

\begin{table*}
\renewcommand{\arraystretch}{1.5}
\centering
\caption{Clifford algebra $\mathcal{C}\ell (d,0)$, invariant group $G_{\mathcal{C}\text{-}\mathcal{R}\text{-}\mathcal{T}\text{-internal}}$ and symmetry operators for free Majorana fermion with space dimension $d=0,1,...,8$. The invariant group is 8-fold periodic. $\mathbb{Z}_2^F$, $\mathbb{Z}_2^\chi$, $\mathbb{Z}_2^{\mathcal{R}_1}$, $\mathbb{Z}_2^\mathcal{T}$ denotes fermion parity, fermion chiral symmetry, reflection symmetry on the first direction, and time-reversal symmetry, respectively. $\mathcal{J}_i$ denotes the generators of the corresponding Lie algebra.}
  \begin{tabular}{c|cc|ccccc}
  \hline
    $d$       & $\mathcal{C}\ell(d,0)$     & $G_{\mathcal{C}\text{-}\mathcal{R}\text{-}\mathcal{T}\text{-internal}}$    & $\mathbb{Z}_2^F$   & $\mathbb{Z}_2^\chi$  & Lie Algebra  & $\mathbb{Z}_2^{\mathcal{R}_1}$ & $\mathbb{Z}_2^{\mathcal{T}}$               \\ \hline
    0         & $\mathbb{R}(1)$                    & $\mathbb{Z}_2^F\times\mathbb{Z}_2^{\mathcal{T}}$                                   & $-1$               &                       &                                               &                           & $\mathcal{K}$                        \\
    1         & $\mathbb{R}(1)\oplus\mathbb{R}(1)$ & $\mathbb{D}_8^{\mathcal{T},\chi}\times\mathbb{Z}_2^{\mathcal{R}_1\mathcal{T}\chi}$ & $-\sigma^{0}$    & $\sigma^3$      &                                               & $\sigma^1$                & $\mathcal{K}\mathrm{i}\sigma^2$      \\
    2         & $\mathbb{R}(2)$                    & $\mathbb{D}_8^{\mathcal{T},\mathcal{R}_1}$                                         & $-\sigma^{0}$    &                       &                                               & $\sigma^3$                & $\mathcal{K}\mathrm{i}\sigma^2$      \\
    3         & $\mathbb{C}(2)$                    & $\text{Pin}_-^{\mathcal{T}}(2)\times_{\mathbb{Z}_2^F}\mathbb{Z}_4^{\mathcal{JR}_1\mathcal{T}}$      & $-\sigma^{00}$   &                       & $\sigma^{02}$                                 & $\sigma^{33}$             & $\mathcal{K}\mathrm{i}\sigma^{23}$   \\
    4         & $\mathbb{H}(2)$                    & $\text{Spin}(3)\times_{\mathbb{Z}_2^F}\mathbb{D}_8^{\mathcal{J}_1\mathcal{R}_1,\mathcal{J}_1\mathcal{T}}$    & $-\sigma^{000}$  &                       & ($\sigma^{002},\sigma^{021},\sigma^{023}$)    & $\sigma^{330}$  & $\mathcal{K}\mathrm{i}\sigma^{230}$            \\
    5         & $\mathbb{H}(2)\oplus\mathbb{H}(2)$ & $\text{Spin}(4)\rtimes_{\mathbb{Z}_2^F}\mathbb{Z}_4^{\mathcal{T}}\times\mathbb{Z}_2^{\mathcal{R}_1\mathcal{T}\chi}$   & $-\sigma^{0000}$ & $\sigma^{3000}$ & ($\sigma^{0002},\sigma^{0021},\sigma^{0023}$) & $\sigma^{1100}$ & $\mathcal{K}\mathrm{i}\sigma^{2000}$ \\
    6         & $\mathbb{H}(4)$                    & $\text{Spin}(3)\times_{\mathbb{Z}_2^F}\mathbb{D}_8^{\mathcal{T},\mathcal{R}_1}$    & $-\sigma^{0000}$ &                       & ($\sigma^{0002},\sigma^{0021},\sigma^{0023}$) & $\sigma^{3000}$           & $\mathcal{K}\mathrm{i}\sigma^{2000}$ \\
    7         & $\mathbb{C}(8)$                    & $\text{Pin}_+^{\mathcal{T}}(2)\times\mathbb{Z}_4^{\mathcal{JR}_1\mathcal{T}}$      & $-\sigma^{0000}$ &                       & $\sigma^{0002}$                               & $\mathrm{i}\sigma^{3023}$ & $\mathcal{K}\sigma^{2023}$           \\
    8         & $\mathbb{R}(16)$                   & $\mathbb{D}_8^{\mathcal{R}_1,\mathcal{T}}$                                         & $-\sigma^{0000}$ &                       &                                               & $\mathrm{i}\sigma^{3023}$ & $\mathcal{K}\sigma^{2023}$           \\
    \hline
    \end{tabular}\label{tab: CPT-M}
\end{table*}

\subsection{$0+1$d Symmetric Mass Generation}\label{Sec: 0+1d}

To obtain the SMG for $0+1$d spacetime, we need to promote our previous free model to the many-body version. A convenient choice for our Majorana operator basis is given by the Jordan-Wigner transformation, i.e. $\chi_{2n-1}=(\otimes_{\nu=1}^{n-1}\sigma^{3})\otimes \sigma^1$, $\chi_{2n}=(\otimes_{\nu=1}^{n-1}\sigma^{3})\otimes \sigma^2$, or, alternatively, $\mathcal{K}\chi_{\nu}\mathcal{K}=(-)^{\nu+1}\chi_{\nu}$. \footnote{Note that, unlike the complex conjugation in the free model, $\mathcal{K}$ here also conjugates the representation of Majorana operators in the second quantization level.}

In the $0+1$d case, we can assign the following symmetries analogous to those in the free model:

Fermion parity $\mathbb{Z}_2^F$ is generated by operator $(-)^F$ acting as:

\begin{equation}
    (-)^F\chi_{\nu}=-\chi_{\nu}(-)^F
\end{equation}
where $\nu$ labels Majorana operators of different copies.

Time-reversal symmetry $\mathbb{Z}_2^{\mathcal{T}}$ is generated by operator $\mathcal{T}$ acting as:

\begin{equation}
    \mathcal{T}\chi_{\nu}=\chi_{\nu}\mathcal{T},\ \mathcal{T}\mathrm{i}=-\mathrm{i}\mathcal{T}.
\end{equation}

For one copy of our Majorana system, we cannot pair them up to form an integer number of Dirac (complex) fermions. As discussed in some references~\cite{2024arXiv240104223F,2021PhRvL.127b6402T}, we can still consider the system of odd fermions as the boundary modes of one higher dimension, the fermion parity $\mathbb{Z}_2^F$ will rule out all possible interaction terms to drive the system gapped.

We then start with two copies of the Majorana system. The fermion parity operator $(-)^F$ is defined to flip the signs of Majorana operators, and one can easily check that 

\begin{equation}
    (-)^F=\mathrm{i}\chi_1\chi_2
\end{equation}
is a qualified operator that gives rise to a 2-fold symmetry. The time-reversal operator $\mathcal{T}$ is defined to apply complex conjugation and keep Majorana operators $\chi_\nu$ invariant. However, the complex conjugation operator $\mathcal{K}$ in many-body Hilbert space spontaneously acts on the representation of Majorana operators. Thus, in order to realize the action of $\mathcal{T}$, we should assign an additional $\chi_1$ operator to rectify the unnecessary sign change for $\chi_2$. In this sense, the time-reversal operator $\mathcal{T}$ is written in the form

\begin{equation}
    \mathcal{T}=\mathcal{K}\chi_1.
\end{equation}
However, the anti-commutation law between these two operators makes a non-Abelian invariant group $\mathbb{D}_8^{F\mathcal{T},\mathcal{T}}$. 
The $\mathbb{Z}_4^{F\mathcal{T}}$ subgroup ensures Kramers' degeneracy and rules out unique ground states. Alternatively, as we'll detail later, the fractionalized group structure shows the anomalous property, making it impossible to generate a symmetric mass gap.

With four copies of the Majorana system, we can similarly assign the operators of $\mathbb{Z}_2^F$ and $\mathbb{Z}_2^{\mathcal{T}}$ as 

\begin{equation}
    (-)^F=\chi_1\chi_2\chi_3\chi_4
\end{equation}
and 

\begin{equation}
    \mathcal{T}=\mathcal{K}\chi_2\chi_4.
\end{equation}
Unfortunately, the time-reversal symmetry here actually forms a $\mathbb{Z}_4^{\mathcal{T}}$ structure, and the invariant group is given by $\mathbb{Z}_2^F\times\mathbb{Z}_4^{\mathcal{T}}$. The 4-fold structure for time-reversal symmetry leads to Kramers' degeneracy, setting an obstruction for gapping.

By again doubling the copy number to eight copies, the anomalous characteristic for time-reversal symmetry cancels. Following the same process, we can assign 

\begin{equation}
    (-)^F=\prod_{\nu=1}^8\chi_\nu
\end{equation}
and

\begin{equation}
    \mathcal{T}=\mathcal{K}\prod_{\nu=even}\chi_\nu
\end{equation}
this time. The invariant group recovers the original $\mathbb{Z}_2^F\times\mathbb{Z}_2^{\mathcal{T}}$ structure, and we expect the existence of a symmetric gapping interaction.

Previous research~\cite{PhysRevB.81.134509,2015PhRvB..91l5147Y,PhysRevB.42.6523,2007APS..OSF.C2003T,PhysRevB.79.064515,PhysRevLett.103.010404,Berg2009,PhysRevB.82.134511,PhysRevB.85.245123,PhysRevB.95.241103,You_2014,PhysRevB.103.085130} shows copious possible symmetric interaction terms for eight copies, including Fidkowski-Kitaev interaction~\cite{PhysRevB.81.134509}, charge-4e superconducting interaction~\cite{PhysRevB.42.6523,2007APS..OSF.C2003T,PhysRevB.79.064515,PhysRevLett.103.010404,Berg2009,PhysRevB.82.134511,PhysRevB.85.245123,PhysRevB.95.241103} and spin-spin interaction~\cite{You_2014,PhysRevB.103.085130}. In this work, we provide interaction terms through direct searching of stabilizers. The result of straightforward searching is listed in the Appendix.~\ref{Ap.search_M}, and we can pick out four independent ones that commute with each other (i.e., four stabilizers) from these interaction terms. Notably, the interaction terms included in Ref.~\cite{PhysRevB.81.134509,2015PhRvB..91l5147Y,PhysRevB.42.6523,2007APS..OSF.C2003T,PhysRevB.79.064515,PhysRevLett.103.010404,Berg2009,PhysRevB.82.134511,PhysRevB.85.245123,PhysRevB.95.241103,You_2014,PhysRevB.103.085130} can be spanned by these stabilizers. For convenience, we'll choose $\chi_1\chi_2\chi_3\chi_4,$ $\chi_1\chi_2\chi_5\chi_6$, $\chi_1\chi_3\chi_5\chi_7$, $\chi_2\chi_3\chi_5\chi_8$ as our stabilizers to form an interaction term
\begin{equation}
H_{int}=\chi_1\chi_2\chi_3\chi_4+\chi_1\chi_2\chi_5\chi_6+\chi_1\chi_3\chi_5\chi_7+\chi_2\chi_3\chi_5\chi_8.\label{eq:int}
\end{equation}

Written in the qubit representation, these four stabilizers enjoy the same properties of $Z_1$, $Z_2$, $Z_3$, and $Z_4$, so we can easily obtain that the degeneracy of the eigenstates is 1, 4, 6, 4, 1 and the ground state is unique.

\subsection{$1+1$d Symmetric Mass Generation}\label{Sec:1+1d_M}

The lattice anomaly for $1+1$d Majorana chain has been carefully investigated in Ref.~\cite{10.21468/SciPostPhys.16.3.064}, and a $\mathbb{Z}_8$ anomaly for low-energy translational symmetry is found. However, the lattice anomaly in the high-energy limit possesses a richer structure, and the full information of anomaly cannot be directly extracted from the discussion of a single copy. In this section, we find a $\mathbb{Z}_8$ anomaly for time-reversal $\mathcal{T}$, which is not discernible in the sole discussion of one copy case. This $\mathbb{Z}_8$ anomaly is also discussed in Ref.~\cite{2025arXiv250817115S}.

\subsubsection{Free Majorana Chain and Lattice Realization}\label{Sec:1+1d_M_1}

In the continuum model, where we have assumed that the wavelength is far longer than the lattice constant, we can use 2 Majorana fermions to describe the Hamiltonian. Here, we set these two Majorana fermions as $\chi_L$ and $\chi_R$ describing left- and right-moving Majorana modes at low energy. The Hamiltonian of the free Majorana chain is then given by

\begin{equation}
  H=\frac{1}{2}\int \dd x (\chi_L \mathrm{i}\partial_x\chi_L-\chi_R \mathrm{i}\partial_x\chi_R).
\end{equation}

To realize the system on a lattice, we can set a Majorana chain with $L$ sites 
(assuming $L\to\infty$), with Majorana fermions $\chi_l$ ($l\in\mathbb{Z}_L$, 
$\chi_{L+l}\sim\pm\chi_{l}$) on each site, defined by

\begin{equation}
  \{\chi_l,\chi_{l'}\}=2\delta_{l,l'},
\end{equation}
with nearest-neighbor hopping Hamiltonian given by

\begin{equation}
  H=\frac{\mathrm{i}}{2}\sum_{l}\chi_l\chi_{l+1}.
\end{equation}
By transforming the Hamiltonian into the momentum space $\chi_k=\sum_{l}e^{\mathrm{i}kl}\chi_l$, we can rewrite the lattice Hamiltonian as

\begin{equation}
  H=\frac{1}{2}\sum_k\chi_{-k}(\sin k)\chi_k.
\end{equation}
The low-energy Majorana modes at $k=0,\pi$ corresponds to 

\begin{equation}
  \chi_R\sim \chi_{k=0}=\sum_l \chi_l,
\end{equation}

\begin{equation}
  \chi_L\sim \chi_{k=\pi}=\sum_l (-)^l \chi_l.
\end{equation}

\subsubsection{$\mathcal{C}$-$\mathcal{R}$-$\mathcal{T}$-Internal Symmetry in Continuum Model}\label{Sec:1+1d_M_2}

Before promoting our $\mathcal{C}$-$\mathcal{R}$-$\mathcal{T}$-internal symmetry to the lattice, we first briefly review their definition in the continuum limit. As demonstrated in Tab.~\ref{tab: CPT-M}, we can assign fermion parity $\mathbb{Z}_2^F$, fermion chiral symmetry $\mathbb{Z}_2^\chi$, reflection symmetry $\mathbb{Z}_2^{\mathcal{R}}$ and time-reversal symmetry $\mathbb{Z}_4^{\mathcal{T}}$ for $1+1$d Majorana chain.

In continuum model, the fermion parity $\mathbb{Z}_2^F$ is generated by operator $(-)^F$ defined to flip the signs of all Majorana modes:

\begin{equation}
  (-)^F\chi_R=-\chi_R(-)^F,\ (-)^F\chi_L=-\chi_L(-)^F.
\end{equation}

The fermion chiral symmetry $\mathbb{Z}_2^\chi$ is generated by $(-)^{F_L}$ defined to attach an additional sign on the left-moving mode:

\begin{equation}\label{Eq:chi}
  (-)^{F_L}\chi_R=\chi_R(-)^{F_L},\ (-)^{F_L}\chi_L=-\chi_L(-)^{F_L}.
\end{equation}

The reflection symmetry $\mathbb{Z}_2^{\mathcal{R}}$ is generated by $\mathcal{R}$ defined to inverse the coordinate and swap the left- and right-moving modes:

\begin{equation}\label{Eq:R}
  \mathcal{R}\chi_R=\chi_L\mathcal{R},\ \mathcal{R}\chi_L=\chi_R\mathcal{R},\ \mathcal{R}\partial_x=-\partial_x\mathcal{R}.
\end{equation}

The time-reversal symmetry $\mathbb{Z}_4^{\mathcal{T}}$ is generated by 
$\mathcal{T}$ defined to swap the left- and right-moving modes and do complex conjugation in the first quantization level:

\begin{equation}\label{Eq:T}
  \mathcal{T}\chi_R=-\chi_L\mathcal{T},\ \mathcal{T}\chi_L=\chi_R\mathcal{T},\ \mathcal{T}\mathrm{i}=-\mathrm{i}\mathcal{T}.
\end{equation}

In free fermion level, these symmetries form the invariant group $\mathbb{D}_8^{\mathcal{T},\chi}\times \mathbb{Z}_2^{\mathcal{RT}\chi}$.

\subsubsection{$\mathcal{C}$-$\mathcal{R}$-$\mathcal{T}$-Internal Symmetry in Lattice Model}\label{Sec. 1+1d lattice}

In Sec.~\ref{Sec: 0+1d}, we've not distinguished the invariant group of the continuum model and lattice model since the infrared (IR) and ultraviolet (UV) theory share a symmetry group. Nevertheless, in $1+1$d Majorana chain, we'll not find the 2-fold fermion chiral symmetry $\mathbb{Z}_2^\chi$. On the lattice, this symmetry is replaced by the translational symmetry $\mathbb{Z}_L^{T}$ which emanates to the low-energy $\mathbb{Z}_2^\chi$~\cite{10.21468/SciPostPhys.16.3.064}. 

To see this, we first set the translation symmetry $\mathbb{Z}_L^T$ generated by $T$. $T$ is defined by the action:

\begin{equation}
    T\chi_l=\chi_{l+1}T.
\end{equation}

When acting on low-energy modes $\chi_R$ and $\chi_L$, the action becomes

\begin{equation}
\begin{aligned}
    &T\chi_R T^{-1}=T\left(\sum_{l}\chi_l\right)T^{-1}=\chi_R,\\
    &T\chi_L T^{-1}=T\left(\sum_{l}(-)^l\chi_l\right)T^{-1}=-\chi_L,
\end{aligned}
\end{equation}
which is exactly the action of $\mathbb{Z}_2^\chi$ in Eq.~[\ref{Eq:chi}].

Other symmetries in the $1+1$d Majorana chain are similarly assigned. Reflection $\mathbb{Z}_2^{\mathcal{R}}$ is generated by $\mathcal{R}$ defined to reflect Majorana operators with the reflection center on the 0-th site:

\begin{equation}
    \mathcal{R}\chi_l=(-)^l\chi_{-l}\mathcal{R}.
\end{equation}

Acting on low-energy modes, the action of $\mathcal{R}$

\begin{equation}
\begin{aligned}
    &\mathcal{R}\chi_R \mathcal{R}^{-1}=\mathcal{R}\left(\sum_{l}\chi_l\right)\mathcal{R}^{-1}=\chi_L,\\
    &\mathcal{R}\chi_L \mathcal{R}^{-1}=\mathcal{R}\left(\sum_{l}(-)^l\chi_l\right)\mathcal{R}^{-1}=\chi_R,
\end{aligned}
\end{equation}
corresponds to low-energy $\mathcal{R}$ in Eq.~[\ref{Eq:R}].

We can also combine the translation $T$ and site reflection $\mathcal{R}$ to form bond reflection $T\mathcal{R}$ whose reflection center lies on the bond between the 0-th and 1-st sites. Fortunately, bond reflection spontaneously forms a $\mathbb{Z}_4^{T\mathcal{R}}$ symmetry group, so we can derive the operator of fermion parity $\mathbb{Z}_2^F$ by $(-)^F=(T\mathcal{R})^2$.

Finally, time-reversal symmetry $\mathbb{Z}_2^{\mathcal{T}}$ is generated by $\mathcal{T}$ defined as:

\begin{equation}
    \mathcal{T}\chi_l=(-)^l\chi_l\mathcal{T},\ \mathcal{T}\mathrm{i}=-\mathrm{i}\mathcal{T}.\label{Eq: TRS}
\end{equation}

Then we'll find the lattice operator $T\mathcal{T}$ has the same action on low-energy modes as low-energy time-reversion:

\begin{equation}
\begin{aligned}
    &T\mathcal{T}\chi_R (T\mathcal{T})^{-1}=T\mathcal{T}\left(\sum_{l}\chi_l\right)(T\mathcal{T})^{-1}=-\chi_L,\\
    &T\mathcal{T}\chi_L (T\mathcal{T})^{-1}=T\mathcal{T}\left(\sum_{l}(-)^l\chi_l\right)(T\mathcal{T})^{-1}=\chi_R.
\end{aligned}
\end{equation}

At lattice model level, the invariant group is no longer $\mathbb{D}_8^{\mathcal{T},\chi}\times \mathbb{Z}_2^{\mathcal{RT}\chi}$ of order 16, but instead group $G$ of order $4L$ ($G\cong\mathbb{D}_{2L}\times\mathbb{Z}_2\cong\mathbb{D}_{4L}$~\cite{GAP4}), with its presentation:

\begin{equation}
  \begin{aligned}
    &T^L=1,\ \mathcal{R}^2=1,\ \mathcal{T}^2=1,\ (-)^F=(T\mathcal{R})^2,\\
    &(-)^F(-)^F=1,\ \mathcal{R}T=(-)^FT^\dagger\mathcal{R},\\
    &\mathcal{T}T=(-)^FT\mathcal{T},\ \mathcal{RT}=\mathcal{TR}.
  \end{aligned}\label{Eq: 1dG}
\end{equation}

We can also derive other commutation relations from the given group presentation, which is partially listed in Appendix.~\ref{Ap. 1+1dchain}.

\subsubsection{Many-Body Symmetry with 1 Copy: Even-$L$}\label{Sec.1+1d_1copy_e}

With the lattice version of $\mathcal{C}$-$\mathcal{R}$-$\mathcal{T}$-internal symmetry group, we're ready to do the second quantization process and investigate the system in many-body Hilbert space.

To start with, we first focus on even-$L$ case  (with $L$ being the number of lattice sites) where reflection $\mathcal{R}$ is well-defined. \footnote{For odd-$L$, the $(-)^l$, $l\in\mathbb{Z}_L$ factor in the action of reflection is ill-defined.} For convenience, we set the condition $\mathcal{K}\chi_{\nu,l}\mathcal{K}=(-)^{\nu+1+l}\chi_{\nu,l}$, where $\nu=1,2,...$ labels different copies (here $\nu=1$) and $l=0,1,...$ labels the sites.\footnote{The $l$ dependent part of this condition is chosen to match the pattern of time-reversal symmetry in Eq.~[\ref{Eq: TRS}] and it will not affect the result if we change basis (e.g. chosen in real Clifford algebra $\mathcal{C}\ell(L,0)$). However, the $\nu$ dependent part is essential for our result, and for cases of $\nu=4$ that we'll discuss later, we cannot find $\nu$ real matrices as the on-site representation of $\chi_\nu$ in the Hilbert space of dimension $2^{\nu/2}$, i.e. 8=dim$_{\mathbb{R}}\chi_{\mathcal{C}\ell(4,0)}$>dim$_{\mathbb{R}}\chi_{\mathcal{C}\ell(4)}$=4.}

To assign translation $T$, we need to swap these Majorana operators on each site in sequence:

\begin{equation}
  T=\chi_0\prod_{l=0}^{L-2}\frac{1+\chi_l\chi_{l+1}}{\sqrt{2}},
\end{equation}
where $\chi_0$ here is to rectify the sign change of Majorana swapping.

Now, in the first quantization level, our given operator $T$ faithfully represents the action of translation on Majorana operators. However, we still have a phase arbitrariness for any unitary operator in the second quantization level, since $(e^{\mathrm{i}\theta}T)\chi_l(e^{\mathrm{i}\theta}T)^\dagger=T\chi_lT^\dagger$. We can fix this phase by requiring $T^L=1$ as in our first quantization model.

A straightforward calculation gives

\begin{equation}
  T^L=\begin{cases}
    1 & L=0,2\mod 8,\\
    -1 & L=4,6\mod 8,
  \end{cases}
\end{equation}
so we need to apply an extra phase factor to rectify this sign. The unitary operator $T$ of the translational symmetry $\mathbb{Z}_L^{T}$ is assigned as:

\begin{equation}
    T=e^{\frac{\mathrm{i}\pi}{8}(L-2)}\chi_0 \prod_{l=0}^{L-2} \frac{1+\chi_l\chi_{l+1}}{\sqrt{2}}.
\end{equation}

Following the same procedure, we can assign the rest of the symmetries:

Reflection symmetry $\mathbb{Z}_2^{\mathcal{R}}$ is generated by unitary operator $\mathcal{R}$:

\begin{equation}
  \mathcal{R}=e^{\frac{\mathrm{i}\pi}{16}(L-2)L}\chi_{L/2}^{L/2-1}\left(\prod_{l=0}^{L/2-1}\chi_l\right)\prod_{l=1}^{L/2-1}\frac{1+(-)^l\chi_l\chi_{-l}}{\sqrt{2}}.
\end{equation}

Fermion parity $\mathbb{Z}_2^F$ is generated by unitary operator $(-)^F$\footnote{As discussed in Sec.~\ref{Sec. 1+1d lattice}, we only have 3 independent operators $T$, $\mathcal{R}$ and $\mathcal{T}$, so the phase factor of $(-)^F$ is determined by the phases of $T$ and $\mathcal{R}$.}:

\begin{equation}
  (-)^F=(T\mathcal{R})^2=\mathrm{i}^{L/2+1}\prod_{l=0}^{L-1}\chi_l.
\end{equation}

Time-reversal symmetry $\mathbb{Z}_2^{\mathcal{T}}$ is generated by anti-unitary operator $\mathcal{T}$:

\begin{equation}
  \mathcal{T}=\mathcal{K}.
\end{equation}

Equipped with all these symmetry operators, we can check if we reproduce our expected invariant group in Eq.~[\ref{Eq: 1dG}], and find that the invariant group is realized projectively \footnote{Throughout this article,
we use the red color phase factor 
${\color{red}-1}$, ${\color{red} \pm  \ii}$, 
${\color{red} \exp( \ii \theta(L))}$
etc. where $\theta(L)$ could have the system size 
$L$ dependence
to
indicate the source of the anomaly of the lattice symmetry.
}:

\begin{equation}
  \begin{aligned}
    &T^L=1,\ \mathcal{R}^2=1,\ \mathcal{T}^2=1,\ (-)^F=(T\mathcal{R})^2,\\
    &(-)^F(-)^F={\color{red}-1},\ \mathcal{R}T={\color{red}-}(-)^FT^\dagger\mathcal{R},\\
    &\mathcal{T}T={\color{red}-}(-)^FT\mathcal{T},\ \mathcal{RT}=\mathcal{TR}.
  \end{aligned}
\end{equation}

In conclusion, we have

\begin{itemize}
    \item $\mathbb{Z}_2$ anomaly of fermion parity $(-)^F$.
    \item mixed $\mathbb{Z}_2$ anomaly between translational symmetry $\mathbb{Z}_L^T$ and reflection symmetry $\mathbb{Z}_2^{\mathcal{R}}$,
    \item mixed $\mathbb{Z}_2$ anomaly between translational symmetry $\mathbb{Z}_L^T$ and time-reversal symmetry $\mathbb{Z}_2^{\mathcal{T}}$.
\end{itemize}

More anomalies lying in the commutation relations are listed in Appendix.~\ref{Ap. 1+1dchain}. The anomalies will form an obstruction towards gapping~\cite{10.21468/SciPostPhys.16.3.064}.

\subsubsection{Many-Body Symmetry with 1 Copy: Odd-$L$}

In odd-$L$ case, reflection $\mathcal{R}$ is ill-defined, but it still suffices to define the symmetries in the $0+1$d invariant group (i.e. fermion parity $\mathbb{Z}_2^F$ and time-reversal symmetry $\mathbb{Z}_2^{\mathcal{T}}$). We'll only focus on this subgroup, and for more detailed discussions of other symmetries like translation $T$, see Ref.~\cite{10.21468/SciPostPhys.16.3.064}.

With the lattice number odd, the previously defined time-reversion $\mathcal{T}$ acts on the Hamiltonian as

\begin{equation}
  \mathcal{T}H\mathcal{T}^{-1}=\frac{\mathrm{i}}{2}\sum_{l=0}^{L-2}\chi_l\chi_{l+1}-\frac{\mathrm{i}}{2}\chi_{L-1}\chi_{0}\neq H,
\end{equation}
which is instead the twisted version of Hamiltonian $H_{tw}$, with an effective anti-periodic boundary condition.

To discuss time-reversal symmetry in our framework, we need to consider the enlarged Hilbert space with enlarged Hamiltonian

\begin{equation}
    \tilde{H}=H\oplus H_{tw}=\begin{pmatrix}
        H&0\\
        0&H_{tw}
    \end{pmatrix},
\end{equation}
and the time-reversion $\mathcal{T}$ becomes an off-diagonal operator. This time, we can assign on-site symmetries in the enlarged Hilbert space:

Fermion parity $\mathbb{Z}_2^F$ is generated by unitary operator $(-)^F$:

\begin{equation}
  (-)^F=\chi_L\otimes \sigma^0=\begin{pmatrix}
      \chi_L&0\\
      0&\chi_L
  \end{pmatrix},
\end{equation}
with $\chi_L$ an extra imaginary Majorana operator that anti-commutes with all other $\chi_l$, $l=0,1,...,L-1$. We can always find such $\chi_L$ since for odd-$L$, $\mathcal{C}\ell(L)$ and $\mathcal{C}\ell(L+1)$ have the same dimensions.

Time-reversal symmetry $\mathbb{Z}_2^{\mathcal{T}}$ is generated by anti-unitary operator $\mathcal{T}$:

\begin{equation}
  \mathcal{T}=\mathcal{K}\otimes \sigma^1=\begin{pmatrix}
      0&\mathcal{K}\\
      \mathcal{K}&0
  \end{pmatrix}.
\end{equation}

The anti-commuting law between $(-)^F$ and $\mathcal{T}$ gives the projectively realized invariant subgroup $\mathbb{D}_8^{F\mathcal{T},\mathcal{T}}$. 

In conclusion, we have

\begin{itemize}
    \item mixed $\mathbb{Z}_2$ anomaly between fermion parity $\mathbb{Z}_2^F$ and time-reversal symmetry $\mathbb{Z}_2^{\mathcal{T}}$.
\end{itemize}

\subsubsection{Many-Body Symmetry with 2 Copies: Even-$L$}\label{Sec:2even}

Since translation $T$ and reflection $\mathcal{R}$ are all fermionic operators (i.e. containing an odd number of Majorana operators), they will act on the other copy and provide an additional minus sign. Due to this reason, the operators of 2 copies are not simple multiplication of operators for each copy. Instead, they have the following form:

Translational symmetry $\mathbb{Z}_L^T$ is generated by unitary operator $T$:

\begin{equation}
  T=e^{\frac{\mathrm{i}\pi}{4}(L-4)}\prod_{\nu=1}^2 \chi_{\nu,0}\prod_{l=0}^{L-2}\frac{1-\chi_{\nu,l}\chi_{\nu,l+1}}{\sqrt{2}}.
\end{equation}

Reflection symmetry $\mathbb{Z}_2^{\mathcal{R}}$ is generated by unitary operator $\mathcal{R}$:

\begin{equation}
  \mathcal{R}=\mathrm{i}\prod_{\nu=1}^2\chi_{\nu,L/2}^{L/2}\left(\prod_{l=L/2+1}^{L-1}\chi_{\nu,l}\right)\prod_{l=1}^{L/2-1}\frac{1+(-)^l\chi_{\nu,l}\chi_{\nu,-l}}{\sqrt{2}}.
\end{equation}

Fermion parity $\mathbb{Z}_2^F$ is generated by unitary operator $(-)^F$:

\begin{equation}
  (-)^F=(T\mathcal{R})^2=(-)^{L/2-1}\prod_{\nu=1}^2\prod_{l=0}^{L-1}\chi_{\nu,l}.
\end{equation}

Time-reversal symmetry $\mathbb{Z}_2^{\mathcal{T}}$ is generated by anti-unitary operator $\mathcal{T}$:

\begin{equation}
  \mathcal{T}=\mathcal{K}\prod_{l=0}^{L-1}\chi_{2,l}.
\end{equation}

Unfortunately, we still have a projectively realized invariant group with presentation:

\begin{equation}
  \begin{aligned}
    &T^L=1,\ \mathcal{R}^2=1,\ \mathcal{T}^2=1,\ (-)^F=(T\mathcal{R})^2,\\
    &(-)^F(-)^F=1,\ \mathcal{R}T=(-)^FT^\dagger\mathcal{R},\\
    &\mathcal{T}T=(-)^FT\mathcal{T},\ \mathcal{RT}=\mathcal{TR}.
  \end{aligned}
\end{equation}

As expected, the $\mathbb{Z}_2$ anomaly in Sec.~\ref{Sec.1+1d_1copy_e} cancels, and we reproduce the original invariant group in Eq.~\ref{Eq: 1dG}.

\subsubsection{Many-Body Symmetry with 2 Copies: Odd-$L$}

Though for the even-$L$ case, we successfully reproduce the original invariant group without anomalies, we'll find in this section that we're still in an anomalous theory with 2 copies. We similarly set symmetries in the $0+1$d subgroup within the enlarged Hilbert space $\tilde{\mathcal{H}}=\mathcal{H}\oplus \mathcal{H}_{tw}$ as follows:

Fermion parity $\mathbb{Z}_2^F$ is generated by unitary operator $(-)^F$:

\begin{equation}
  (-)^F=\left(\mathrm{i}\prod_{\nu=1}^2\prod_{l=0}^{L-1}\chi_{\nu,l}\right)\otimes\sigma^0.
\end{equation}

Time-reversal symmetry $\mathbb{Z}_2^{\mathcal{T}}$ is generated by anti-unitary operator $\mathcal{T}$:

\begin{equation}
  \mathcal{T}=\left(\mathcal{K}\prod_{l=0}^{L-1}\chi_{1,l}\right)\otimes \sigma^1.
\end{equation}

Again, the anti-commuting law between $(-)^F$ and $\mathcal{T}$ gives rise to the projective invariant subgroup $\mathbb{D}_8^{F\mathcal{T},\mathcal{T}}$. Remarkably, we find that the invariant subgroup shows the same anomaly as we've seen in the $0+1$d case discussed in Sec.~\ref{Sec: 0+1d} with two copies.

In conclusion, we have

\begin{itemize}
    \item mixed $\mathbb{Z}_2$ anomaly between fermion parity $\mathbb{Z}_2^F$ and time-reversal symmetry $\mathbb{Z}_2^{\mathcal{T}}$.
\end{itemize}

\subsubsection{Many-Body Symmetry with 4 Copies: Even-$L$}\label{Sec:4copy_e}

With all operators discussed in Sec.~\ref{Sec:2even} bosonic operators, we can simply duplicate them as:

Translational symmetry $\mathbb{Z}_L^T$ is generated by unitary operator $T$:

\begin{equation}
  T= \prod_{\nu=1}^4\chi_{\nu,0}\prod_{l=0}^{L-2}\frac{1-\chi_{\nu,l}\chi_{\nu,l+1}}{\sqrt{2}}.
\end{equation}

Reflection symmetry $\mathbb{Z}_2^{\mathcal{R}}$ is generated by unitary operator $\mathcal{R}$:

\begin{equation}
  \mathcal{R}=\prod_{\nu=1}^4\chi_{\nu,L/2}^{L/2}\left(\prod_{l=L/2+1}^{L-1}\chi_{\nu,l}\right)\prod_{l=0}^{L/2-1}\frac{1+(-)^l\chi_{\nu,l}\chi_{\nu,-l}}{\sqrt{2}},
\end{equation}

Fermion parity $\mathbb{Z}_2^F$ is generated by unitary operator $(-)^F$:

\begin{equation}
  (-)^F=(T\mathcal{R})^2=\prod_{\nu=1}^4\prod_{l=0}^{L-1}\chi_{\nu,l}.
\end{equation}

Time-reversal symmetry $\mathbb{Z}_2^{\mathcal{T}}$ is generated by anti-unitary operator $\mathcal{T}$:

\begin{equation}
  \mathcal{T}=\mathcal{K}\prod_{\nu=1}^2\prod_{l=0}^{L-1}\chi_{2\nu,l}.
\end{equation}

As expected, we reproduce the original invariant group in Eq.~\ref{Eq: 1dG}.

\subsubsection{Many-Body Symmetry with 4 Copies: Odd-$L$}\label{Sec:4copy_o}

Again, we can assign symmetries in the $0+1$d invariant subgroup:

Fermion parity $\mathbb{Z}_2^F$ is generated by unitary operator $(-)^F$:

\begin{equation}
  (-)^F=\left(\prod_{\nu=1}^4\prod_{l=0}^{L-1}\chi_{\nu,l}\right)\otimes\sigma^0.
\end{equation}

Time-reversal symmetry $\mathbb{Z}_2^{\mathcal{T}}$ is generated by anti-unitary operator $\mathcal{T}$:

\begin{equation}
  \mathcal{T}=\left(\mathcal{K}\prod_{\nu=1}^2\prod_{l=0}^{L-1}\chi_{2\nu,l}\right)\otimes \sigma^1.
\end{equation}

The invariant subgroup spanned by these symmetries is $\mathbb{Z}_2^F\times\mathbb{Z}_4^{\mathcal{T}}$ since $\mathcal{T}^2=-1$. We again show that the invariant subgroup for the odd-$L$ case coincides with that of the $0+1$d system.

In conclusion, we have

\begin{itemize}
    \item $\mathbb{Z}_2$ anomaly for time-reversal symmetry $\mathbb{Z}_2^{\mathcal{T}}$.
\end{itemize}

\subsubsection{Many-Body Symmetry with 8 Copies}\label{Sec.8copy_D}

By doubling the system discussed in Sec.~\ref{Sec:4copy_e} and Sec.~\ref{Sec:4copy_o}, the $\mathbb{Z}_2$ anomalies successfully cancel, and there are no obstructions towards gapping. Indeed, we again set the interaction found in Sec.~\ref{Sec: 0+1d} on each site of the Majorana chain. The interaction term is written as

\begin{equation}
\begin{aligned}
  H_{int}=&\sum_{l} \chi_{1,l}\chi_{2,l}\chi_{3,l}\chi_{4,l}+\chi_{1,l}\chi_{2,l}\chi_{5,l}\chi_{6,l}\\
  &+\chi_{1,l}\chi_{3,l}\chi_{5,l}\chi_{7,l}+\chi_{2,l}\chi_{3,l}\chi_{5,l}\chi_{8,l}.
\end{aligned}
\end{equation}

\subsection{$2+1$d Symmetric Mass Generation}\label{Sec:2+1d_M}

\subsubsection{Staggered Fermion Model}

\begin{figure}[h]
\includegraphics[width=.7\linewidth]{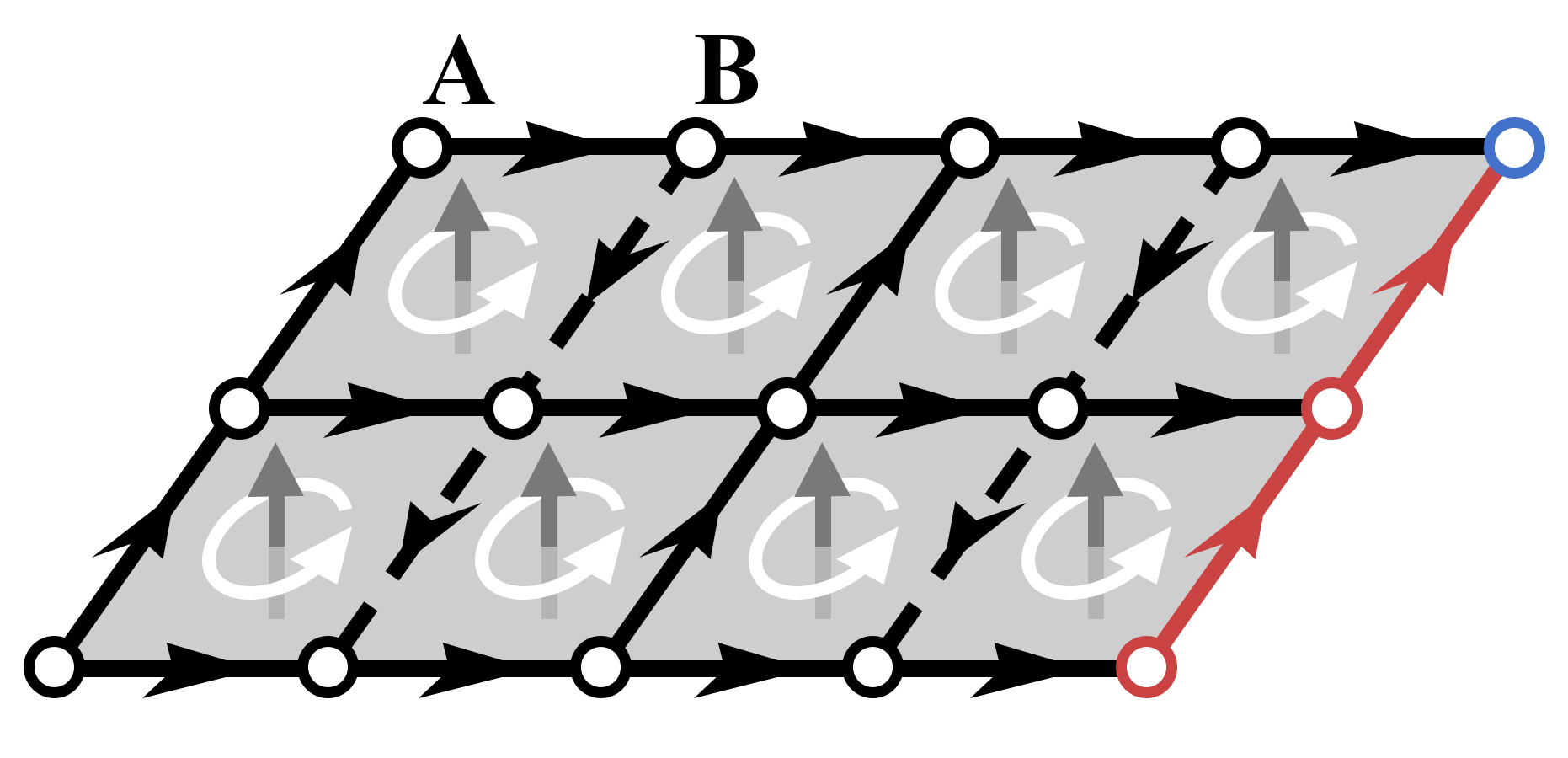}
    
\caption{Staggered fermion model on $2d$ lattice with staggered hopping amplitudes. There is an effective $\pi$-flux in each plaquette. Each unit cell contains A and B sublattices. By setting $L_1$ odd, the anomalies in the 1d chain (shaded in red) are exposed. By further setting $L_2$ odd, the anomalies in the 0d site (shaded in blue) are exposed.
}
\label{fig:2d_lattice}
\end{figure}

For higher space dimension $d>1$, a convenient choice of lattice realization for Majorana fermion is the staggered fermion model, with the staggered Hamiltonian

\begin{equation}
    H=\frac{\mathrm{i}}{2}\sum_{l,\hat{\mu}}(-)^{\sum_{\nu<\mu}l_\nu}\chi_l\chi_{l+\hat{\mu}},
\end{equation}
with the same hopping amplitude in the first direction, staggered hopping amplitude $(-)^{l_1}$ in the second direction, etc. Here $\mu=1,2,...,d$ denotes the spatial direction, and $\hat{\mu}$ denotes the unit vector on the $\mu$-th direction. $l$ is a $d$-component vector describing the coordinates of sites. By applying staggered phase $(-)^{\sum_{\nu<\mu}l_\nu}$, we've actually inserted a $\pi$-flux in each plaquette, and the unit cell is extended to $2\times...\times 2\times 1$, with sublattices $(0,...,0,0,0)$, $(0,...,0,1,0)$, $(0,...,1,0,0)$, $(0,...,1,1,0)$, ..., $(1,...,1,1,0)$. This sublattice pattern ensures that we can minimize the fermion doubling problem by reducing its Brillouin zone. Now the only doubler is the two Weyl points $(...,0,0)$ and $(...,0,\pi)$.

We can further Fourier transform our Hamiltonian into momentum space $\chi_{k,i}=\sum_{l\mod 2^{d-1}=i}e^{\mathrm{i}k\cdot l}\chi_l$ ($i$ labels sublattices) and write a vector $\chi_{k}$ that collects these $\chi_{k,i}$ components. The Hamiltonian can then be written as

\begin{equation}
    H=\sum_k \chi_k^\dagger \left(\sum_{i=1}^d\sin k_i\alpha_i\right)\chi_k,\label{Eq:stagbasis}
\end{equation}
with $\alpha_i$ the staggered matrices of dimension $2^{d-1}$, given by $\alpha_1=\sigma^{10...00}$, $\alpha_2=\sigma^{31...00}$, ..., $\alpha_{d-1}=\sigma^{33...31}$, and $\alpha_d=\sigma^{33...33}$.

The total number of components for the staggered fermion model (including two Weyl points) is $2^d$, and the number of components for the root states written as the representation of Clifford algebra is given by dim$_{\mathbb{R}}\chi_{\mathcal{C}\ell(d,0)}$. Therefore, the staggered fermion model intrinsically gives $\nu_{stag}=2^d/$dim$_{\mathbb{R}}\chi_{\mathcal{C}\ell(d,0)}$ copies of root states. The intrinsic doubling is demonstrated in Tab.~\ref{tab: stag}.

\begin{table}[h]
\renewcommand{\arraystretch}{1.5}
\centering
\caption{The minimum real degree of freedom for free Majorana fermion in $d$ spatial dimension is calculated by dim$_{\mathbb{R}}\chi_{\mathcal{C}\ell(d,0)}$, and the real degree of freedom ($2^{d-1}$ sublattices and $2$ Weyl points) for staggered Majorana fermion is $2^d$. The intrinsic doubling for staggered fermion is $\nu_{stag}(d)=2^d/$dim$_{\mathbb{R}}\chi_{\mathcal{C}\ell(d,0)}$ in different spatial dimensions.}
  \begin{tabular}{c|cc|c}
  \hline
    $d$       & $\mathcal{C}\ell(d,0)$     & $\text{dim}_{\mathbb{R}}\chi_{\mathcal{C}\ell(d,0)}$    & $\nu_{stag}(d)$       \\ \hline
    1         & $\mathbb{R}(1)\oplus\mathbb{R}(1)$       & 2                                                     & 1                     \\
    2         & $\mathbb{R}(2)$                          & 2                                                     & 2                     \\
    3         & $\mathbb{C}(2)$                          & 4                                                     & 2                     \\
    4         & $\mathbb{H}(2)$                          & 8                                                     & 2                     \\
    5         & $\mathbb{H}(2)\oplus\mathbb{H}(2)$       & 16                                                    & 2                     \\
    6         & $\mathbb{H}(4)$                          & 16                                                    & 4                     \\
    7         & $\mathbb{C}(8)$                          & 16                                                    & 8                     \\
    8         & $\mathbb{R}(16)$                         & 16                                                    & 16                    \\
$d$+8     & $\mathbb{R}(16)\otimes\mathcal{C}\ell(d,0)$  & 16$\times\text{dim}_{\mathbb{R}}\chi_{\mathcal{C}\ell(d,0)}$ & 16$\times\nu_{stag}(d)$\\
\hline
    \end{tabular}\label{tab: stag}
\end{table}

\subsubsection{$\mathcal{C}$-$\mathcal{R}$-$\mathcal{T}$-Internal Symmetry in Lattice Model}\label{Sec. 2+1d lattice}

To be more specific, we denote the Majorana operators on different sublattices ($A(0,0)$ and $B(1,0)$) as $\chi_{A;l_1,l_2}=\chi_{2l_1,l_2}$ and $\chi_{B;l_1,l_2}=\chi_{2l_1+1,l_2}$. The Hamiltonian is rewritten as

\begin{equation}
\begin{aligned}
    H=&\frac{\mathrm{i}}{2}\sum_{l_1,l_2}\left(\chi_{A;l_1,l_2}\chi_{B;l_1,l_2}+\chi_{B;l_1,l_2}\chi_{A;l_1+1,l_2}\right)\\
    +&\frac{\mathrm{i}}{2}\sum_{l_1,l_2}\left(\chi_{A;l_1,l_2}\chi_{A;l_1,l_2+1}-\chi_{B;l_1,l_2}\chi_{B;l_1,l_2+1}\right).
\end{aligned}
\end{equation}

The Majorana operators satisfy

\begin{equation}
    \{\chi_{\nu,l},\chi_{\nu',l'}\}=2\delta_{\nu,\nu'}\delta_{l,l'}
\end{equation}
and

\begin{equation}
    \mathcal{K}\chi_{\nu,l}\mathcal{K}=(-)^{\nu+1+\sum_{i=1}^2l_i}\chi_{\nu,l}.
\end{equation}
where $\nu$ labels different copies and $l$ labels sites.

With staggered Hamiltonian, the assignment for symmetry operators should involve staggered phases to preserve the Hamiltonian:

We first assign translational symmetries on $x-$ and $y-$ directions $\mathbb{Z}_{L_i}^{T_i}$ ($i=1,2$), generated by unitary operators $T_1$ and $T_2$ acting as:

\begin{equation}
\begin{aligned}
    &T_1 \chi_{A;l_1,l_2}=(-)^{l_2}\chi_{B;l_1,l_2}T_1,\\
    &T_1 \chi_{B;l_1,l_2}=(-)^{l_2} \chi_{A;l_1+1,l_2}T_1,
\end{aligned}
\end{equation}

\begin{equation}
\begin{aligned}
    &T_2 \chi_{A;l_1,l_2}=\chi_{A;l_1,l_2+1}T_2,\\
    &T_2 \chi_{B;l_1,l_2}=\chi_{B;l_1,l_2+1}T_2.
\end{aligned}
\end{equation}

Reflection symmetries on $x-$ and $y-$ directions $\mathbb{Z}_2^{\mathcal{R}_i}$ is generated by unitary operators $\mathcal{R}_1$ and $\mathcal{R}_2$ acting as:

\begin{equation}
\begin{aligned}
    &\mathcal{R}_1 \chi_{A;l_1,l_2}=\chi_{A;-l_1,l_2}\mathcal{R}_1,\\
    &\mathcal{R}_1 \chi_{B;l_1,l_2}=-\chi_{B;-l_1-1,l_2}\mathcal{R}_1,
\end{aligned}
\end{equation}

\begin{equation}
\begin{aligned}
    &\mathcal{R}_2 \chi_{A;l_1,l_2}=(-)^{l_2}\chi_{A;l_1,-l_2}\mathcal{R}_2,\\
    &\mathcal{R}_2 \chi_{B;l_1,l_2}=(-)^{l_2}\chi_{B;l_1,-l_2}\mathcal{R}_2.
\end{aligned}
\end{equation}

Finally, we can assign time-reversal symmetry $\mathbb{Z}_2^{\mathcal{T}}$, generated by anti-unitary operator $\mathcal{T}$ acting as:

\begin{equation}
\begin{aligned}
    \mathcal{T}\chi_{A;l_1,l_2}&=(-)^{l_2}\chi_{A;l_1,l_2}\mathcal{T},\\
    \mathcal{T}\chi_{B;l_1,l_2}&=(-)^{l_2+1}\chi_{B;l_1,l_2}\mathcal{T},\\
    \mathcal{T}\mathrm{i}&=-\mathrm{i}\mathcal{T}.
\end{aligned}
\end{equation}

By checking the action of these operators on low-energy modes, we can find the correspondence of low-energy symmetry on the lattice:

Low-energy fermion parity generator $(-)^F$ corresponds to the lattice fermion parity generator $(-)^F$, which can also be written in lattice translation and reflection operators as $(T_1\mathcal{R}_1)^2$ or $(T_2\mathcal{R}_2)^2$.

Low-energy reflection generator $\mathcal{R}_1$ and $\mathcal{R}_2$ correspond to lattice $\mathcal{R}_1$ and $\mathcal{R}_2$, while low-energy time-reversion operator $\mathcal{T}$ correspond to lattice $T_1\mathcal{T}$.

In lattice model level, the invariant group is $G$ of order $8L_1L_2$ ($G\cong \mathbb{D}_{2L_1}\times\mathbb{D}_{2L_1}\times\mathbb{Z}_2$\cite{GAP4}), with its presentation:

\begin{equation}
  \begin{aligned}
    &T_i^{L_i}=1,\ \mathcal{R}_i^2=1,\ \mathcal{T}^2=1,\ (-)^F=(T_i\mathcal{R}_i)^2,\\
    &(-)^F(-)^F=1,\ \mathcal{R}_iT_i=(-)^FT_i^\dagger\mathcal{R}_i,\\
    &\mathcal{T}T_i=(-)^FT_i\mathcal{T},\ \mathcal{R}_i\mathcal{T}=\mathcal{TR}_i,\ \forall i=1,2,\\
    &T_iT_j=(-)^FT_jT_i,\ \mathcal{R}_i\mathcal{R}_j=\mathcal{R}_j\mathcal{R}_i,\\
    &\mathcal{R}_iT_j=T_j\mathcal{R}_i,\ \forall i\neq j.
  \end{aligned}\label{Eq: G_2}
\end{equation}

\subsubsection{Many-Body Symmetry with 2 Copies: Even-$L_1$, Even-$L_2$}

We start our discussion with $L_1$ and $L_2$ even, where all reflection symmetries $\mathbb{Z}_2^{\mathcal{R}_i}$ ($i=1,2$) is well-defined. With the intrinsic doubling $\nu_{stag}(2)=2^2/\text{dim}_{\mathbb{R}}\chi_{\mathcal{C}\ell(2,0)}=2$, we automatically start with 2 copies on the lattice model.

Translational symmetry on $x-$ and $y-$ directions $\mathbb{Z}_{L_i}^{T_i}$ is generated by unitary operators $T_1$ and $T_2$ defined as:

\begin{equation}
    T_1=e^{\frac{\mathrm{i}\pi}{8}(L_1-4)L_2}\prod_{l_2}\chi_{0,l_2}\prod_{l_1=0}^{L_1-2}\frac{1-(-)^{l_2}\chi_{l_1,l_2}\chi_{l_1+1,l_2}}{\sqrt{2}},
\end{equation}

\begin{equation}
    T_2=e^{\frac{\mathrm{i}\pi}{8}(L_2-4)L_1}\prod_{l_1}\chi_{l_1,0}\prod_{l_2=0}^{L_2-2}\frac{1-\chi_{l_1,l_2}\chi_{l_1,l_2+1}}{\sqrt{2}}.
\end{equation}

Reflection symmetry on $x-$ and $y-$ directions $\mathbb{Z}_2^{\mathcal{R}_i}$ is generated by unitary operators $\mathcal{R}_1$ and $\mathcal{R}_2$ defined as:

\begin{equation}
\begin{aligned}
    \mathcal{R}_1=&\mathrm{i}^{L_2/2}\prod_{l_2}\chi_{L_1/2,l_2}^{L_1/2}\left(\prod_{l_1=L_1/2+1}^{L_1-1}\chi_{l_1,l_2}\right)\\
    &\prod_{l_1=1}^{L_1/2-1}\frac{1+(-)^{l_1}\chi_{l_1,l_2}\chi_{-l_1,l_2}}{\sqrt{2}},
\end{aligned}
\end{equation}

\begin{equation}
\begin{aligned}
    \mathcal{R}_2=&\mathrm{i}^{L_1/2}\prod_{l_1}\chi_{l_1,L_2/2}^{L_2/2}\left(\prod_{l_2=L_2/2+1}^{L_2-1}\chi_{l_1,l_2}\right)\\
    &\prod_{l_2=1}^{L_2/2-1}\frac{1+(-)^{l_2}\chi_{l_1,l_2}\chi_{l_1,-l_2}}{\sqrt{2}}.
\end{aligned}
\end{equation}

Fermion parity $\mathbb{Z}_2^F$ is generated by unitary operator $(-)^F$ defined as:

\begin{equation}
    (-)^F=(T_1\mathcal{R}_1)^2=(-)^{(L_1/2+1)L_2/2}\prod_{l_1,l_2}\chi_{l_1,l_2}.
\end{equation}

Time-reversal symmetry $\mathbb{Z}_2^{\mathcal{T}}$ is generated by anti-unitary operator $\mathcal{T}$ defined as:

\begin{equation}
    \mathcal{T}=\mathcal{K}.
\end{equation}

Equipped with all these symmetry operators, we can find that the invariant group in Eq.~(\ref{Eq: G_2}) is realized projectively:

\begin{equation}
  \begin{aligned}
    &T_1^{L_1}=1,\ T_2^{L_2}=1,\ \mathcal{R}_1^2=1,\ \mathcal{R}_2^2=1,\ \mathcal{T}^2=1,\\
    &(-)^F=(T_1\mathcal{R}_1)^2={\color{red}(-)^{(L_1+L_2)/2}}(T_2\mathcal{R}_2)^2,\ (-)^F(-)^F=1\\
    &\mathcal{R}_1T_1=(-)^FT_1^\dagger\mathcal{R}_1,\ \mathcal{R}_2T_2={\color{red}(-)^{(L_1+L_2)/2}}(-)^FT_2^\dagger\mathcal{R}_2,\\
    &\mathcal{T}T_1=(-)^FT_1\mathcal{T},\ \mathcal{T}T_2={\color{red}(-)^{(L_1+L_2)/2}}(-)^FT_2\mathcal{T},\\
    &\mathcal{R}_1\mathcal{T}=\mathcal{TR}_1,\ \mathcal{R}_2\mathcal{T}=\mathcal{TR}_2,\\
    &T_1T_2={\color{red}(-)^{(L_1/2+1)L_2/2+1}}(-)^FT_2T_1,\ \mathcal{R}_1\mathcal{R}_2={\color{red}-}\mathcal{R}_2\mathcal{R}_1,\\
    &\mathcal{R}_1T_2={\color{red}-}T_2\mathcal{R}_1,\ \mathcal{R}_2T_1={\color{red}-}T_1\mathcal{R}_2.
  \end{aligned}
\end{equation}

In conclusion, we have

\begin{itemize}
    \item $\mathbb{Z}_2$ anomaly for fermion parity $\mathbb{Z}_2^F$,
    \item mixed $\mathbb{Z}_2$ anomaly between translational symmetry $\mathbb{Z}_{L_2}^{T_2}$ and reflection symmetry $\mathbb{Z}_2^{\mathcal{R}_2}$,
    \item mixed $\mathbb{Z}_2$ anomaly between translational symmetry $\mathbb{Z}_{L_2}^{T_2}$ and time-reversal symmetry $\mathbb{Z}_2^{\mathcal{T}}$,
    \item mixed $\mathbb{Z}_2$ anomaly between translational symmetries $\mathbb{Z}_{L_1}^{T_1}$ and $\mathbb{Z}_{L_2}^{T_2}$.
    \item mixed $\mathbb{Z}_2$ anomaly between reflection symmetries $\mathbb{Z}_{2}^{\mathcal{R}_1}$ and $\mathbb{Z}_{2}^{\mathcal{R}_2}$,
    \item mixed $\mathbb{Z}_2$ anomaly between reflection symmetry $\mathbb{Z}_{2}^{\mathcal{R}_1}$ and translational symmetry $\mathbb{Z}_{L_2}^{T_2}$,
    \item mixed $\mathbb{Z}_2$ anomaly between translational symmetry $\mathbb{Z}_{L_1}^{T_1}$ and reflection symmetry $\mathbb{Z}_{2}^{\mathcal{R}_2}$.
\end{itemize}

\subsubsection{Many-Body Symmetry with 2 Copies: Odd-$L_1$, Even-$L_2$}

We then reduce $L_1$ to be odd by setting $T_1$ topological defect. With the full invariant group above, we find that there's an invariant subgroup

\begin{equation}
  \begin{aligned}
    &T_2^{L_2}=1,\ \mathcal{R}_2^2=1,\ \mathcal{T}^2=1,\ (-)^F=(T_2\mathcal{R}_2)^2,\\
    &(-)^F(-)^F=1,\ \mathcal{R}_2T_2=(-)^FT_2^\dagger\mathcal{R}_2,\\
    &\mathcal{T}T_2=(-)^FT_2\mathcal{T},\ \mathcal{R}_2\mathcal{T}=\mathcal{TR}_2,
  \end{aligned}
\end{equation}
which is isomorphic to the invariant group in the $1+1$d case in Eq.~[\ref{Eq: 1dG}]. This motivates us to use the dimensional reduction method to reduce the invariant group together with the anomaly structure to $1+1$d by setting $L_1$ to be odd. Odd-$L_1$ will expose the anomaly of $y-$ Majorana chain, as we'll show below.

To obtain the Hamiltonian for odd-$L_1$, we can set a $T_1$ topological defect~\cite{10.21468/SciPostPhys.16.3.064,2023ScPP...15...51C} on the even-$L_1$ system. The corresponding Hamiltonian is twisted by the defect:

\begin{equation}
\begin{aligned}
    H=&\frac{\mathrm{i}}{2}\sum_{\nu,l_2}(\sum_{l_1=0}^{L_1-2}\chi_{\nu,l_1,l_2}\chi_{\nu,l_1+1,l_2}\\
    +&(-)^{l_2}\chi_{\nu,L_1-1,l_2}\chi_{\nu,0,l_2})\\
    +&\frac{\mathrm{i}}{2}\sum_{\nu,l_1,l_2}(-)^{l_1}\chi_{\nu,l_1,l_2}\chi_{\nu,l_1,l_2+1}.
\end{aligned}
\end{equation}

We can further twist the Hamiltonian to obtain an anti-periodic version:

\begin{equation}
\begin{aligned}
    H_{tw}=&\frac{\mathrm{i}}{2}\sum_{\nu,l_2}(\sum_{l_1=0}^{L_1-2}\chi_{\nu,l_1,l_2}\chi_{\nu,l_1+1,l_2}\\
    -&(-)^{l_2}\chi_{\nu,L_1-1,l_2}\chi_{\nu,0,l_2})\\
    +&\frac{\mathrm{i}}{2}\sum_{\nu,l_1,l_2}(-)^{l_1}\chi_{\nu,l_1,l_2}\chi_{\nu,l_1,l_2+1}.
\end{aligned}
\end{equation}

Within the enlarged Hilbert space $\tilde{\mathcal{H}}=\mathcal{H}\oplus \mathcal{H}_{tw}$, time-reversal symmetry $\mathbb{Z}_2^{\mathcal{T}}$ is well-defined, and we can assign the symmetries in the invariant subgroup:

Translational symmetry $\mathbb{Z}_{L_2}^{T_2}$ is generated by unitary operator $T_2$ defined as:

\begin{equation}
    T_2=e^{\frac{\mathrm{i}\pi}{8}(L_2-2)L_1}\left(\prod_{l_1=0}^{L_1-1}\chi_{l_1,0}\prod_{l_2=0}^{L_2-2}\frac{1+\chi_{l_1,l_2}\chi_{l_1,l_2+1}}{\sqrt{2}}\right)\otimes \sigma^0.
\end{equation}

Reflection symmetry $\mathbb{Z}_2^{\mathcal{R}_2}$ is generated by unitary operator $\mathcal{R}_2$ defined as:

\begin{equation}
\begin{aligned}
    \mathcal{R}_2=&e^{\frac{\mathrm{i}\pi}{16}(L_2-2)L_2L_1}\left(\prod_{l_1=0}^{L_1-1}\chi_{l_1,L_2/2}^{L_2/2-1}\left(\prod_{l_2=0}^{L_2/2-1}\chi_{l_1,l_2}\right)\right.\\
    &\left.\prod_{l_2=1}^{L_2/2-1}\frac{1+(-)^{l_2}\chi_{l_1,l_2}\chi_{l_1,-l_2}}{\sqrt{2}}\right)\otimes \sigma^0.
\end{aligned}
\end{equation}

Fermion parity $\mathbb{Z}_2^F$ is generated by unitary operator $(-)^F$ defined as:

\begin{equation}
    (-)^F=(T_2\mathcal{R}_2)^2=e^{\frac{\mathrm{i}\pi}{4}(L_2+2)L_1}\left(\prod_{l_1,l_2}\chi_{l_1,l_2}\right)\otimes\sigma^0.
\end{equation}

Time-reversal symmetry $\mathbb{Z}_2^{\mathcal{T}}$ is generated by anti-unitary operator $\mathcal{T}$ defined as:

\begin{equation}
    \mathcal{T}=\mathcal{K}\otimes\sigma^1.
\end{equation}

We can check that the $1+1$d invariant subgroup is projectively realized as:

\begin{equation}
  \begin{aligned}
    &T_2^{L_2}=1,\ \mathcal{R}_2^2=1,\ \mathcal{T}^2=1,\ (-)^F=(T_2\mathcal{R}_2)^2,\\
    &(-)^F(-)^F={\color{red}-1},\ \mathcal{R}_2T_2={\color{red}-}(-)^FT_2^\dagger\mathcal{R}_2,\\
    &\mathcal{T}T_2={\color{red}-}(-)^FT_2\mathcal{T},\ \mathcal{R}_2\mathcal{T}=\mathcal{TR}_2.
  \end{aligned}
\end{equation}

In conclusion, we have

\begin{itemize}
    \item $\mathbb{Z}_2$ anomaly for fermion parity $\mathbb{Z}_2^F$,
    \item mixed $\mathbb{Z}_2$ anomaly between translational symmetry $\mathbb{Z}_{L_2}^{T_2}$ and reflection symmetry $\mathbb{Z}_2^{\mathcal{R}_2}$,
    \item mixed $\mathbb{Z}_2$ anomaly between translational symmetry $\mathbb{Z}_{L_2}^{T_2}$ and time-reversal symmetry $\mathbb{Z}_2^{\mathcal{T}}$,
\end{itemize}
which aligns with the anomaly structure of the $1+1$d Majorana chain.

\subsubsection{Many-Body Symmetry with 8 Copies: Odd-$L_1$, Odd-$L_2$}

We further show that the symmetry group is still anomalous with eight copies of root states. Obviously, we can expect the time-reversal anomaly in the $0+1$d invariant subgroup. To do this, we further reduce $L_2$ to be odd by setting $T_2$ topological defect and focus on the $0+1$d invariant subgroup $\mathbb{Z}_2^F\times\mathbb{Z}_2^{\mathcal{T}}$. The Hamiltonian and twisted Hamiltonian are correspondingly given by:

\begin{equation}
\begin{aligned}
    H=&\frac{\mathrm{i}}{2}\sum_{\nu,l_2}(\sum_{l_1=0}^{L_1-2}\chi_{\nu,l_1,l_2}\chi_{\nu,l_1+1,l_2}\\
    +&(-)^{l_2}\chi_{\nu,L_1-1,l_2}\chi_{\nu,0,l_2})\\
    +&\frac{\mathrm{i}}{2}\sum_{\nu,l_1,l_2}(-)^{l_1}\chi_{\nu,l_1,l_2}\chi_{\nu,l_1,l_2+1},
\end{aligned}
\end{equation}

\begin{equation}
\begin{aligned}
    H_{tw}=&\frac{\mathrm{i}}{2}\sum_{\nu,l_2}(\sum_{l_1=0}^{L_1-2}\chi_{\nu,l_1,l_2}\chi_{\nu,l_1+1,l_2}\\
    -&(-)^{l_2}\chi_{\nu,L_1-1,l_2}\chi_{\nu,0,l_2})\\
    +&\frac{\mathrm{i}}{2}\sum_{\nu,l_1,l_2}(-)^{l_1}(\chi_{\nu,l_1,l_2}\chi_{\nu,l_1,l_2+1}\\
    -&\chi_{\nu,l_1,L_2-1}\chi_{\nu,l_1,0}).
\end{aligned}
\end{equation}

In the enlarged Hilbert space $\tilde{\mathcal{H}}=\mathcal{H}\oplus\mathcal{H}_{tw}$, we can assign symmetry operators in $0+1$d subgroup:

Fermion parity $\mathbb{Z}_2^F$ is generated by unitary operator $(-)^F$ defined as:

\begin{equation}
    (-)^F=\left(\prod_{\nu,l_1,l_2}\chi_{\nu,l_1,l_2}\right)\otimes\sigma^0.
\end{equation}

Time-reversal symmetry $\mathbb{Z}_2^{\mathcal{T}}$ is generated by anti-unitary operator $\mathcal{T}$ defined as:

\begin{equation}
    \mathcal{T}=\left(\mathcal{K}\prod_{\nu=even,l_1,l_2}\chi_{\nu,l_1,l_2}\right)\otimes\sigma^1.
\end{equation}

The invariant subgroup spanned by these symmetries is $\mathbb{Z}_2^F\times\mathbb{Z}_4^{\mathcal{T}}$ since $\mathcal{T}^2=-1$.

\subsubsection{Many-Body Symmetry with 16 Copies: Even-$L_1$, Even-$L_2$}

With sixteen copies of root states (eight copies of $2d$ lattice), the anomalies in the $1+1$d and $0+1$d invariant subgroup cancel. Our next task is to check the whole invariant group with $L_1$ and $L_2$ even.

Translational symmetries on $x-$ and $y-$ directions $\mathbb{Z}_{L_1}^{T_1}$ and $\mathbb{Z}_{L_2}^{T_2}$ are generated by unitary operators $T_1$ and $T_2$ defined as:

\begin{equation}
    T_1=\prod_{\nu,l_2}\chi_{\nu,0,l_2}\prod_{l_1=0}^{L_1-2}\frac{1-(-)^{l_2}\chi_{\nu,l_1,l_2}\chi_{\nu,l_1+1,l_2}}{\sqrt{2}},
\end{equation}

\begin{equation}
    T_2=\prod_{\nu,l_1}\chi_{\nu,l_1,0}\prod_{l_2=0}^{L_2-2}\frac{1-\chi_{\nu,l_1,l_2}\chi_{\nu,l_1,l_2+1}}{\sqrt{2}}.
\end{equation}

Reflection symmetry on $x-$ and $y-$ directions $\mathbb{Z}_{2}^{\mathcal{R}_1}$ and $\mathbb{Z}_{2}^{\mathcal{R}_2}$ are generated by unitary operators $\mathcal{R}_1$ and $\mathcal{R}_2$ defined as:

\begin{equation}
\begin{aligned}
    \mathcal{R}_1=&\prod_{\nu,l_2}\chi_{\nu,L_1/2,l_2}^{L_1/2}\left(\prod_{l_1=L_1/2+1}^{L_1-1}\chi_{\nu,l_1,l_2}\right)\\
    &\prod_{l_1=1}^{L_1/2-1}\frac{1+(-)^{l_1}\chi_{\nu,l_1,l_2}\chi_{\nu,-l_1,l_2}}{\sqrt{2}},
\end{aligned}
\end{equation}

\begin{equation}
\begin{aligned}
    \mathcal{R}_2=&\prod_{\nu,l_1}\chi_{\nu,l_1,L_2/2}^{L_2/2}\left(\prod_{l_2=L_2/2+1}^{L_2-1}\chi_{\nu,l_1,l_2}\right)\\
    &\prod_{l_2=1}^{L_2/2-1}\frac{1+(-)^{l_2}\chi_{\nu,l_1,l_2}\chi_{\nu,l_1,-l_2}}{\sqrt{2}}.
\end{aligned}
\end{equation}

Time-reversal symmetry $\mathbb{Z}_{2}^{\mathcal{T}}$ is generated by anti-unitary operator $\mathcal{T}$ defined as:

\begin{equation}
    \mathcal{T}=\mathcal{K}\prod_{\nu=even,l_1,l_2}\chi_{\nu,l_1,l_2}.
\end{equation}

One can check by straightforward calculation that with sixteen copies of root states (eight copies of lattice), we can faithfully reproduce the invariant group defined in Eq.~[\ref{Eq: G_2}].

With the $\mathbb{Z}_{16}$ SMG classification, we can assign on-site interaction term found in Sec.~\ref{Sec: 0+1d} on a given site, and generate the full interaction Hamiltonian by translations in all directions:

\begin{equation}
\begin{aligned}
  H_{int}=&\sum_{l} \chi_{1,l}\chi_{2,l}\chi_{3,l}\chi_{4,l}+\chi_{1,l}\chi_{2,l}\chi_{5,l}\chi_{6,l}\\
  &+\chi_{1,l}\chi_{3,l}\chi_{5,l}\chi_{7,l}+\chi_{2,l}\chi_{3,l}\chi_{5,l}\chi_{8,l}.
\end{aligned}
\end{equation}

\subsection{$3+1$d Symmetric Mass Generation}\label{Sec:3+1d_M}

\subsubsection{$\mathcal{C}$-$\mathcal{R}$-$\mathcal{T}$-Internal Symmetry in Lattice Model}\label{Sec. 3+1d lattice}

\begin{figure}[h]
\includegraphics[width=.6\linewidth]{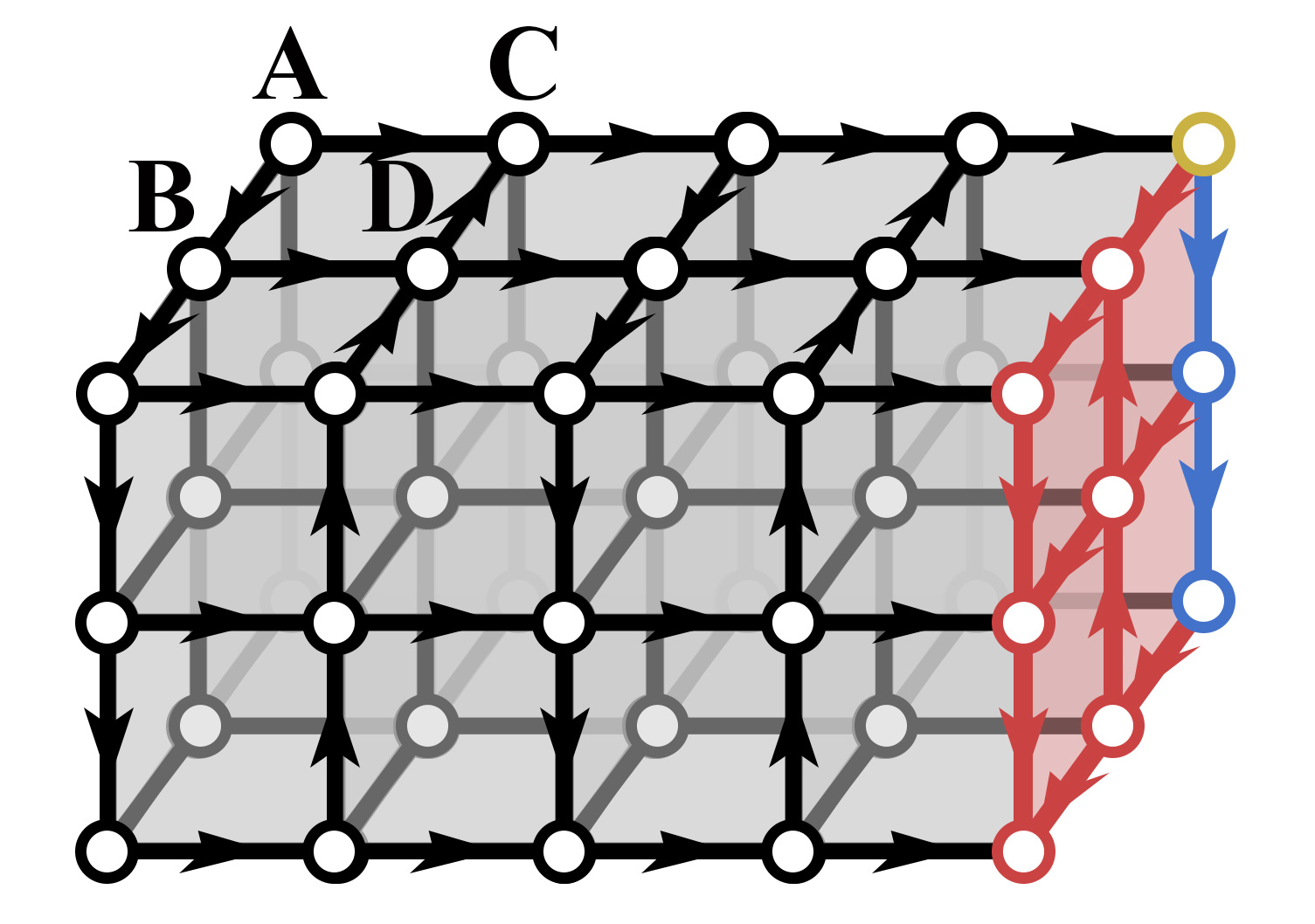}
    
\caption{Staggered fermion model on 3d lattice with staggered hopping amplitudes. Each unit cell contains $A$, $B$, $C$, and $D$ sublattices. By setting $L_1$ odd, the anomalies in the $2d$ plane (shaded in red) are exposed. By subsequently setting $L_2$ odd, the anomalies in the 1d chain (shaded in blue) are exposed. By further setting $L_3$ odd, the anomalies in the 0d site (shaded in yellow) are exposed.
}
\label{fig:3d_lattice}
\end{figure}

We proceed to the investigation of $3+1$d staggered fermion model with staggered Hamiltonian

\begin{equation}
\begin{aligned}
    H=&\frac{\mathrm{i}}{2}\sum_{l_1,l_2,l_3}\chi_{l_1,l_2,l_3}\chi_{l_1+1,l_2,l_3}\\
    +&\frac{\mathrm{i}}{2}\sum_{l_1,l_2,l_3}(-)^{l_1}\chi_{l_1,l_2,l_3}\chi_{l_1,l_2+1,l_3}\\
    +&\frac{\mathrm{i}}{2}\sum_{l_1,l_2,l_3}(-)^{l_1+l_2}\chi_{l_1,l_2,l_3}\chi_{l_1,l_2,l_3+1}.
\end{aligned}
\end{equation}

The unit cell of 3+1d staggered fermion model is $2\times 2\times 1$, with four sublattices $A(0,0,0)$, $B(0,1,0)$, $C(1,0,0)$, $D(1,1,0)$. By Fourier transformation into momentum space, we constructed the lattice Majorana fermion with a minimum doubling problem by reducing its Brillouin zone. Now the only doubler is provided by two Weyl points $(0,0,0)$ and $(0,0,\pi)$ within the momentum space.

To see the low-energy behavior of symmetry operators, it's intuitive for us to rewrite the Hamiltonian in sublattice basis $\chi_{A;l_1,l_2,l_3}=\chi_{2l_1,2l_2,l_3},\ \chi_{B;l_1,l_2,l_3}=\chi_{2l_1,2l_2+1,l_3},\ \chi_{C;l_1,l_2,l_3}=\chi_{2l_1+1,2l_2,l_3},\ \chi_{D;l_1,l_2,l_3}=\chi_{2l_1+1,2l_2+1,l_3}$:

\begin{equation}
\begin{aligned}
    H=&\frac{\mathrm{i}}{2}\sum_{l_1,l_2,l_3}(\chi_{A;l_1,l_2,l_3}\chi_{C;l_1,l_2,l_3}+\chi_{C;l_1,l_2,l_3}\chi_{A;l_1+1,l_2,l_3}\\
    +&\chi_{B;l_1,l_2,l_3}\chi_{D;l_1,l_2,l_3}+\chi_{D;l_1,l_2,l_3}\chi_{B;l_1+1,l_2,l_3})\\
    +&\frac{\mathrm{i}}{2}\sum_{l_1,l_2,l_3}(\chi_{A;l_1,l_2,l_3}\chi_{B;l_1,l_2,l_3}+\chi_{B;l_1,l_2,l_3}\chi_{A;l_1,l_2+1,l_3}\\
    -&\chi_{C;l_1,l_2,l_3}\chi_{D;l_1,l_2,l_3}-\chi_{D;l_1,l_2,l_3}\chi_{C;l_1,l_2+1,l_3})\\
    +&\frac{\mathrm{i}}{2}\sum_{l_1,l_2,l_3}(\chi_{A;l_1,l_2,l_3}\chi_{A;l_1,l_2,l_3+1}-\chi_{B;l_1,l_2,l_3}\chi_{B;l_1,l_2,l_3+1}\\
    -&\chi_{C;l_1,l_2,l_3}\chi_{C;l_1,l_2,l_3+1}+\chi_{D;l_1,l_2,l_3}\chi_{D;l_1,l_2,l_3+1}).
\end{aligned}
\end{equation}

We first assign translational symmetries on $x-$, $y-$ and $z-$ directions $\mathbb{Z}_{L_i}^{T_i}$ ($i=1,2,3$), generated by unitary operators $T_1$, $T_2$, $T_3$ acting as:

\begin{equation}
\begin{aligned}
    &T_1 \chi_{A;l_1,l_2,l_3}=(-)^{l_3}\chi_{C;l_1,l_2,l_3}T_1,\\
    &T_1 \chi_{B;l_1,l_2,l_3}=(-)^{l_3+1} \chi_{D;l_1,l_2,l_3}T_1,\\
    &T_1 \chi_{C;l_1,l_2,l_3}=(-)^{l_3}\chi_{A;l_1+1,l_2,l_3}T_1,\\
    &T_1 \chi_{D;l_1,l_2,l_3}=(-)^{l_3+1} \chi_{B;l_1+1,l_2,l_3}T_1,
\end{aligned}
\end{equation}

\begin{equation}
\begin{aligned}
    &T_2 \chi_{A;l_1,l_2,l_3}=(-)^{l_3}\chi_{B;l_1,l_2,l_3}T_2,\\
    &T_2 \chi_{B;l_1,l_2,l_3}=(-)^{l_3} \chi_{A;l_1,l_2+1,l_3}T_2,\\
    &T_2 \chi_{C;l_1,l_2,l_3}=(-)^{l_3}\chi_{D;l_1,l_2,l_3}T_2,\\
    &T_2 \chi_{D;l_1,l_2,l_3}=(-)^{l_3} \chi_{C;l_1,l_2+1,l_3}T_2,
\end{aligned}
\end{equation}

\begin{equation}
\begin{aligned}
    &T_3 \chi_{A;l_1,l_2,l_3}=\chi_{A;l_1,l_2,l_3+1}T_3,\\
    &T_3 \chi_{B;l_1,l_2,l_3}=\chi_{B;l_1,l_2,l_3+1}T_3,\\
    &T_3 \chi_{C;l_1,l_2,l_3}=\chi_{C;l_1,l_2,l_3+1}T_3,\\
    &T_3 \chi_{D;l_1,l_2,l_3}=\chi_{D;l_1,l_2,l_3+1}T_3.
\end{aligned}
\end{equation}

Reflection symmetries on $x-$, $y-$, and $z-$ directions $\mathbb{Z}_2^{\mathcal{R}_i}$ ($i=1,2,3$) is generated by unitary operators $\mathcal{R}_1$, $\mathcal{R}_2$, $\mathcal{R}_3$ acting as:

\begin{equation}
\begin{aligned}
    &\mathcal{R}_1 \chi_{A;l_1,l_2,l_3}=\chi_{A;-l_1,l_2,l_3}\mathcal{R}_1,\\
    &\mathcal{R}_1 \chi_{B;l_1,l_2,l_3}=\chi_{B;-l_1,l_2,l_3}\mathcal{R}_1,\\
    &\mathcal{R}_1 \chi_{C;l_1,l_2,l_3}=-\chi_{C;-l_1-1,l_2,l_3}\mathcal{R}_1,\\
    &\mathcal{R}_1 \chi_{D;l_1,l_2,l_3}=-\chi_{D;-l_1-1,l_2,l_3}\mathcal{R}_1,
\end{aligned}
\end{equation}

\begin{equation}
\begin{aligned}
    &\mathcal{R}_2 \chi_{A;l_1,l_2,l_3}=\chi_{A;l_1,-l_2,l_3}\mathcal{R}_2,\\
    &\mathcal{R}_2 \chi_{B;l_1,l_2,l_3}=-\chi_{B;l_1,-l_2-1,l_3}\mathcal{R}_2,\\
    &\mathcal{R}_2 \chi_{C;l_1,l_2,l_3}=\chi_{C;l_1,-l_2,l_3}\mathcal{R}_2,\\
    &\mathcal{R}_2 \chi_{D;l_1,l_2,l_3}=-\chi_{D;l_1,-l_2-1,l_3}\mathcal{R}_2,
\end{aligned}
\end{equation}

\begin{equation}
\begin{aligned}
    &\mathcal{R}_3 \chi_{A;l_1,l_2,l_3}=(-)^{l_3}\chi_{A;l_1,l_2,-l_3}\mathcal{R}_3,\\
    &\mathcal{R}_3 \chi_{B;l_1,l_2,l_3}=(-)^{l_3}\chi_{B;l_1,l_2,-l_3}\mathcal{R}_3,\\
    &\mathcal{R}_3 \chi_{C;l_1,l_2,l_3}=(-)^{l_3}\chi_{C;l_1,l_2,-l_3}\mathcal{R}_3,\\
    &\mathcal{R}_3 \chi_{D;l_1,l_2,l_3}=(-)^{l_3}\chi_{D;l_1,l_2,-l_3}\mathcal{R}_3.
\end{aligned}
\end{equation}

Finally, we can assign time-reversal symmetry $\mathbb{Z}_2^{\mathcal{T}}$, generated by anti-unitary operator $\mathcal{T}$ acting as:

\begin{equation}
\begin{aligned}
    \mathcal{T}\chi_{A;l_1,l_2,l_3}&=(-)^{l_3}\chi_{A;l_1,l_2,l_3}\mathcal{T},\\
    \mathcal{T}\chi_{B;l_1,l_2,l_3}&=(-)^{l_3+1}\chi_{B;l_1,l_2,l_3}\mathcal{T},\\
    \mathcal{T}\chi_{C;l_1,l_2,l_3}&=(-)^{l_3+1}\chi_{C;l_1,l_2,l_3}\mathcal{T},\\
    \mathcal{T}\chi_{D;l_1,l_2,l_3}&=(-)^{l_3}\chi_{D;l_1,l_2,l_3}\mathcal{T},\\
    \mathcal{T}\mathrm{i}&=-\mathrm{i}\mathcal{T}.
\end{aligned}
\end{equation}

By checking the action of these operators on low-energy modes, we can find the correspondence of low-energy symmetry on the lattice:

Low-energy fermion parity generator $(-)^F$ corresponds to the lattice fermion parity generator $(-)^F$, which can also be written in lattice translation and reflection operators as $(T_1\mathcal{R}_1)^2$, $(T_2\mathcal{R}_2)^2$ or $(T_3\mathcal{R}_3)^2$.

Low-energy reflection generator $\mathcal{R}_1$, $\mathcal{R}_2$, and $\mathcal{R}_3$ correspond to lattice $\mathcal{R}_1$, $\mathcal{R}_2$, and $\mathcal{R}_3$, while low-energy time-reversion operator $\mathcal{T}$ correspond to lattice $T_1\mathcal{T}$.

Low-energy internal $U(1)$ symmetry is an emergent symmetry. The broken $\mathbb{Z}_4^{\mathcal{J}}$ symmetry is an emanant symmetry, whose generator $\mathcal{J}$ corresponds to lattice $T_2T_1$.

In lattice model level, the invariant group is $G$ of order $16L_1L_2L_3$ ($G\cong \mathbb{D}_{2L_1}\times\mathbb{D}_{2L_1}\times\mathbb{D}_{2L_3}\times\mathbb{Z}_2$\cite{GAP4}), with its presentation:

\begin{equation}
  \begin{aligned}
    &T_i^{L_i}=1,\ \mathcal{R}_i^2=1,\ \mathcal{T}^2=1,\ (-)^F=(T_i\mathcal{R}_i)^2,\\
    &(-)^F(-)^F=1,\ \mathcal{R}_iT_i=(-)^FT_i^\dagger\mathcal{R}_i,\\
    &\mathcal{T}T_i=(-)^FT_i\mathcal{T},\ \mathcal{R}_i\mathcal{T}=\mathcal{TR}_i,\ \forall i=1,2,3,\\
    &T_iT_j=(-)^FT_jT_i,\ \mathcal{R}_i\mathcal{R}_j=\mathcal{R}_j\mathcal{R}_i,\\
    &\mathcal{R}_iT_j=T_j\mathcal{R}_i,\ \forall i\neq j.
  \end{aligned}\label{Eq: G_3}
\end{equation}

\subsubsection{Many-Body Symmetry with 2 Copies: Even-$L_1$, Even-$L_2$, Even-$L_3$}

We start our discussion with $L_1$, $L_2$, and $L_3$ even, where all reflection symmetries $\mathbb{Z}_2^{\mathcal{R}_i}$ ($i=1,2,3$) is well-defined. With the intrinsic doubling $\nu_{stag}(2)=2^3/\text{dim}_{\mathbb{R}}\chi_{\mathcal{C}\ell(3,0)}=2$, we automatically start with 2 copies on the lattice model.

Translational symmetry on $x-$, $y-$, and $z-$ directions $\mathbb{Z}_{L_i}^{T_i}$ is generated by unitary operators $T_1$, $T_2$, and $T_3$ defined as:

\begin{equation}
    T_1=\prod_{l_2,l_3}\chi_{0,l_2,l_3}\prod_{l_1=0}^{L_1-2}\frac{1-(-)^{l_2+l_3}\chi_{l_1,l_2,l_3}\chi_{l_1+1,l_2,l_3}}{\sqrt{2}},
\end{equation}

\begin{equation}
    T_2=\prod_{l_1,l_3}\chi_{l_1,0,l_3}\prod_{l_2=0}^{L_2-2}\frac{1-(-)^{l_3}\chi_{l_1,l_2,l_3}\chi_{l_1,l_2+1,l_3}}{\sqrt{2}}.
\end{equation}

\begin{equation}
    T_3=\prod_{l_1,l_2}\chi_{l_1,l_2,0}\prod_{l_3=0}^{L_3-2}\frac{1-\chi_{l_1,l_2,l_3}\chi_{l_1,l_2,l_3+1}}{\sqrt{2}},
\end{equation}

Reflection symmetry on $x-$, $y-$, and $z-$ directions $\mathbb{Z}_2^{\mathcal{R}_i}$ is generated by unitary operators $\mathcal{R}_1$, $\mathcal{R}_2$, and $\mathcal{R}_3$ defined as:

\begin{equation}
\begin{aligned}
    \mathcal{R}_1=&\prod_{l_2,l_3}\chi_{L_1/2,l_2,l_3}^{L_1/2}\left(\prod_{l_1=L_1/2+1}^{L_1-1}\chi_{l_1,l_2,l_3}\right)\\
    &\prod_{l_1=1}^{L_1/2-1}\frac{1+(-)^{l_1}\chi_{l_1,l_2,l_3}\chi_{-l_1,l_2,l_3}}{\sqrt{2}},
\end{aligned}
\end{equation}

\begin{equation}
\begin{aligned}
    \mathcal{R}_2=&\prod_{l_1,l_3}\chi_{l_1,L_2/2,l_3}^{L_2/2}\left(\prod_{l_2=L_2/2+1}^{L_2-1}\chi_{l_1,l_2,l_3}\right)\\
    &\prod_{l_2=1}^{L_2/2-1}\frac{1+(-)^{l_2}\chi_{l_1,l_2,l_3}\chi_{l_1,-l_2,l_3}}{\sqrt{2}},
\end{aligned}
\end{equation}

\begin{equation}
\begin{aligned}
    \mathcal{R}_3=&\prod_{l_1,l_2}\chi_{l_1,l_2,L_3/2}^{L_3/2}\left(\prod_{l_3=L_3/2+1}^{L_3-1}\chi_{l_1,l_2,l_3}\right)\\
    &\prod_{l_3=1}^{L_3/2-1}\frac{1+(-)^{l_3}\chi_{l_1,l_2,l_3}\chi_{l_1,l_2,-l_3}}{\sqrt{2}}.
\end{aligned}
\end{equation}

Fermion parity $\mathbb{Z}_2^F$ is generated by unitary operator $(-)^F$ defined as:

\begin{equation}
    (-)^F=(T_1\mathcal{R}_1)^2=\prod_{l_1,l_2,l_3}\chi_{l_1,l_2,l_3}.
\end{equation}

Time-reversal symmetry $\mathbb{Z}_2^{\mathcal{T}}$ is generated by anti-unitary operator $\mathcal{T}$ defined as:

\begin{equation}
    \mathcal{T}=\mathcal{K}.
\end{equation}

Equipped with all these symmetry operators, we can check the anomaly structure in the invariant group. Surprisingly, we faithfully reproduce the original invariant group this time. However, this doesn't mean we're in an anomaly-free system. Instead, the anomalies hide in odd-$L$ cases, as we'll discuss in the following.

\subsubsection{Many-Body Symmetry with 2 Copies: Odd-$L_1$, Even-$L_2$, Even-$L_3$}

To expose the hidden anomalies, we reduce $L_1$ to be odd by setting $T_1$ topological defect. With the full invariant group above, we find that there's an invariant subgroup

\begin{equation}
  \begin{aligned}
    &T_i^{L_i}=1,\ \mathcal{R}_i^2=1,\ \mathcal{T}^2=1,\ (-)^F=(T_i\mathcal{R}_i)^2,\\
    &(-)^F(-)^F=1,\ \mathcal{R}_iT_i=(-)^FT_i^\dagger\mathcal{R}_i,\\
    &\mathcal{T}T_i=(-)^FT_i\mathcal{T},\ \mathcal{R}_i\mathcal{T}=\mathcal{TR}_i,\ \forall i=2,3,\\
    &T_iT_j=(-)^FT_jT_i,\ \mathcal{R}_i\mathcal{R}_j=\mathcal{R}_j\mathcal{R}_i,\\
    &\mathcal{R}_iT_j=T_j\mathcal{R}_i,\ \forall i\neq j>1.
  \end{aligned}
\end{equation}
which is isomorphic to the invariant group in the $2+1$d case in Eq.~[\ref{Eq: G_2}]. This, again, motivates us to use the dimensional reduction method to reduce the invariant group together with the anomaly structure to $2+1$d by setting $L_1$ to be odd. Odd-$L_1$ will expose the anomaly of $yz-$plane Majorana system, as we'll show below.

To obtain the Hamiltonian for odd-$L_1$, we can set a $T_1$ topological defect~\cite{10.21468/SciPostPhys.16.3.064,2023ScPP...15...51C} on the even-$L_1$ system. The corresponding Hamiltonian is twisted by the defect:

\begin{equation}
\begin{aligned}
    H=&\frac{\mathrm{i}}{2}\sum_{\nu,l_2,l_3}(\sum_{l_1=0}^{L_1-2}\chi_{\nu,l_1,l_2,l_3}\chi_{\nu,l_1+1,l_2,l_3}\\
    +&(-)^{l_2+l_3}\chi_{\nu,L_1-1,l_2,l_3}\chi_{\nu,0,l_2,l_3})\\
    +&\frac{\mathrm{i}}{2}\sum_{\nu,l_1,l_2,l_3}(-)^{l_1}\chi_{\nu,l_1,l_2,l_3}\chi_{\nu,l_1,l_2+1,l_3}\\
    +&\frac{\mathrm{i}}{2}\sum_{\nu,l_1,l_2,l_3}(-)^{l_1+l_2}\chi_{\nu,l_1,l_2,l_3}\chi_{\nu,l_1,l_2,l_3+1}.
\end{aligned}
\end{equation}

We can further twist the Hamiltonian to obtain an anti-periodic version:

\begin{equation}
\begin{aligned}
    H_{tw}=&\frac{\mathrm{i}}{2}\sum_{\nu,l_2,l_3}(\sum_{l_1=0}^{L_1-2}\chi_{\nu,l_1,l_2,l_3}\chi_{\nu,l_1+1,l_2,l_3}\\
    -&(-)^{l_2+l_3}\chi_{\nu,L_1-1,l_2,l_3}\chi_{\nu,0,l_2,l_3})\\
    +&\frac{\mathrm{i}}{2}\sum_{\nu,l_1,l_2,l_3}(-)^{l_1}\chi_{\nu,l_1,l_2,l_3}\chi_{\nu,l_1,l_2+1,l_3}\\
    +&\frac{\mathrm{i}}{2}\sum_{\nu,l_1,l_2,l_3}(-)^{l_1+l_2}\chi_{\nu,l_1,l_2,l_3}\chi_{\nu,l_1,l_2,l_3+1}.
\end{aligned}
\end{equation}

Within the enlarged Hilbert space $\tilde{\mathcal{H}}=\mathcal{H}\oplus \mathcal{H}_{tw}$, time-reversal symmetry $\mathbb{Z}_2^{\mathcal{T}}$ is well-defined, and we can assign the symmetries in the invariant subgroup:

Translational symmetries on $y-$ and $z-$ directions $\mathbb{Z}_{L_i}^{T_i}$ is generated by unitary operators $T_2$ and $T_3$ defined as:

\begin{equation}
\begin{aligned}
    T_2=&e^{\frac{\mathrm{i}\pi}{8}(L_2-4)L_3L_1}\left(\prod_{l_1,l_3}\chi_{l_1,0,l_3}\right.\\
    &\left.\prod_{l_2=0}^{L_2-2}\frac{1-(-)^{l_3}\chi_{l_1,l_2,l_3}\chi_{l_1,l_2+1,l_3}}{\sqrt{2}}\right)\otimes \sigma^0,
\end{aligned}
\end{equation}

\begin{equation}
\begin{aligned}
    T_3=&e^{\frac{\mathrm{i}\pi}{8}(L_3-4)L_2L_1}\left(\prod_{l_1,l_2}\chi_{l_1,l_2,0}\right.\\
    &\left.\prod_{l_3=0}^{L_3-2}\frac{1-\chi_{l_1,l_2,l_3}\chi_{l_1,l_2,l_3+1}}{\sqrt{2}}\right)\otimes \sigma^0.
\end{aligned}
\end{equation}

Reflection symmetries on $y-$ and $z-$ directions $\mathbb{Z}_2^{\mathcal{R}_i}$ is generated by unitary operator $\mathcal{R}_2$ and $\mathcal{R}_3$ defined as:

\begin{equation}
\begin{aligned}
    \mathcal{R}_2=&\mathrm{i}^{L_3L_1/2}\left(\prod_{l_1,l_3}\chi_{l_1,L_2/2,l_3}^{L_2/2}\left(\prod_{l_2=L_2/2+1}^{L_2-1}\chi_{l_1,l_2,l_3}\right)\right.\\
    &\left.\prod_{l_2=1}^{L_2/2-1}\frac{1+(-)^{l_2}\chi_{l_1,l_2,l_3}\chi_{l_1,-l_2,l_3}}{\sqrt{2}}\right)\otimes \sigma^0,
\end{aligned}
\end{equation}

\begin{equation}
\begin{aligned}
    \mathcal{R}_3=&\mathrm{i}^{L_2L_1/2}\left(\prod_{l_1,l_2}\chi_{l_1,l_2,L_3/2}^{L_3/2}\left(\prod_{l_3=L_3/2+1}^{L_3-1}\chi_{l_1,l_2,l_3}\right)\right.\\
    &\left.\prod_{l_3=1}^{L_3/2-1}\frac{1+(-)^{l_3}\chi_{l_1,l_2,l_3}\chi_{l_1,l_2,-l_3}}{\sqrt{2}}\right)\otimes \sigma^0.
\end{aligned}
\end{equation}

Fermion parity $\mathbb{Z}_2^F$ is generated by unitary operator $(-)^F$ defined as:

\begin{equation}
    (-)^F=(T_2\mathcal{R}_2)^2=(-)^{(L_2/2+1)L_3L_1/2}\left(\prod_{l_1,l_2,l_3}\chi_{l_1,l_2,l_3}\right)\otimes\sigma^0.
\end{equation}

Time-reversal symmetry $\mathbb{Z}_2^{\mathcal{T}}$ is generated by anti-unitary operator $\mathcal{T}$ defined as:

\begin{equation}
    \mathcal{T}=\mathcal{K}\otimes\sigma^1.
\end{equation}

We can check that the $2+1$d invariant subgroup is projectively realized as:

\begin{equation}
  \begin{aligned}
    &T_2^{L_2}=1,\ T_3^{L_3}=1,\ \mathcal{R}_2^2=1,\ \mathcal{R}_3^2=1,\ \mathcal{T}^2=1,\\
    &(-)^F=(T_2\mathcal{R}_2)^2={\color{red}(-)^{(L_2+L_3)L_1/2}}(T_3\mathcal{R}_3)^2,\ (-)^F(-)^F=1\\
    &\mathcal{R}_2T_2=(-)^FT_2^\dagger\mathcal{R}_2,\ \mathcal{R}_3T_3={\color{red}(-)^{(L_2+L_3)L_1/2}}(-)^FT_3^\dagger\mathcal{R}_3,\\
    &\mathcal{T}T_2=(-)^FT_2\mathcal{T},\ \mathcal{T}T_3={\color{red}(-)^{(L_2+L_3)L_1/2}}(-)^FT_3\mathcal{T},\\
    &\mathcal{R}_2\mathcal{T}=\mathcal{TR}_2,\ \mathcal{R}_3\mathcal{T}=\mathcal{TR}_3,\\
    &T_2T_3={\color{red}(-)^{((L_2/2+1)L_3/2+1)L_1}}(-)^FT_3T_2,\ \mathcal{R}_2\mathcal{R}_3={\color{red}(-)^{L_1}}\mathcal{R}_3\mathcal{R}_2,\\
    &\mathcal{R}_2T_3={\color{red}(-)^{L_1}}T_3\mathcal{R}_2,\ \mathcal{R}_3T_2={\color{red}(-)^{L_1}}T_2\mathcal{R}_3.
  \end{aligned}
\end{equation}

In conclusion, we have

\begin{itemize}
    \item $\mathbb{Z}_2$ anomaly for fermion parity $\mathbb{Z}_2^F$,
    \item mixed $\mathbb{Z}_2$ anomaly between translational symmetry $\mathbb{Z}_{L_3}^{T_3}$ and reflection symmetry $\mathbb{Z}_2^{\mathcal{R}_3}$,
    \item mixed $\mathbb{Z}_2$ anomaly between translational symmetry $\mathbb{Z}_{L_3}^{T_3}$ and time-reversal symmetry $\mathbb{Z}_2^{\mathcal{T}}$,
    \item mixed $\mathbb{Z}_2$ anomaly between translational symmetries $\mathbb{Z}_{L_2}^{T_2}$ and $\mathbb{Z}_{L_3}^{T_3}$,
    \item mixed $\mathbb{Z}_2$ anomaly between reflection symmetries $\mathbb{Z}_{2}^{\mathcal{R}_2}$ and $\mathbb{Z}_{2}^{\mathcal{R}_3}$,
    \item mixed $\mathbb{Z}_2$ anomaly between reflection symmetry $\mathbb{Z}_{2}^{\mathcal{R}_2}$ and translational symmetry $\mathbb{Z}_{L_3}^{T_3}$,
    \item mixed $\mathbb{Z}_2$ anomaly between translational symmetry $\mathbb{Z}_{L_2}^{T_2}$ and reflection symmetry $\mathbb{Z}_{2}^{\mathcal{R}_3}$.
\end{itemize}
which aligns with the anomaly structure of the $2+1$d Majorana system.

\subsubsection{Many-Body Symmetry with 8 Copies: Odd-$L_1$, Odd-$L_2$, Odd-$L_3$}

According to the discussion in $1+1$d and $2+1$d, we expect to find the $0+1$d anomaly by setting $L_1$, $L_2$, and $L_3$ odd. To do this, we first set $T_1$, $T_2$, and $T_3$ topological defects to the Hamiltonian:

\begin{equation}
\begin{aligned}
    H=&\frac{\mathrm{i}}{2}\sum_{\nu,l_2,l_3}(\sum_{l_1=0}^{L_1-2}\chi_{\nu,l_1,l_2,l_3}\chi_{\nu,l_1+1,l_2,l_3}\\
    +&(-)^{l_2+l_3}\chi_{\nu,L_1-1,l_2,l_3}\chi_{\nu,0,l_2,l_3})\\
    +&\frac{\mathrm{i}}{2}\sum_{\nu,l_1,l_3}(\sum_{l_2=0}^{L_2-2}(-)^{l_1}\chi_{\nu,l_1,l_2,l_3}\chi_{\nu,l_1,l_2+1,l_3}\\
    +&(-)^{l_1+l_3}\chi_{\nu,l_1,L_2-1,l_3}\chi_{\nu,l_1,0,l_3})\\
    +&\frac{\mathrm{i}}{2}\sum_{\nu,l_1,l_2}(\sum_{l_3=0}^{L_3-2}(-)^{l_1+l_2}\chi_{\nu,l_1,l_2,l_3}\chi_{\nu,l_1,l_2,l_3+1}\\
    +&(-)^{l_1+l_2}\chi_{\nu,l_1,l_2,L_3-1}\chi_{\nu,l_1,l_2,0}).
\end{aligned}
\end{equation}

We can further twist the Hamiltonian to obtain an anti-periodic version:

\begin{equation}
\begin{aligned}
    H_{tw}=&\frac{\mathrm{i}}{2}\sum_{\nu,l_2,l_3}(\sum_{l_1=0}^{L_1-2}\chi_{\nu,l_1,l_2,l_3}\chi_{\nu,l_1+1,l_2,l_3}\\
    -&(-)^{l_2+l_3}\chi_{\nu,L_1-1,l_2,l_3}\chi_{\nu,0,l_2,l_3})\\
    +&\frac{\mathrm{i}}{2}\sum_{\nu,l_1,l_3}(\sum_{l_2=0}^{L_2-2}(-)^{l_1}\chi_{\nu,l_1,l_2,l_3}\chi_{\nu,l_1,l_2+1,l_3}\\
    -&(-)^{l_1+l_3}\chi_{\nu,l_1,L_2-1,l_3}\chi_{\nu,l_1,0,l_3})\\
    +&\frac{\mathrm{i}}{2}\sum_{\nu,l_1,l_2}(\sum_{l_3=0}^{L_3-2}(-)^{l_1+l_2}\chi_{\nu,l_1,l_2,l_3}\chi_{\nu,l_1,l_2,l_3+1}\\
    -&(-)^{l_1+l_2}\chi_{\nu,l_1,l_2,L_3-1}\chi_{\nu,l_1,l_2,0}),
\end{aligned}
\end{equation}
and the enlarged Hamiltonian is given by $\tilde{H}=H\oplus H_{tw}$.

With four copies of lattice fermion, we can assign the symmetry operators in the $0+1$d invariant subgroup as follows:

Fermion parity $\mathbb{Z}_2^F$ is generated by unitary operator $(-)^F$ defined as:

\begin{equation}
    (-)^F=\left(\prod_{\nu,l_1,l_2,l_3}\chi_{\nu,l_1,l_2,l_3}\right)\otimes\sigma^0.
\end{equation}

Time-reversal symmetry $\mathbb{Z}_2^{\mathcal{T}}$ is generated by anti-unitary operator $\mathcal{T}$ defined as:

\begin{equation}
    \mathcal{T}=\left(\mathcal{K}\prod_{\nu=even,l_1,l_2,l_3}\chi_{\nu,l_1,l_2,l_3}\right)\otimes\sigma^1.
\end{equation}

The projectively realized invariant subgroup is $\mathbb{Z}_2^F\times\mathbb{Z}_4^{\mathcal{T}}$ as we expected, and this anomaly protects the gapless regime of our given system.

\subsubsection{Many-Body Symmetry with 16 Copies: Even-$L_1$, Even-$L_2$, Even-$L_3$}

With sixteen copies of root states (eight copies of lattice system), the anomaly in the $0+1$d invariant subgroup cancels. Our next task is to check the whole invariant group with $L_1$, $L_2$, and $L_3$ even.

Translational symmetries on $x-$, $y-$, and $z-$ directions $\mathbb{Z}_{L_1}^{T_1}$, $\mathbb{Z}_{L_2}^{T_2}$, and $\mathbb{Z}_{L_3}^{T_3}$ are generated by unitary operators $T_1$, $T_2$, and $T_3$ defined as:

\begin{equation}
    T_1=\prod_{\nu,l_2,l_3}\chi_{\nu,0,l_2,l_3}\prod_{l_1=0}^{L_1-2}\frac{1-(-)^{l_2+l_3}\chi_{\nu,l_1,l_2,l_3}\chi_{\nu,l_1+1,l_2,l_3}}{\sqrt{2}},
\end{equation}

\begin{equation}
    T_2=\prod_{\nu,l_1,l_3}\chi_{\nu,l_1,0,l_3}\prod_{l_2=0}^{L_2-2}\frac{1-(-)^{l_3}\chi_{\nu,l_1,l_2,l_3}\chi_{\nu,l_1,l_2+1,l_3}}{\sqrt{2}},
\end{equation}

\begin{equation}
    T_3=\prod_{\nu,l_1,l_2}\chi_{\nu,l_1,l_2,0}\prod_{l_3=0}^{L_3-2}\frac{1-\chi_{\nu,l_1,l_2,l_3}\chi_{\nu,l_1,l_2,l_3+1}}{\sqrt{2}}.
\end{equation}

Reflection symmetry on $x-$, $y-$, and $z-$ directions $\mathbb{Z}_{2}^{\mathcal{R}_1}$, $\mathbb{Z}_{2}^{\mathcal{R}_2}$, and $\mathbb{Z}_{2}^{\mathcal{R}_3}$ are generated by unitary operators $\mathcal{R}_1$, $\mathcal{R}_2$, and $\mathcal{R}_3$ defined as:

\begin{equation}
\begin{aligned}
    \mathcal{R}_1=&\prod_{\nu,l_2,l_3}\chi_{\nu,L_1/2,l_2,l_3}^{L_1/2}\left(\prod_{l_1=L_1/2+1}^{L_1-1}\chi_{\nu,l_1,l_2,l_3}\right)\\
    &\prod_{l_1=1}^{L_1/2-1}\frac{1+(-)^{l_1}\chi_{\nu,l_1,l_2,l_3}\chi_{\nu,-l_1,l_2,l_3}}{\sqrt{2}},
\end{aligned}
\end{equation}

\begin{equation}
\begin{aligned}
    \mathcal{R}_2=&\prod_{\nu,l_1,l_3}\chi_{\nu,l_1,L_2/2,l_3}^{L_2/2}\left(\prod_{l_2=L_2/2+1}^{L_2-1}\chi_{\nu,l_1,l_2,l_3}\right)\\
    &\prod_{l_2=1}^{L_2/2-1}\frac{1+(-)^{l_2}\chi_{\nu,l_1,l_2,l_3}\chi_{\nu,l_1,-l_2,l_3}}{\sqrt{2}},
\end{aligned}
\end{equation}

\begin{equation}
\begin{aligned}
    \mathcal{R}_3=&\prod_{\nu,l_1,l_2}\chi_{\nu,l_1,l_2,L_3/2}^{L_3/2}\left(\prod_{l_3=L_3/2+1}^{L_3-1}\chi_{\nu,l_1,l_2,l_3}\right)\\
    &\prod_{l_3=1}^{L_3/2-1}\frac{1+(-)^{l_3}\chi_{\nu,l_1,l_2,l_3}\chi_{\nu,l_1,l_2,-l_3}}{\sqrt{2}}.
\end{aligned}
\end{equation}

Time-reversal symmetry $\mathbb{Z}_{2}^{\mathcal{T}}$ is generated by anti-unitary operator $\mathcal{T}$ defined as:

\begin{equation}
    \mathcal{T}=\mathcal{K}\prod_{\nu=even,l_1,l_2,l_3}\chi_{\nu,l_1,l_2,l_3}.
\end{equation}

One can check by straightforward calculation that with sixteen copies of root states (eight copies of lattice), we can faithfully reproduce the invariant group defined in Eq.~[\ref{Eq: G_3}].

With the $\mathbb{Z}_{16}$ SMG classification, we can assign on-site interaction term found in Sec.~\ref{Sec: 0+1d} on a given site, and generate the full interaction Hamiltonian by translations in all directions:

\begin{equation}
\begin{aligned}
  H_{int}=&\sum_{l} \chi_{1,l}\chi_{2,l}\chi_{3,l}\chi_{4,l}+\chi_{1,l}\chi_{2,l}\chi_{5,l}\chi_{6,l}\\
  &+\chi_{1,l}\chi_{3,l}\chi_{5,l}\chi_{7,l}+\chi_{2,l}\chi_{3,l}\chi_{5,l}\chi_{8,l}.
\end{aligned}
\end{equation}

\subsection{Symmetric Mass Generation for Higher Dimensions}\label{Sec:SMG_M_higher}

\subsubsection{$\mathcal{C}$-$\mathcal{R}$-$\mathcal{T}$-Internal Symmetry in Lattice Model}

Similar to the discussion in Secs.~\ref{Sec. 1+1d lattice}-\ref{Sec. 3+1d lattice}, we can promote these symmetries to a general dimension. With staggered Hamiltonian this time, we should apply some additional staggered phases in the definition of symmetry operators to preserve Hamiltonian:

We first assign translational symmetries on each direction $\mathbb{Z}_{L_i}^{T_i}$ ($i=1,2,...,d$), generated by unitary operator $T_i$ acting as:

\begin{equation}
    T_i\chi_l=(-)^{\sum_{\mu>i}l_\mu}\chi_{\tilde{T}_i(l)}T_i,\label{eq:dT}
\end{equation}
where $\tilde{T}_i(l_1,...,l_i,...,l_d)=(l_1,...,l_i+1,...,l_d)$ translate $l$ in the $i$-th direction by one site.

Reflection symmetries on each direction $\mathbb{Z}_2^{\mathcal{R}_i}$ is generated by unitary operator $\mathcal{R}_i$ acting as:

\begin{equation}
    \mathcal{R}_i\chi_l=(-)^{l_i}\chi_{\tilde{\mathcal{R}}_i(l)}\mathcal{R}_i,
\end{equation}
where $\tilde{R}_i(l_1,...,l_i,...,l_d)=(l_1,...,-l_i,...,l_d)$ reflect $l$ in the $i$-th direction about reflection plane $l_i=0$.

Finally, we can assign time-reversal symmetry $\mathbb{Z}_2^{\mathcal{T}}$, generated by anti-unitary operator $\mathcal{T}$ acting as:

\begin{equation}
    \mathcal{T}\chi_l=(-)^{\sum_\mu l_\mu}\chi_l\mathcal{T},\ \mathcal{T}\mathrm{i}=-\mathrm{i}\mathcal{T}.
\end{equation}

By checking the action of these operators on low-energy modes (i.e. $\chi_{k=0,i}$ written in sublattice basis), we can find the correspondence of low-energy symmetries on the lattice. In the continuum model, we'll find some continuous internal symmetry in the invariant group (e.g. Lie algebra in $d$=3,4,5,6,7$\mod 8$ as listed in Tab.~\ref{tab: CPT-M}). In the lattice model, these continuous symmetries break into discrete translational symmetries (the action on low-energy modes is 4-fold). More details about the correspondence between lattice and free model are collected in Tab.~\ref{Tab:cor_M}. The unitary transformations connecting staggered basis $\alpha_{stag}$ defined in Eq.~[\ref{Eq:stagbasis}] and free basis $\alpha_{free}$ defined in Eq.~[\ref{Eq:freebasis}] are listed in Appendix.~\ref{Ap. bastrans}.

\begin{table*}
\renewcommand{\arraystretch}{1.5}
\centering
\caption{The lattice correspondence of low-energy fermion parity operator $(-)^F$, fermion chiral operator $(-)^{F_L}$, operators of the $\mathbb{Z}_4^{\mathcal{J}_i}$ subgroup $\mathcal{J}_i$ broken from Lie algebra, reflection operator $\mathcal{R}_1$, time-reversal operator $\mathcal{T}$.}
    \begin{tabular}{c|ccccccc}
    \hline
      $d$ & $(-)^F$  & $(-)^{F_L}$ & $\mathcal{J}_1$  & $\mathcal{J}_2$  & $\mathcal{J}_3$  & $\mathcal{R}_1$   & $\mathcal{T}$         \\ \hline
      1         & $(T_1\mathcal{R}_1)^2$ & $T_1$           &                      &          &          & $\mathcal{R}_1$                 & $T_1\mathcal{T}$                 \\
      2         & $(T_i\mathcal{R}_i)^2$ &                 &                      &          &          & $\mathcal{R}_1$                 & $T_1\mathcal{T}$                 \\
      3         & $(T_i\mathcal{R}_i)^2$ &                 & $T_2T_1$             &          &          & $\mathcal{R}_1$                 & $T_1\mathcal{T}$                 \\
      4         & $(T_i\mathcal{R}_i)^2$ &                 & $T_2T_1$             & $T_3T_2$ & $T_3T_1$ & $T_3T_1\mathcal{R}_1$  & $T_3\mathcal{T}$  \\
      5         & $(T_i\mathcal{R}_i)^2$ & $\prod_{i=1}^4T_i $ & $T_4T_1$             & $T_4T_3$ & $T_3T_1$ & $T_2T_1\mathcal{R}_1$                 & $T_1\mathcal{T}$                 \\
      6         & $(T_i\mathcal{R}_i)^2$ &                 & $T_3T_1$             & $T_5T_1$ & $T_5T_3$ & $(\prod_{i=2}^5T_i)\mathcal{R}_1$  & $(\prod_{i=1}^5T_i)\mathcal{T}$  \\
      7         & $(T_i\mathcal{R}_i)^2$ &                 & $\prod_{i=1}^6T_i$   &          &          & $(\prod_{i=1}^4T_i)\mathcal{R}_1$           & $(\prod_{i=2}^4T_i)\mathcal{T}$           \\
      8         & $(T_i\mathcal{R}_i)^2$ &                 &                      &          &          & $(\prod_{i=2}^7T_i)\mathcal{R}_1$  & $(\prod_{i=1}^7T_i)\mathcal{T}$  \\
      \hline
    \end{tabular}\label{Tab:cor_M}
\end{table*}

In lattice model level, the invariant group is $G$ of order $2^{d+1}\prod_{i=1}^dL_i$ ($G\cong\mathbb{D}_{2L_1}\times...\times\mathbb{D}_{2L_d}\times\mathbb{Z}_2$~\cite{GAP4}), with its presentation:

\begin{equation}
  \begin{aligned}
    &T_i^{L_i}=1,\ \mathcal{R}_i^2=1,\ \mathcal{T}^2=1,\ (-)^F=(T_i\mathcal{R}_i)^2,\\
    &(-)^F(-)^F=1,\ \mathcal{R}_iT_i=(-)^FT_i^\dagger\mathcal{R}_i,\\
    &\mathcal{T}T_i=(-)^FT_i\mathcal{T},\ \mathcal{R}_i\mathcal{T}=\mathcal{TR}_i,\ \forall i=1,...,d,\\
    &T_iT_j=(-)^FT_jT_i,\ \mathcal{R}_i\mathcal{R}_j=\mathcal{R}_j\mathcal{R}_i,\\
    &\mathcal{R}_iT_j=T_j\mathcal{R}_i,\ \forall i\neq j.
  \end{aligned}\label{Eq: G}
\end{equation}

We can always find an invariant subgroup $\tilde{G}$ isomorphic to the invariant group of the $d-1$ dimensional system:

\begin{equation}
  \begin{aligned}
    &T_i^{L_i}=1,\ \mathcal{R}_i^2=1,\ \mathcal{T}^2=1,\ (-)^F=(T_i\mathcal{R}_i)^2,\\
    &(-)^F(-)^F=1,\ \mathcal{R}_iT_i=(-)^FT_i^\dagger\mathcal{R}_i,\\
    &\mathcal{T}T_i=(-)^FT_i\mathcal{T},\ \mathcal{R}_i\mathcal{T}=\mathcal{TR}_i,\ \forall i=2,...,d,\\
    &T_iT_j=(-)^FT_jT_i,\ \mathcal{R}_i\mathcal{R}_j=\mathcal{R}_j\mathcal{R}_i,\\
    &\mathcal{R}_iT_j=T_j\mathcal{R}_i,\ \forall i\neq j\geq 2,
  \end{aligned}\label{Eq: G}
\end{equation}
which encourages us to use dimensional reduction way in the classification of SMG in higher dimensions.

\subsubsection{Dimensional Reduction with $2^{d+2}/\text{dim}_{\mathbb{R}}\chi_{\mathcal{C}\ell(d,0)}$ Copies: Odd-$L$}\label{Sec:dim_reduc}

Following the discussion from $1+1$d to $3+1$d, we'll do the same dimensional reduction process by setting translational defects~\cite{10.21468/SciPostPhys.16.3.064,2023ScPP...15...51C} $T_i$ on an even-$L_i$ ($\forall i=1,...,d$) lattice to get an odd-$L_i$ ($\forall i=1,...,d$) case, with Hamiltonian:

\begin{equation}
\begin{aligned}
    H&=\frac{\mathrm{i}}{2}\sum_{l_j\neq l_i}(\sum_{l_i=0}^{L_i-2}(-)^{\sum_{\mu<i}l_\mu}\chi_{l}\chi_{\tilde{T}_i(l)}\\
    &+(-)^{\sum_{\mu\neq i}l_\mu}\chi_{l_i=L-1,l_j}\chi_{l_i=0,l_j}),
\end{aligned}
\end{equation}
with effective boundary condition $\chi_{L_i+l_i}\sim (-)^{\sum_{\mu>i}l_\mu}\chi_{l_i}$ from the action of translation $T_i$ defined in Eq.~[\ref{eq:dT}].

To construct the enlarged Hilbert space, we need to define the twisted Hamiltonian:

\begin{equation}
\begin{aligned}
    H_{tw}&=\frac{\mathrm{i}}{2}\sum_{l_j\neq l_i}(\sum_{l_i=0}^{L_i-2}(-)^{\sum_{\mu<i}l_\mu}\chi_{l}\chi_{\tilde{T}_i(l)}\\
    &-(-)^{\sum_{\mu\neq i}l_\mu}\chi_{l_i=L-1,l_j}\chi_{l_i=0,l_j}),
\end{aligned}
\end{equation}
and the enlarged Hamiltonian is given by $\tilde{H}=H\oplus H_{tw}$.

With four copies of lattice fermion and all $L_i$'s odd, we can assign the symmetry operators in the 0+1d invariant subgroup as follows:

Fermion parity $\mathbb{Z}_2^F$ is generated by unitary operator $(-)^F$ defined as:

\begin{equation}
    (-)^F=\left(\prod_{\nu,l}\chi_{\nu,l}\right)\otimes \sigma^0.
\end{equation}

Time-reversal symmetry $\mathbb{Z}_2^{\mathcal{T}}$ is generated by anti-unitary operator $\mathcal{T}$ defined as:

\begin{equation}
    \mathcal{T}=\left(\mathcal{K}\prod_{\nu=even,l}\chi_{\nu,l}\right)\otimes \sigma^1.
\end{equation}

The projectively realized invariant subgroup is $\mathbb{Z}_2^F\times\mathbb{Z}_4^{\mathcal{T}}$ as we expected, and this anomalous nature protects the gapless regime of our given system.

\subsubsection{Classification for Symmetric Mass Generation}\label{Sec:class}

With eight copies of the lattice system (i.e. 8$\times 2^d/\text{dim}_{\mathbb{R}}\chi_{\mathcal{C}\ell(d,0)}$ copies of root states), the anomaly in the 0+1d invariant subgroup cancels. Our next task is to check the whole invariant group with all $L_i$ even.

Translational symmetry on each direction $\mathbb{Z}_{L_i}^{T_i}$ is generated by unitary operator $T_i$ defined as:

\begin{equation}
    T_i=\prod_{\nu,l_j\neq l_i}\chi_{\nu,l_i=0,l_j}\prod_{l_i=0}^{L_i-2}\frac{1-(-)^{\sum_{\mu>i}l_\mu}\chi_{\nu,l}\chi_{\nu,\tilde{T}_i(l)}}{\sqrt{2}}.
\end{equation}

Reflection symmetry on each direction $\mathbb{Z}_{2}^{\mathcal{R}_i}$ is generated by unitary operator $\mathcal{R}_i$ defined as:

\begin{equation}
\begin{aligned}
    \mathcal{R}_i=&\prod_{\nu,l_j\neq l_i}\chi_{\nu,l_i=L_i/2,l_j}^{L_i/2}\left(\prod_{l_i=L_i/2+1}^{L_i-1}\chi_{\nu,l}\right)\\
    &\prod_{l_i=1}^{L_i/2-1}\frac{1+(-)^{l_i}\chi_{\nu,l}\chi_{\nu,\tilde{\mathcal{R}}_i(l)}}{\sqrt{2}}.
\end{aligned}
\end{equation}

Time-reversal symmetry $\mathbb{Z}_{2}^{\mathcal{T}}$ is generated by anti-unitary operator $\mathcal{T}$ defined as:

\begin{equation}
    \mathcal{T}=\mathcal{K}\prod_{\nu=even,l}\chi_{\nu,l}.
\end{equation}

One can check by straightforward calculation that with eight copies of lattice, we can faithfully reproduce the invariant group defined in Eq.~[\ref{Eq: G}].

With eight copies of Majorana operators on each site, we can assign on-site interaction term found in Sec.~\ref{Sec: 0+1d} on a given site, and generate the full interaction Hamiltonian by translations on all directions:

\begin{equation}
\begin{aligned}
  H_{int}=&\sum_{l} \chi_{1,l}\chi_{2,l}\chi_{3,l}\chi_{4,l}+\chi_{1,l}\chi_{2,l}\chi_{5,l}\chi_{6,l}\\
  &+\chi_{1,l}\chi_{3,l}\chi_{5,l}\chi_{7,l}+\chi_{2,l}\chi_{3,l}\chi_{5,l}\chi_{8,l}.
\end{aligned}
\end{equation}

In conclusion, we show that SMG happens with $8\times 2^d/\text{dim}_{\mathbb{R}}\mathcal{C}\ell(d,0)$ copies of root states or 8 copies of staggered fermion (2 copies of K\"ahler-Dirac fermion~\cite{2023PhRvB.108k5139G}). The classification is listed in Tab.~\ref{Tab: class}.

\begin{table}[h]
\renewcommand{\arraystretch}{1.5}
\centering
\caption{The classification of root state, staggered fermion and K\"ahler-Dirac fermion~\cite{2023PhRvB.108k5139G}. The staggered fermion is intrinsically $\nu_{stag}=2^d/$dim$_{\mathbb{R}}\chi_{\mathcal{C}\ell(d,0)}$ copies of root states and 1/4 copies of K\"ahler-Dirac fermion. $2\mathbb{R}$ and $2\mathbb{H}$ are abbreviations of $\mathbb{R}\oplus\mathbb{R}$ and $\mathbb{H}\oplus\mathbb{H}$, respectively.}
  \begin{tabular}{c|c|ccc}
  \hline
    $d\mod 8$       & $\mathcal{C}\ell(d,0)$     & root state    & staggered & K\"ahler-Dirac      \\ \hline
    0         & $\mathbb{R}(2^{\frac{d}{2}})$ & $\mathbb{Z}_{2^{\frac{d+6}{2}}}$ & $\mathbb{Z}_8$ & $\mathbb{Z}_2$ \\
    1         & $2\mathbb{R}(2^{\frac{d-1}{2}})$& $\mathbb{Z}_{2^{\frac{d+5}{2}}}$ & $\mathbb{Z}_8$ & $\mathbb{Z}_2$ \\
    2         & $\mathbb{R}(2^{\frac{d}{2}})$ & $\mathbb{Z}_{2^{\frac{d+6}{2}}}$ & $\mathbb{Z}_8$ & $\mathbb{Z}_2$ \\
    3         & $\mathbb{C}(2^{\frac{d-1}{2}})$ & $\mathbb{Z}_{2^{\frac{d+5}{2}}}$ & $\mathbb{Z}_8$ & $\mathbb{Z}_2$ \\
    4         & $\mathbb{H}(2^{\frac{d-2}{2}})$ & $\mathbb{Z}_{2^{\frac{d+4}{2}}}$ & $\mathbb{Z}_8$ & $\mathbb{Z}_2$ \\
    5         & $2\mathbb{H}(2^{\frac{d-3}{2}})$& $\mathbb{Z}_{2^{\frac{d+3}{2}}}$ & $\mathbb{Z}_8$ & $\mathbb{Z}_2$ \\
    6         & $\mathbb{H}(2^{\frac{d-2}{2}})$ & $\mathbb{Z}_{2^{\frac{d+4}{2}}}$ & $\mathbb{Z}_8$ & $\mathbb{Z}_2$ \\
    7         & $\mathbb{C}(2^{\frac{d-1}{2}})$ & $\mathbb{Z}_{2^{\frac{d+5}{2}}}$ & $\mathbb{Z}_8$ & $\mathbb{Z}_2$ \\
\hline
    \end{tabular}\label{Tab: class}
\end{table}

\section{Symmetric Mass Generation for Dirac Fermion}

\subsection{Free Dirac Fermion Model}\label{Sec:free_Dirac}

A similar discussion can be applied to Dirac fermion. In order to write down a free massless Dirac Hamiltonian in $d+1$ dimensional spacetime, we need to find $d$ Hermitian matrices $\alpha_i$ that anticommute with each other (i.e., gamma matrices in the conventional nomenclature). By coupling these matrices with momentum (i.e., $\mathrm{i}\partial_i$), we can write the Hamiltonian for free Dirac fermion as:

\begin{equation}
    H=\frac{1}{2}\int \dd^d x \psi^\dagger\left(\sum_{i=1}^d\alpha_i\mathrm{i}\partial_i\right)\psi,\label{Eq.freebas_D}
\end{equation}
where $\alpha_i$ is chosen in the complex Clifford algebra $\mathcal{C}\ell(d)$ (see Appendix.~\ref{Ap. bastrans_D} for the explicit representation), and $\psi$ and $\psi^*$ are Dirac spinor (and its anti-excitation counterpart) with dim$_{\mathbb{C}}\psi_{\mathcal{C}\ell(d)}$ flavors. 

The unitary transformation of the Dirac spinor preserving the complex Clifford algebra of $\alpha_i$ gives the internal symmetry of the Dirac system. Also, we can assign $\mathcal{C}$-$\mathcal{R}$-$\mathcal{T}$ symmetry to the Hamiltonian (For more mathematically rigorous discussion of $\mathcal{C}$-$\mathcal{R}$-$\mathcal{T}$ action on $\psi$ space and $\psi^*$ space, see Ref.~\cite{2023arXiv231217126W}), where the charge conjugation operator $\mathcal{C}$ acts as

\begin{equation}
    \mathcal{C}\psi\mathcal{C}^{-1}=M_\mathcal{C}\psi^*,\ \mathcal{C}\psi^*\mathcal{C}^{-1}=M_\mathcal{C}^*\psi,
\end{equation}
where $M_\mathcal{C}\in U(\text{dim}_{\mathbb{C}}\psi_{\mathcal{C}\ell(d)})$ is defined as

\begin{equation}
    M_\mathcal{C}^\dagger \alpha_i M_\mathcal{C}=\alpha_i^T=\alpha_i^*,\ \forall i=1,...,d.
\end{equation}
For convenience, we denote the action $\psi\to\psi^*$ (or $\psi^*\to\psi$) by $\mathcal{K}_{\mathcal{C}}$. $\mathcal{K}_{\mathcal{C}}$ is a complex conjugation applied inside of the spinor while keeping the charge conjugation a unitary transformation rather than an anti-unitary one. The action of $\mathcal{C}$ on Dirac spinors can therefore be rewritten as

\begin{equation}
    \mathcal{C}\psi\mathcal{C}^{-1}=\mathcal{K}_{\mathcal{C}}M_\mathcal{C}\psi,\ \mathcal{C}\psi^*\mathcal{C}^{-1}=\mathcal{K}_{\mathcal{C}}M_\mathcal{C}^*\psi^*.
\end{equation}

Reflection operator on the first direction $\mathcal{R}_1$ acts as

\begin{equation}
    \mathcal{R}_1\partial_i\mathcal{R}_1^{-1}=\begin{cases}
        -\partial_i,&i=1,\\
        \partial_i,&i\neq 1,
    \end{cases}
\end{equation}
and

\begin{equation}
    \mathcal{R}_1\psi\mathcal{R}_1^{-1}=M_{\mathcal{R}_1}\psi,\ \mathcal{R}_1\psi^*\mathcal{R}_1^{-1}=M_{\mathcal{R}_1}^*\psi^*,
\end{equation}
where $M_{\mathcal{R}_1}\in U(\text{dim}_{\mathbb{C}}\psi_{\mathcal{C}\ell(d)})$ is defined as

\begin{equation}
    M_{\mathcal{R}_1}^\dagger\alpha_i M_{\mathcal{R}_1}=\begin{cases}
        -\alpha_i,&i=1,\\
        \alpha_i,&i\neq 1,
    \end{cases}
\end{equation}
and time-reversal operator $\mathcal{T}$ acts as

\begin{equation}
    \mathcal{T}\psi\mathcal{T}^{-1}=\mathcal{K}M_{\mathcal{T}}\psi,\ \mathcal{T}\psi^*\mathcal{T}^{-1}=\mathcal{K}M_{\mathcal{T}}^*\psi^*,
\end{equation}
and

\begin{equation}
    \mathcal{T}\mathrm{i}\mathcal{T}^{-1}=-\mathrm{i},
\end{equation}
where $M_{\mathcal{T}}\in U(\text{dim}_{\mathbb{C}}\psi_{\mathcal{C}\ell(d)})$ is defined as

\begin{equation}
    M_{\mathcal{T}}^\dagger\alpha_i M_{\mathcal{T}}=-\alpha_i^*,\ \forall i=1,...,d.
\end{equation}

By combining the internal symmetry group and $\mathcal{C}$-$\mathcal{R}$-$\mathcal{T}$ symmetry group, we can provide the invariant group $G_{\mathcal{C}\text{-}\mathcal{R}\text{-}\mathcal{T}\text{-internal}}$, listed in Tab.~\ref{tab: CPT-D}. More details about the invariant group for free Dirac fermion are discussed in Ref.~\cite{singleparticle}.

\begin{table*}
    \renewcommand{\arraystretch}{1.5}
    \centering
    \caption{Clifford algebra $\mathcal{C}\ell (d)$, invariant group $G_{\mathcal{C}\text{-}\mathcal{R}\text{-}\mathcal{T}}$ and $G_{\mathcal{C}\text{-}\mathcal{R}\text{-}\mathcal{T}\text{-internal}}$, and symmetry operators for free Dirac fermion with space dimension $d=0,1,...,8$. The invariant group is 8-fold periodic. $U(1)^F$, $U(1)^\chi$, $\mathbb{Z}_2^{\mathcal{C}}$, $\mathbb{Z}_2^{\mathcal{R}_1}$, $\mathbb{Z}_2^\mathcal{T}$ denotes vector U(1) symmetry (with generator $(-)^F$), axial U(1) symmetry (with generator $(-)^\chi$), charge conjugation symmetry, reflection symmetry on the first direction, and time-reversal symmetry, respectively. $\mathcal{Q}$ is the charge operator acting as $\mathcal{Q}\psi=\psi$, $\mathcal{Q}\psi^*=-\psi^*$ to assign different $U(1)$ charge on $\psi$ and its anti-excitation $\psi^*$. Note that the semi-direct product in $G_{\mathcal{C}\text{-}\mathcal{R}\text{-}\mathcal{T}\text{-internal}}$ is ambiguous, see Appendix.~\ref{Ap. CRTonU(1)} for the full group presentation of $G_{\mathcal{C}\text{-}\mathcal{R}\text{-}\mathcal{T}\text{-internal}}$.}
      \begin{tabular}{c|ccc|ccccc}
      \hline
    $d$ & $\mathcal{C}\ell (d)$ & $G_{\mathcal{C}\text{-}\mathcal{R}\text{-}\mathcal{T}}$ & $G_{\mathcal{C}\text{-}\mathcal{R}\text{-}\mathcal{T}\text{-internal}}$ & $U(1)^F$ & $U(1)^{\chi}$ & $\mathbb{Z}_2^{\mathcal{C}}$ &  $\mathbb{Z}_2^{\mathcal{R}_1}$ & $\mathbb{Z}_2^{\mathcal{T}}$ \\
    \hline
     0 & $\mathbb{C}(1)$ &$\mathbb{Z}_2^\mathcal{C}\times\mathbb{Z}_2^\mathcal{T}$  &$U(1)^F\rtimes G_{\mathcal{C}\text{-}\mathcal{R}\text{-}\mathcal{T}}$  & $\mathcal{Q}$ &  & $\mathcal{K}_{\mathcal{C}}$ &    & $\mathcal{K}$  \\
     1 & $\mathbb{C}(1)\oplus \mathbb{C}(1)$ & $\mathbb{D}_8^{\mathcal{CT,C}}\times\mathbb{Z}_2^{\mathcal{R}_1\mathcal{T}}$ & $(U(1)^F\times U(1)^\chi)\rtimes_{\mathbb{Z}_2^F} G_{\mathcal{C}\text{-}\mathcal{R}\text{-}\mathcal{T}}$ & $\mathcal{Q}\sigma ^0$ & $\mathcal{Q}\sigma ^3$
     & $\mathcal{K}_{\mathcal{C}}\sigma^3$  & $\sigma^1$ & $\mathcal{K} \sigma ^1$ \\
     2 & $\mathbb{C}(2)$ & $\mathbb{D}_8^{\mathcal{T,R}_1}\times\mathbb{Z}_2^{\mathcal{C}}$ & $U(1)^F\rtimes_{\mathbb{Z}_2^F} G_{\mathcal{C}\text{-}\mathcal{R}\text{-}\mathcal{T}}$ & $\mathcal{Q}\sigma ^0$ &  & $\mathcal{K}_{\mathcal{C}}\sigma^1$ &     $\sigma^2$ & $\mathcal{K} \sigma ^2$ \\
     3 & $\mathbb{C}(2)\oplus \mathbb{C}(2)$ & $\mathbb{D}_8^{\mathcal{T,R}_1}\times\mathbb{Z}_2^{\mathcal{C}}$ & $(U(1)^F\times U(1)^\chi)\rtimes_{\mathbb{Z}_2^F} G_{\mathcal{C}\text{-}\mathcal{R}\text{-}\mathcal{T}}$ & $\mathcal{Q}\sigma ^{00}$ & $\mathcal{Q}\sigma^{30}$ & $\mathcal{K}_{\mathcal{C}}\sigma^{11}$ &    $\sigma^{12}$ & $\mathcal{K} \sigma ^{02}$ \\
     4 & $\mathbb{C}(4)$ & $\mathbb{D}_8^{\mathcal{T},\mathcal{CR}_1}\times_{\mathbb{Z}_2^F}\mathbb{Z}_4^{\mathcal{C}F}$ & $U(1)^F\rtimes_{\mathbb{Z}_2^F} G_{\mathcal{C}\text{-}\mathcal{R}\text{-}\mathcal{T}}$ & $\mathcal{Q}\sigma ^{00}$ &  & $\mathcal{K}_{\mathcal{C}}\sigma^{21}$ &    $\mathrm{i}\sigma^{32}$ & $\mathcal{K} \sigma ^{12}$ \\
     5 & $\mathbb{C}(4)\oplus \mathbb{C}(4)$ & $\mathbb{D}_8^{\mathcal{R}_1,\mathcal{CR}_1\mathcal{T}}\times_{\mathbb{Z}_2^F}\mathbb{Z}_4^{\mathcal{C}F}$ & $(U(1)^F\times U(1)^\chi)\rtimes_{\mathbb{Z}_2^F} G_{\mathcal{C}\text{-}\mathcal{R}\text{-}\mathcal{T}}$ & $\mathcal{Q}\sigma ^{000}$ &  $\mathcal{Q}\sigma^{300}$ & $\mathcal{K}_{\mathcal{C}}\sigma^{021}$ &   $\mathrm{i}\sigma ^{132}$  & $\mathcal{K} \sigma ^{112}$ \\
       6 & $\mathbb{C}(8)$ & $\mathbb{D}_8^{\mathcal{CR}_1,\mathcal{T}}\times_{\mathbb{Z}_2^F}\mathbb{Z}_4^{\mathcal{C}F}$ & $U(1)^F\rtimes_{\mathbb{Z}_2^F} G_{\mathcal{C}\text{-}\mathcal{R}\text{-}\mathcal{T}}$ & $\mathcal{Q}\sigma ^{000}$ &   & $\mathcal{K}_{\mathcal{C}}\sigma^{121}$ &   $\sigma ^{332}$  & $\mathcal{K} \sigma ^{212}$ \\
       7 & $\mathbb{C}(8)\oplus \mathbb{C}(8)$ & $\mathbb{D}_8^{\mathcal{CR}_1,\mathcal{T}}\times_{\mathbb{Z}_2^F}\mathbb{Z}_4^{\mathcal{C}F}$ & $(U(1)^F\times U(1)^\chi)\rtimes_{\mathbb{Z}_2^F} G_{\mathcal{C}\text{-}\mathcal{R}\text{-}\mathcal{T}}$ & $\mathcal{Q}\sigma ^{0000}$ &  $\mathcal{Q}\sigma^{3000}$ & $\mathcal{K}_{\mathcal{C}}\sigma^{1121}$ &   $\sigma ^{1332}$  & $\mathcal{K} \sigma ^{0212}$ \\
        8 & $\mathbb{C}(16)$ & $\mathbb{D}_8^{\mathcal{CR}_1,\mathcal{CT}}\times\mathbb{Z}_2^\mathcal{C}$ & $U(1)^F\rtimes_{\mathbb{Z}_2^F} G_{\mathcal{C}\text{-}\mathcal{R}\text{-}\mathcal{T}}$ & $\mathcal{Q}\sigma ^{0000}$ &   & $\mathcal{K}_{\mathcal{C}}\sigma^{2121}$ &   $\mathrm{i}\sigma ^{3332}$  & $\mathcal{K} \sigma ^{1212}$  \\
        \hline
        \end{tabular}\label{tab: CPT-D}
    \end{table*}

\subsection{$0+1$d Symmetric Mass Generation}\label{Sec.0+1d_D}

For Dirac fermion, SMG is more complicated than the previous case of Majorana fermion, where all the continuous Lie group symmetries break into discrete symmetry. As we'll show in the following section, the full $U(1)$ symmetry involves perturbative anomaly in the many-body system, hence driving the system into $\mathbb{Z}$ classification and obscuring the mechanism of SMG. To reconcile this, we need to break our $U(1)$ symmetry into $\mathbb{Z}_{2^n}^F$ symmetry.

\subsubsection{With Full $U(1)$ Symmetry}

We'll start up with the full invariant group in the many-body Hilbert space and promote our previous Dirac operator $\psi$ and $\psi^\dagger$ into a matrix. To do this, it's intuitive for us to decompose them into Majorana operators $\psi=(\chi^{(1)}+\mathrm{i}\chi^{(2)})/2$, $\psi^\dagger=(\chi^{(1)}-\mathrm{i}\chi^{(2)})/2$. A convenient choice for the Majorana operator basis for different copies is given by the Jordan-Wigner transformation, i.e. $\chi_n^{(1)}=(\otimes_{\nu=1}^{n-1}\sigma^3)\otimes\sigma^1$, $\chi_n^{(2)}=(\otimes_{\nu=1}^{n-1}\sigma^3)\otimes\sigma^2$, or, alternatively, $\mathcal{K}\chi_\nu^{(i)}\mathcal{K}=(-)^{i+1}\chi_\nu^{(i)}$. Now the representation of $\psi_\nu$ and $\psi^\dagger_\nu$ are all real.

In Tab.~\ref{tab: CPT-D}, we set $U(1)$ symmetry, charge conjugation symmetry $\mathbb{Z}_2^\mathcal{C}$, and time-reversal symmetry $\mathbb{Z}_2^\mathcal{T}$ in the single-particle model. In the many-body system, we need to carefully design the definition of $\mathcal{Q}$, $\mathcal{K}_\mathcal{C}$ and $\mathcal{K}$ to promote the single-particle model to a many-body version.

\subsubsection{Many-Body Symmetry with 1 Copy}

The $U(1)$ symmetry with unitary operator $\mathcal{U}(\theta)$ is defined by the action

\begin{equation}
    \mathcal{U}(\theta)\psi=e^{\mathrm{i}\theta}\psi\mathcal{U}(\theta),\ \mathcal{U}(\theta)\psi^\dagger=e^{-\mathrm{i}\theta}\psi^\dagger\mathcal{U}(\theta),
\end{equation}
which can be realized by setting charge operator $\mathcal{Q}=\psi^\dagger\psi$. The operator $\mathcal{U}(\theta)$ in the many-body Hilbert space can be written as:

\begin{equation}
\begin{aligned}
    \mathcal{U}(\theta)=&e^{-\mathrm{i}\theta\psi^\dagger\psi}=1-(1-e^{-\mathrm{i}\theta})\psi^\dagger\psi\\
    =&\frac{1+e^{-\mathrm{i}\theta}}{2}-\frac{1-e^{-\mathrm{i}\theta}}{2}\mathrm{i}\chi^{(1)}\chi^{(2)},
\end{aligned}
\end{equation}
where we've used the property $(\psi^\dagger\psi)^n=\psi^\dagger\psi$ for $\forall n>0$.

Fermion parity $\mathbb{Z}_2^F$ is generated by unitary operator $(-)^F=1-2\psi^\dagger\psi=-\mathrm{i}\chi^{(1)}\chi^{(2)}$ which can be derived from the $U(1)$ symmetry operator by $(-)^F=\mathcal{U}(\pi)$.

Charge conjugation symmetry $\mathbb{Z}_2^\mathcal{C}$ is generated by unitary operator $\mathcal{C}$ defined to flip the charge $\mathcal{Q}=\psi^\dagger\psi$. This can be realized by the following action:

\begin{equation}
    \mathcal{C}\psi=\psi^\dagger\mathcal{C},\ \mathcal{C}\psi^\dagger=\psi\mathcal{C}.\label{Eq.actC}
\end{equation}
Therefore, we can derive the following unitary operator $\mathcal{C}$ in the many-body Hilbert space as:

\begin{equation}
    \mathcal{C}=\psi^\dagger+\psi=\chi^{(1)}.
\end{equation}
Furthermore, the action given in Eq.~[\ref{Eq.actC}] can also be interpreted as a complex conjugation inside of the representation of $\chi$, which is compatible with $\mathcal{K}_\mathcal{C}$ in the single-particle model.

Time-reversal symmetry $\mathbb{Z}_2^\mathcal{T}$ is generated by anti-unitary operator $\mathcal{T}$ defined as a complex conjugation preserving the representation of $\psi$:

\begin{equation}
    \mathcal{T}\psi=\psi\mathcal{T},\ \mathcal{T}\psi^\dagger=\psi^\dagger\mathcal{T},\ \mathcal{T}\mathrm{i}=-\mathrm{i}\mathcal{T},
\end{equation}
which can be realized by

\begin{equation}
    \mathcal{T}=\mathcal{K}.
\end{equation}

Now the invariant group is given by presentation:

\begin{equation}
\begin{aligned}
    &\mathcal{U}(2\pi)=1,\ \mathcal{C}^2=1,\ \mathcal{T}^2=1,\ (-)^F=\mathcal{U}(\pi)\\
    &\mathcal{C}\mathcal{U}(\theta)={\color{red}e^{-\mathrm{i}\theta}}\mathcal{U}(-\theta)\mathcal{C},\ \mathcal{T}\mathcal{U}(\theta)=\mathcal{U}(-\theta)\mathcal{T},\\
    &\mathcal{CT}=\mathcal{TC}.
\end{aligned}
\end{equation}

In conclusion, we have

\begin{itemize}
    \item perturbative anomaly between $U(1)$ symmetry and charge conjugation symmetry $\mathbb{Z}_2^\mathcal{C}$.
\end{itemize}

\subsubsection{Many-Body Symmetry with 2 Copies}

With 2 copies of the Dirac fermion system, the operator form of charge conjugation $\mathcal{C}$ changes, so we need to check the anomaly pattern again with 2 copies. To be more specific, the operators are defined as follows:

$U(1)$ symmetry is characterized by unitary operator $\mathcal{U}(\theta)$ defined as:

\begin{equation}
\begin{aligned}
    \mathcal{U}(\theta)=&e^{-\mathrm{i}\theta\sum_{\nu=1}^2\psi_\nu^\dagger\psi_\nu}=\prod_{\nu=1}^2 \left(1-(1-e^{-\mathrm{i}\theta})\psi_\nu^\dagger\psi_\nu\right)\\
    =&\prod_{\nu=1}^2\left(\frac{1+e^{-\mathrm{i}\theta}}{2}-\frac{1-e^{-\mathrm{i}\theta}}{2}\mathrm{i}\chi_\nu^{(1)}\chi_\nu^{(2)}\right).
\end{aligned}
\end{equation}

Charge conjugation symmetry $\mathbb{Z}_2^\mathcal{C}$ is generated by unitary operator $\mathcal{C}$ defined as:

\begin{equation}
    \mathcal{C}=\mathrm{i}\prod_{\nu=1}^2\left(\psi^\dagger-\psi\right)=-\mathrm{i}\prod_{\nu=1}^2\chi_\nu^{(2)}.
\end{equation}

Time-reversal symmetry $\mathbb{Z}_2^\mathcal{T}$ is generated by anti-unitary operator $\mathcal{T}$ defined as:

\begin{equation}
    \mathcal{T}=\mathcal{K}.
\end{equation}

With 2 copies, the presentation of the invariant group becomes:

\begin{equation}
\begin{aligned}
    &\mathcal{U}(2\pi)=1,\ \mathcal{C}^2=1,\ \mathcal{T}^2=1,\ (-)^F=\mathcal{U}(\pi)\\
    &\mathcal{C}\mathcal{U}(\theta)={\color{red}e^{-2\mathrm{i}\theta}}\mathcal{U}(-\theta)\mathcal{C},\ \mathcal{T}\mathcal{U}(\theta)=\mathcal{U}(-\theta)\mathcal{T},\\
    &\mathcal{CT}={\color{red}-}\mathcal{TC}.
\end{aligned}
\end{equation}

In conclusion, we have

\begin{itemize}
    \item perturbative anomaly between $U(1)$ symmetry and charge conjugation symmetry $\mathbb{Z}_2^\mathcal{C}$,
    \item mixed $\mathbb{Z}_2$ anomaly between charge conjugation symmetry $\mathbb{Z}_2^\mathcal{C}$ and time-reversal symmetry $\mathbb{Z}_2^\mathcal{T}$.
\end{itemize}

For more copies, the mixed $\mathbb{Z}_2$ anomaly between charge conjugation symmetry $\mathbb{Z}_2^\mathcal{C}$ and time-reversal symmetry $\mathbb{Z}_2^\mathcal{T}$ cancels, while the perturbative anomaly between $U(1)$ symmetry and charge conjugation symmetry $\mathbb{Z}_2^\mathcal{C}$ persists. The perturbative anomaly brought by continuous $U(1)$ symmetry drives the system into classification $\mathbb{Z}$. To see how SMG plays a role here, it's necessary for us to break the $U(1)$ symmetry into discrete $\mathbb{Z}_{2^n}^F$ symmetry.

\subsubsection{With $\mathbb{Z}_2^{F}$ or $\mathbb{Z}_4^F$ Symmetry}

For $n\geq 3$, the result is trivial, given by classification $\mathbb{Z}_{2^n}$ since the $\mathbb{Z}_{2^n}$ anomaly cancels with $\nu=2^n$ copies. However, for $n=1,2$ (i.e. with $\mathbb{Z}_2^F$ and $\mathbb{Z}_4^F$ symmetry), the mixed $\mathbb{Z}_2$ anomaly between charge conjugation symmetry $\mathbb{Z}_2^\mathcal{C}$ and time-reversal symmetry $\mathbb{Z}_2^\mathcal{T}$ plays crucial role, and gives $\mathbb{Z}_4$ classification for SMG.

To find explicit interaction terms preserving $\mathbb{Z}_2^F$, $\mathbb{Z}_2^\mathcal{C}$, and $\mathbb{Z}_2^\mathcal{T}$, we can directly search for possible terms as listed in the Appendix.~\ref{Ap.search_D} and pick out four independent ones that anti-commute with each other (i.e. stabilizers) from these interaction terms. For convenience, we'll choose $\chi_1^{(1)}\chi_1^{(2)}\chi_2^{(1)}\chi_2^{(2)}$, $\chi_1^{(1)}\chi_1^{(2)}\chi_3^{(1)}\chi_3^{(2)}$, $\chi_1^{(1)}\chi_2^{(1)}\chi_3^{(1)}\chi_4^{(1)}$, $\chi_1^{(2)}\chi_2^{(1)}\chi_3^{(1)}\chi_4^{(2)}$ as our stabilizers to form an interaction term

\begin{equation}
\begin{aligned}
    H_{int}=&\chi_1^{(1)}\chi_1^{(2)}\chi_2^{(1)}\chi_2^{(2)}+\chi_1^{(1)}\chi_1^{(2)}\chi_3^{(1)}\chi_3^{(2)}\\
    +&\chi_1^{(1)}\chi_2^{(1)}\chi_3^{(1)}\chi_4^{(1)}+\chi_1^{(2)}\chi_2^{(1)}\chi_3^{(1)}\chi_4^{(2)}.
\end{aligned}
\end{equation}

Similar to the discussion in Sec.~\ref{Sec: 0+1d}, we can show that its eigenstates' degeneracy is 1, 4, 6, 4, 1, and the ground state is unique.

To find explicit interaction terms preserving $\mathbb{Z}_4^F$, $\mathbb{Z}_2^\mathcal{C}$, and $\mathbb{Z}_2^\mathcal{T}$, we can also directly search for possible terms as listed in the Appendix.~\ref{Ap.search_D}. For convenience, we'll choose the interaction term to be the well-known charge-4e superconducting interaction:

\begin{equation}
    H_{int}=\psi_1\psi_2\psi_3\psi_4+\psi_1^\dagger\psi_2^\dagger\psi_3^\dagger\psi_4^\dagger,
\end{equation}
and it's obvious that its eigenstates $(\ket{0000}+\ket{1111})/\sqrt{2}$ and $(\ket{0000}-\ket{1111})/\sqrt{2}$ have eigenvalues 1 and -1, while all other states have 0 eigenvalues. Therefore its eigenstates degeneracy is given by 1, 14, 1 and the ground state is unique.

In the following discussion of SMG for higher dimensions, we'll break $U(1)$ into discrete $\mathbb{Z}_4^F$ symmetry.

\subsection{$1+1$d Symmetric Mass Generation}\label{Sec:1+1d_D}

\subsubsection{Free Dirac Chain and Lattice Realization}

In the continuum model, we can use 2 Dirac fermions to describe the Hamiltonian. The two Dirac fermions are set to be $\psi_L$ and $\psi_R$, describing the left- and right-moving Dirac modes at low energy. The Hamiltonian of the free Dirac chain is then given by

\begin{equation}
    H=\frac{1}{2}\int \dd x(\psi_L^\dagger \mathrm{i}\partial_x \psi_L-\psi_R^\dagger \mathrm{i}\partial_x \psi_R)+h.c.
\end{equation}

To realize the system on a lattice, we can set a Dirac chain with $L$ sites (assuming $L\to \infty$), with Dirac fermions $\psi_l$ ($l\in\mathbb{Z}_L$, $\psi_{L+l}\sim\pm\psi_l$) on each site, defined by

\begin{equation}
    \{\psi_l,\psi_{l'}\}=\{\psi^\dagger_l,\psi^\dagger_{l'}\}=0,\ \{\psi_l,\psi^\dagger_{l'}\}=\delta_{l,l'},
\end{equation}
with nearest-neighbor hopping Hamiltonian given by

\begin{equation}
    H=\frac{\mathrm{i}}{2}\sum_{l}\psi^\dagger_l\psi_{l+1}+h.c.
\end{equation}

By Fourier transformation $\psi_k=\sum_l e^{ikl}\psi_l$, we can rewrite the lattice Hamiltonian as

\begin{equation}
    H=\sum_{k}\psi^\dagger_k(\sin k)\psi_k.
\end{equation}

The low-energy Dirac modes at $k=0,\pi$ corresponds to

\begin{equation}
    \psi_R\sim\psi_{k=0}=\sum_l \psi_l,\ \psi^\dagger_R\sim\psi^\dagger_{k=0}=\sum_l \psi^\dagger_l,
\end{equation}

\begin{equation}
    \psi_L\sim\psi_{k=\pi}=\sum_l (-)^l\psi_l,\ \psi^\dagger_L\sim\psi^\dagger_{k=\pi}=\sum_l (-)^l\psi^\dagger_l.
\end{equation}

\subsubsection{$\mathcal{C}$-$\mathcal{R}$-$\mathcal{T}$-Internal Symmetry in Continuum Model}

Before promoting our $\mathcal{C}$-$\mathcal{R}$-$\mathcal{T}$-internal symmetry to the lattice, we first briefly review their definition in the continuum limit. As demonstrated in Tab.~\ref{tab: CPT-D}, we can assign vector $U(1)^F$ symmetry, axial $U(1)^\chi$ symmetry, charge conjugation symmetry $\mathbb{Z}_2^\mathcal{C}$, reflection symmetry $\mathbb{Z}_2^\mathcal{R}$ and time-reversal symmetry $\mathbb{Z}_2^\mathcal{T}$ for $1+1$d Dirac chain.

In continuum model, the vector $U(1)^F$ symmetry is characterized by operator $\mathcal{U}^F(\theta)$ defined to apply opposite phase factors to $\psi$ and $\psi^\dagger$:

\begin{equation}
    \mathcal{U}^F(\theta)\psi_R=e^{\mathrm{i}\theta}\psi_R\mathcal{U}^F(\theta),\ \mathcal{U}^F(\theta)\psi_L=e^{\mathrm{i}\theta}\psi_L\mathcal{U}^F(\theta),
\end{equation}

\begin{equation}
    \mathcal{U}^F(\theta)\psi^\dagger_R=e^{-\mathrm{i}\theta}\psi^\dagger_R\mathcal{U}^F(\theta),\ \mathcal{U}^F(\theta)\psi^\dagger_L=e^{-\mathrm{i}\theta}\psi^\dagger_L\mathcal{U}^F(\theta).
\end{equation}

The axial $U(1)^\chi$ symmetry is characterized by operator $\mathcal{U}^\chi(\theta)$ defined to apply opposite phase factors to $\psi$ and $\psi^\dagger$, and different modes:

\begin{equation}
    \mathcal{U}^\chi(\theta)\psi_R=e^{\mathrm{i}\theta}\psi_R\mathcal{U}^\chi(\theta),\ \mathcal{U}^\chi(\theta)\psi_L=e^{-\mathrm{i}\theta}\psi_L\mathcal{U}^\chi(\theta),
\end{equation}

\begin{equation}
    \mathcal{U}^\chi(\theta)\psi^\dagger_R=e^{-\mathrm{i}\theta}\psi^\dagger_R\mathcal{U}^\chi(\theta),\ \mathcal{U}^\chi(\theta)\psi^\dagger_L=e^{\mathrm{i}\theta}\psi^\dagger_L\mathcal{U}^\chi(\theta).
\end{equation}

The charge conjugation symmetry $\mathbb{Z}_2^{\mathcal{C}}$ is generated by $\mathcal{C}$ defined to flip the charge of Dirac modes and attach an additional sign on the left-moving mode:

\begin{equation}
    \mathcal{C}\psi_R=\psi^\dagger_R\mathcal{C},\ \mathcal{C}\psi_L=-\psi^\dagger_L\mathcal{C},\label{Eq.C_D_1}
\end{equation}

\begin{equation}
    \mathcal{C}\psi^\dagger_R=\psi_R\mathcal{C},\ \mathcal{C}\psi^\dagger_L=-\psi_L\mathcal{C}.\label{Eq.C_D_2}
\end{equation}

The reflection symmetry $\mathbb{Z}_2^{\mathcal{R}}$ is generated by $\mathcal{R}$ defined to inverse the coordinate and swap the left- and right-moving modes:

\begin{equation}
    \mathcal{R}\psi_R=\psi_L\mathcal{R},\ \mathcal{R}\psi_L=\psi_R\mathcal{R},\ \mathcal{R}\partial_x=-\partial_x\mathcal{R},
\end{equation}

\begin{equation}
    \mathcal{R}\psi^\dagger_R=\psi^\dagger_L\mathcal{R},\ \mathcal{R}\psi^\dagger_L=\psi^\dagger_R\mathcal{R}.
\end{equation}

The time-reversal symmetry $\mathbb{Z}_2^\mathcal{T}$ is generated by $\mathcal{T}$ defined to swap the left- and right-moving modes and do complex conjugation in the first quantization level (i.e., preserving $\psi$'s):

\begin{equation}
    \mathcal{T}\psi_R=\psi_L\mathcal{T},\ \mathcal{T}\psi_L=\psi_R\mathcal{T},\ \mathcal{T}\mathrm{i}=-\mathrm{i}\mathcal{T},
\end{equation}

\begin{equation}
    \mathcal{T}\psi^\dagger_R=\psi^\dagger_L\mathcal{T},\ \mathcal{T}\psi^\dagger_L=\psi^\dagger_R\mathcal{T}.
\end{equation}

In free fermion level, these symmetries form the invariant group $(U(1)^F\times U(1)^\chi)\rtimes_{\mathbb{Z}_2^F}(\mathbb{D}_8^{\mathcal{CT,C}}\times\mathbb{Z}_2^{\mathcal{RT}})$.

\subsubsection{$\mathcal{C}$-$\mathcal{R}$-$\mathcal{T}$-Internal Symmetry in Lattice Model}

As we've discussed in the $0+1$d case in Sec.~\ref{Sec.0+1d_D}, the $U(1)^F$ symmetry should be broken into discrete $\mathbb{Z}_4^F$ symmetry to exposure the role of SMG. Moreover, the fermion chiral symmetry $\mathbb{Z}_2^\chi$ generated by $(-)^{F_L}$ corresponds to the translational symmetry $\mathbb{Z}_L^T$ generated by $T$, which means that trivially promote it to continuous $U(1)^\chi$ symmetry fails on lattice, and we'll only discuss the subgroup $\mathbb{Z}_2^\chi$ in the following sections.

The translational symmetry $\mathbb{Z}_L^T$ is generated by unitary operator $T$ defined by the action:

\begin{equation}
    T\psi_l=\psi_{l+1}T,\ T\psi^\dagger_l=\psi^\dagger_{l+1}T.
\end{equation}

Acting on low-energy modes, the action of $T$

\begin{equation}
    T\psi_RT^{-1}=T\left(\sum_l \psi_l\right)T^{-1}=\psi_R,
\end{equation}

\begin{equation}
    T\psi_LT^{-1}=T\left(\sum_l (-)^l\psi_l\right)T^{-1}=-\psi_L,
\end{equation}

\begin{equation}
    T\psi^\dagger_RT^{-1}=T\left(\sum_l \psi^\dagger_l\right)T^{-1}=\psi^\dagger_R,
\end{equation}

\begin{equation}
    T\psi^\dagger_LT^{-1}=T\left(\sum_l (-)^l\psi^\dagger_l\right)T^{-1}=-\psi^\dagger_L,
\end{equation}
which is exactly the action of $\mathbb{Z}_2^\chi$.

$\mathbb{Z}_4^F$ symmetry is generated by unitary operator $(-)^{F/2}$ defined to attach $\pm\mathrm{i}$ phase factor on each site:

\begin{equation}
    (-)^{F/2}\psi_l[(-)^{F/2}]^{-1}=\ii \psi_l,\ (-)^{F/2}\psi^\dagger_l[(-)^{F/2}]^{-1}=-\ii \psi^\dagger_l.
\end{equation}
Therefore, the generator of $\mathbb{Z}_4^F$ satisfies:

\begin{equation}
    [(-)^{F/2}]^2=(-)^F,\ [(-)^{F/2}]^4=1.
\end{equation}

The corresponding low-energy action is exactly $(-)^{F/2}$ in the free model.

Charge conjugation symmetry $\mathbb{Z}_2^\mathcal{C}$ is generated by unitary operator $\mathcal{C}$ defined as:

\begin{equation}
    \mathcal{C}\psi_l=\psi^\dagger_l\mathcal{C},\ \mathcal{C}\psi^\dagger_l=\psi_l\mathcal{C}.
\end{equation}

Then we'll find the lattice operator $T\mathcal{C}$ has the same action on low-energy modes as low-energy charge conjugation defined in Eq.~[\ref{Eq.C_D_1},\ref{Eq.C_D_2}].

Reflection symmetry $\mathbb{Z}_2^\mathcal{R}$ is generated by unitary operator $\mathcal{R}$ defined to reflect Dirac operators with the reflection center on the 0-th site:

\begin{equation}
    \mathcal{R}\psi_l=(-)^l\psi_{-l}\mathcal{R},\ \mathcal{R}\psi^\dagger_l=(-)^l\psi^\dagger_{-l}\mathcal{R}.
\end{equation}

Acting on low-energy modes, the action of $\mathcal{R}$ corresponds to low-energy $\mathcal{R}$.

Finally, time-reversal symmetry $\mathbb{Z}_2^\mathcal{T}$ is generated by $\mathcal{T}$ defined as:

\begin{equation}
    \mathcal{T}\psi_l=(-)^l\psi_l\mathcal{T},\ \mathcal{T}\psi^\dagger_l=(-)^l\psi^\dagger_l\mathcal{T},\ \mathcal{T}\mathrm{i}=-\mathrm{i}\mathcal{T}.
\end{equation}

Acting on low-energy modes, the action of $\mathcal{T}$ corresponds to low-energy $\mathcal{T}$.

In lattice model level, the invariant group is $G$ of order $16L$ ($G\cong\mathbb{D}_{2L}\times\mathbb{Z}_2^3$~\cite{GAP4}), given by its presentation:

\begin{equation}
  \begin{aligned}
    &T^L=1,\ \mathcal{R}^2=1,\ \mathcal{C}^2=1,\ \mathcal{T}^2=1,\ [(-)^{F/2}]^4=1,\\
    &(-)^F=[(-)^{F/2}]^2=(T\mathcal{R})^2,\ \mathcal{R}T=(-)^FT^\dagger\mathcal{R},\\
    &\mathcal{T}T=(-)^FT\mathcal{T},\ \mathcal{C}(-)^{F/2}=[(-)^{F/2}]^\dagger\mathcal{C},\\
    &\mathcal{T}(-)^{F/2}=[(-)^{F/2}]^\dagger\mathcal{T},\ T(-)^{F/2}=(-)^{F/2}T,\\
    &\mathcal{R}(-)^{F/2}=(-)^{F/2}\mathcal{R},\ \mathcal{C}T=T\mathcal{C},\ \mathcal{RT}=\mathcal{TR},\\
    &\mathcal{CR}=\mathcal{RC},\ \mathcal{CT}=\mathcal{TC}.\\
  \end{aligned}\label{Eq.inv_group_D}
\end{equation}

\subsubsection{Many-Body Symmetry with 1 Copy: Even-$L$}

We're now ready to do the second quantization process to the $\mathcal{C}$-$\mathcal{R}$-$\mathcal{T}$-internal symmetry group on the lattice, and we'll first focus on $L$=even case where reflection $\mathcal{R}$ is well-defined. For convenience, we set the condition $\mathcal{K}\chi_{\nu,l}^{(i)}\mathcal{K}=(-)^{i+1}\chi_{\nu,l}$, where $\nu=1,2,...$ labels different copies (here $\nu=1$) and $l=0,1,...$ labels the sites.

Following the process in Sec.~\ref{Sec.1+1d_1copy_e}, translation symmetry $\mathbb{Z}_L^T$ is generated by unitary operator $T$:

\begin{equation}
    T=e^{\frac{\mathrm{i}\pi}{4}L}\prod_{i=1}^2\chi_0^{(i)} \prod_{l=0}^{L-2} \frac{1-\chi_l^{(i)}\chi_{l+1}^{(i)}}{\sqrt{2}}.
\end{equation}

$\mathbb{Z}_4^F$ symmetry generated by unitary operator $(-)^{F/2}$:

\begin{equation}
\begin{aligned}
    (-)^{F/2}=&\mathrm{i}^{L/2+1}e^{-\mathrm{i}\frac{\pi}{2}\sum_{l}\psi_l^\dagger\psi_l}\\
    =&\mathrm{i}^{L/2+1}\prod_{l=0}^{L-1} \left(1-(1+\mathrm{i})\psi_l^\dagger\psi_l\right)\\
    =&\mathrm{i}^{L/2+1}\prod_{l=0}^{L-1}\left(\frac{1-\mathrm{i}}{2}-\frac{1+\mathrm{i}}{2}\mathrm{i}\chi_l^{(1)}\chi_l^{(2)}\right),
\end{aligned}
\end{equation}
where the extra $\mathrm{i}$ factor is set to satisfy $[(-)^{F/2}]^2=(T\mathcal{R})^2$.

Charge conjugation symmetry $\mathbb{Z}_2^\mathcal{C}$ is generated by unitary operator $\mathcal{C}$:

\begin{equation}
    \mathcal{C}=e^{\frac{\mathrm{i}\pi}{8}L^2}\prod_{l=0}^{L-1} \chi_l^{(2)}.
\end{equation}

Reflection symmetry $\mathbb{Z}_2^\mathcal{R}$ is generated by unitary operator $\mathcal{R}$:

\begin{equation}
    \mathcal{R}=\mathrm{i}\prod_{i=1}^2[\chi_{L/2}^{(i)}]^{L/2}\left(\prod_{l=L/2+1}^{L-1}\chi_{l}^{(i)}\right)\prod_{l=1}^{L/2-1}\frac{1+(-)^l\chi_l^{(i)}\chi_{-l}^{(i)}}{\sqrt{2}}.
\end{equation}

Time-reversal symmetry $\mathbb{Z}_2^\mathcal{T}$ is generated by anti-unitary operator $\mathcal{T}$:

\begin{equation}
    \mathcal{T}=\mathcal{K}\prod_{i=1}^2\prod_{l=odd}\chi_l^{(i)}.
\end{equation}

Equipped with all these symmetry operators, we can check if we reproduce our expected invariant group in Eq.~[\ref{Eq.inv_group_D}], and find that the invariant group is realized projectively:

\begin{equation}
  \begin{aligned}
    &T^L=1,\ \mathcal{R}^2=1,\ \mathcal{C}^2=1,\ \mathcal{T}^2=1,\ [(-)^{F/2}]^4=1,\\
    &(-)^F=[(-)^{F/2}]^2=(T\mathcal{R})^2,\ \mathcal{R}T=(-)^FT^\dagger\mathcal{R},\\
    &\mathcal{T}T=(-)^FT\mathcal{T},\ \mathcal{C}(-)^{F/2}={\color{red}-}[(-)^{F/2}]^\dagger\mathcal{C},\\
    &\mathcal{T}(-)^{F/2}=[(-)^{F/2}]^\dagger\mathcal{T},\ T(-)^{F/2}=(-)^{F/2}T,\\
    &\mathcal{R}(-)^{F/2}=(-)^{F/2}\mathcal{R},\ \mathcal{C}T={\color{red}-}T\mathcal{C},\ \mathcal{RT}=\mathcal{TR},\\
    &\mathcal{CR}={\color{red}-}\mathcal{RC},\ \mathcal{CT}=\mathcal{TC}.\\
  \end{aligned}
\end{equation}

In conclusion, we have

\begin{itemize}
    \item mixed $\mathbb{Z}_2$ anomaly between $\mathbb{Z}_4^F$ symmetry and charge conjugation symmetry $\mathbb{Z}_2^\mathcal{C}$,
    \item mixed $\mathbb{Z}_2$ anomaly between translational symmetry $\mathbb{Z}_L^T$ and charge conjugation symmetry $\mathbb{Z}_2^\mathcal{C}$,
    \item mixed $\mathbb{Z}_2$ anomaly between reflection symmetry $\mathbb{Z}_2^\mathcal{R}$ and charge conjugation symmetry $\mathbb{Z}_2^\mathcal{C}$.
\end{itemize}

\subsubsection{Many-Body Symmetry with 1 Copy: Odd-$L$}

In the odd-$L$ case where reflection $\mathcal{R}$ is ill-defined, we'll again focus on the 0+1d invariant subgroup and promote our original Hilbert space into enlarged one, and the corresponding Hamiltonian becomes

\begin{equation}
    \tilde{H}=H\oplus H_{tw}=\begin{pmatrix}
        H&0\\
        0&H_{tw}
    \end{pmatrix}.
\end{equation}

In the 0+1d subgroup $\mathbb{D}_8^{F,\mathcal{C}}\times\mathbb{Z}_2^{\mathcal{CT}}$, the symmetry operators are assigned as follows:

$\mathbb{Z}_4^F$ symmetry is generated by unitary operator $(-)^{F/2}$ defined as:

\begin{equation}
\begin{aligned}
    (-)^{F/2}=&\left(e^{-\mathrm{i}\frac{\pi}{2}\sum_{l}\psi_l^\dagger\psi_l}\right)\otimes \sigma^0\\
    =&\left(\prod_{l=0}^{L-1} 1-(1+\mathrm{i})\psi_l^\dagger\psi_l\right)\otimes \sigma^0\\
    =&\left(\prod_{l=0}^{L-1}\frac{1-\mathrm{i}}{2}-\frac{1+\mathrm{i}}{2}\mathrm{i}\chi_l^{(1)}\chi_l^{(2)}\right)\otimes \sigma^0.
\end{aligned}
\end{equation}

Charge conjugation symmetry $\mathbb{Z}_2^\mathcal{C}$ is generated by unitary operator $\mathcal{C}$ defined as:

\begin{equation}
    \mathcal{C}=\left(\prod_{l=0}^{L-1}\psi_l^\dagger+\psi_l\right)\otimes\sigma^0=\left(\prod_{l=0}^{L-1}\chi_l^{(1)}\right)\otimes\sigma^0.
\end{equation}

Time-reversal symmetry $\mathbb{Z}_2^\mathcal{T}$ is generated by anti-unitary operator $\mathcal{T}$ defined as:

\begin{equation}
    \mathcal{T}=\left(\mathcal{K}\prod_{i=1}^2\prod_{l=odd}\chi_l^{(i)}\right)\otimes\sigma^1.
\end{equation}

Now these operators give rise to the projectively realized invariant subgroup given by the presentation:

\begin{equation}
\begin{aligned}
    &\mathcal{C}^2=1,\ \mathcal{T}^2=1,\ [(-)^{F/2}]^4=1,\ (-)^F=[(-)^{F/2}]^2,\\
    &\mathcal{C}(-)^{F/2}={\color{red}(-\mathrm{i})^L}[(-)^{F/2}]^\dagger\mathcal{C},\ \mathcal{T}(-)^{F/2}=[(-)^{F/2}]^\dagger\mathcal{T},\\
    &\mathcal{CT}={\color{red}(-)^{\frac{L-1}{2}}}\mathcal{TC}.
\end{aligned}
\end{equation}

In conclusion, we have

\begin{itemize}
    \item mixed $\mathbb{Z}_4$ anomaly between $\mathbb{Z}_4^F$ symmetry and charge conjugation symmetry $\mathbb{Z}_2^\mathcal{C}$,
    \item mixed $\mathbb{Z}_2$ anomaly between charge conjugation symmetry $\mathbb{Z}_2^\mathcal{C}$ and time-reversal symmetry $\mathbb{Z}_2^\mathcal{T}$.
\end{itemize}

Except for the latter $\mathbb{Z}_2$ anomaly that vanishes at $L=1\mod 4$, we reproduce the anomaly we've seen in the previous $0+1$d analysis.

\subsubsection{Many-Body Symmetry with 2 Copies: Even-$L$}

Since all operators in the discussion of 1 copy (even-$L$) are bosonic operators, we can simply duplicate them to form the operator with 2 copies:

Translational symmetry $\mathbb{Z}_L^T$ is generated by unitary operator $T$:

\begin{equation}
    T=\prod_{i=1}^2\prod_{\nu=1}^2\chi_{\nu,0}^{(i)} \prod_{l=0}^{L-2} \frac{1-\chi_{\nu,l}^{(i)}\chi_{\nu,l+1}^{(i)}}{\sqrt{2}}.
\end{equation}

$\mathbb{Z}_4^F$ symmetry generated by unitary operator $(-)^{F/2}$:

\begin{equation}
\begin{aligned}
    (-)^{F/2}=&e^{-\mathrm{i}\frac{\pi}{2}\sum_{\nu,l}\psi_{\nu,l}^\dagger\psi_{\nu,l}}\\
    =&\prod_{\nu,l} \left(1-(1+\mathrm{i})\psi_{\nu,l}^\dagger\psi_{\nu,l}\right)\\
    =&\prod_{\nu,l}\left(\frac{1-\mathrm{i}}{2}-\frac{1+\mathrm{i}}{2}\mathrm{i}\chi_{\nu,l}^{(1)}\chi_{\nu,l}^{(2)}\right).
\end{aligned}
\end{equation}

Charge conjugation symmetry $\mathbb{Z}_2^\mathcal{C}$ is generated by unitary operator $\mathcal{C}$:

\begin{equation}
    \mathcal{C}=\prod_{\nu,l} \chi_{\nu,l}^{(2)}.
\end{equation}

Reflection symmetry $\mathbb{Z}_2^\mathcal{R}$ is generated by unitary operator $\mathcal{R}$:

\begin{equation}
\begin{aligned}
    \mathcal{R}=&\prod_{i=1}^2\prod_{\nu=1}^2[\chi_{\nu,L/2}^{(i)}]^{L/2}\left(\prod_{l=L/2+1}^{L-1}\chi_{\nu,l}^{(i)}\right)\\
    &\prod_{l=1}^{L/2-1}\frac{1+(-)^l\chi_{\nu,l}^{(i)}\chi_{\nu,-l}^{(i)}}{\sqrt{2}}.
\end{aligned}
\end{equation}

Time-reversal symmetry $\mathbb{Z}_2^\mathcal{T}$ is generated by anti-unitary operator $\mathcal{T}$:

\begin{equation}
    \mathcal{T}=\mathcal{K}\prod_{i=1}^2\prod_{\nu=1}^2\prod_{l=odd}\chi_{\nu,l}^{(i)}.
\end{equation}

These symmetry operators faithfully reproduce the expected invariant group:

\begin{equation}
  \begin{aligned}
    &T^L=1,\ \mathcal{R}^2=1,\ \mathcal{C}^2=1,\ \mathcal{T}^2=1,\ [(-)^{F/2}]^4=1,\\
    &(-)^F=[(-)^{F/2}]^2=(T\mathcal{R})^2,\ \mathcal{R}T=(-)^FT^\dagger\mathcal{R},\\
    &\mathcal{T}T=(-)^FT\mathcal{T},\ \mathcal{C}(-)^{F/2}=[(-)^{F/2}]^\dagger\mathcal{C},\\
    &\mathcal{T}(-)^{F/2}=[(-)^{F/2}]^\dagger\mathcal{T},\ T(-)^{F/2}=(-)^{F/2}T,\\
    &\mathcal{R}(-)^{F/2}=(-)^{F/2}\mathcal{R},\ \mathcal{C}T=T\mathcal{C},\ \mathcal{RT}=\mathcal{TR},\\
    &\mathcal{CR}=\mathcal{RC},\ \mathcal{CT}=\mathcal{TC}.\\
  \end{aligned}
\end{equation}

\subsubsection{Many-Body Symmetry with 2 Copies: Odd-$L$}

Though for even-$L$, the invariant group is free of anomalies, we'll find in this section that there are still anomalies underlying in the odd-$L$ case with 2 copies. Again, we can assign symmetries in the $0+1$d invariant subgroup:

$\mathbb{Z}_4^F$ symmetry is generated by unitary operator $(-)^{F/2}$ defined as:

\begin{equation}
\begin{aligned}
    (-)^{F/2}=&\left(e^{-\mathrm{i}\frac{\pi}{2}\sum_{\nu,l}\psi_{\nu,l}^\dagger\psi_{\nu,l}}\right)\otimes \sigma^0\\
    =&\left(\prod_{\nu,l} 1-(1+\mathrm{i})\psi_{\nu,l}^\dagger\psi_{\nu,l}\right)\otimes \sigma^0\\
    =&\left(\prod_{\nu,l}\frac{1-\mathrm{i}}{2}-\frac{1+\mathrm{i}}{2}\mathrm{i}\chi_{\nu,l}^{(1)}\chi_{\nu,l}^{(2)}\right)\otimes \sigma^0.
\end{aligned}
\end{equation}

Charge conjugation symmetry $\mathbb{Z}_2^\mathcal{C}$ is generated by unitary operator $\mathcal{C}$ defined as:

\begin{equation}
    \mathcal{C}=\mathrm{i}\left(\prod_{\nu,l}\psi_{\nu,l}^\dagger-\psi_{\nu,l}\right)\otimes\sigma^0=-\mathrm{i}\left(\prod_{\nu,l}\chi_{\nu,l}^{(2)}\right)\otimes\sigma^0.
\end{equation}

Time-reversal symmetry $\mathbb{Z}_2^\mathcal{T}$ is generated by anti-unitary operator $\mathcal{T}$ defined as:

\begin{equation}
    \mathcal{T}=\left(\mathcal{K}\prod_{i=1}^2\prod_{\nu=1}^2\prod_{l=odd}\chi_{\nu,l}^{(i)}\right)\otimes\sigma^1.
\end{equation}

Now these operators give rise to the projectively realized invariant subgroup given by the presentation:

\begin{equation}
\begin{aligned}
    &\mathcal{C}^2=1,\ \mathcal{T}^2=1,\ [(-)^{F/2}]^4=1,\ (-)^F=[(-)^{F/2}]^2,\\
    &\mathcal{C}(-)^{F/2}={\color{red}-}[(-)^{F/2}]^\dagger\mathcal{C},\ \mathcal{T}(-)^{F/2}=[(-)^{F/2}]^\dagger\mathcal{T},\\
    &\mathcal{CT}={\color{red}-}\mathcal{TC}.
\end{aligned}
\end{equation}

In conclusion, we have

\begin{itemize}
    \item mixed $\mathbb{Z}_2$ anomaly between $\mathbb{Z}_4^F$ symmetry and charge conjugation symmetry $\mathbb{Z}_2^\mathcal{C}$,
    \item mixed $\mathbb{Z}_2$ anomaly between charge conjugation symmetry $\mathbb{Z}_2^\mathcal{C}$ and time-reversal symmetry $\mathbb{Z}_2^\mathcal{T}$.
\end{itemize}

\subsubsection{Many-Body Symmetry with 4 Copies}

By doubling the system with 2 copies, the $\mathbb{Z}_2$ anomalies successfully cancel, and there are no more obstructions towards gapping. Indeed, we can set on-site charge-4e superconducting interaction or other interactions given in Appendix.~\ref{Ap.search_D}. The explicit interaction term is then given by:

\begin{equation}
    H_{int}=\sum_l \psi_{1,l}\psi_{2,l}\psi_{3,l}\psi_{4,l}+\psi^\dagger_{1,l}\psi^\dagger_{2,l}\psi^\dagger_{3,l}\psi^\dagger_{4,l}.
\end{equation}

\subsection{Symmetric Mass Generation for Higher Dimensions}\label{Sec:SMG_D_higher}

\subsubsection{Staggered Fermion Model}

For higher space dimension $d>1$, we can again use the staggered fermion model as lattice realization for Dirac fermion for convenience. The staggered fermion is described by the Hamiltonian:

\begin{equation}
    H=\frac{\mathrm{i}}{2}\sum_{l,\hat{\mu}}(-)^{\sum_{\nu<\mu}l_\nu}\psi^\dagger_l\psi_{l+\hat{\mu}}+h.c.,
\end{equation}
with the same hopping amplitude in the first direction, staggered hopping amplitude $(-)^{l_1}$ in the second direction, etc. Here $\mu=1,2,...,d$ denotes the spatial direction, and $\hat{\mu}$ denotes the unit vector on the $\mu$-th direction. $l$ is a $d$-component vector describing the coordinates of sites. The staggered phase describes a $\pi$-flux in each plaquette, and the unit cell is extended to $2\times...\times 2\times 1$, with sublattices $(0,...,0,0,0)$, $(0,...,0,1,0)$, $(0,...,1,0,0)$, $(0,...,1,1,0)$, ..., $(1,...,1,1,0)$.

We can transform the Hamiltonian in real space into momentum space $\psi^{k,i}=\sum_{l\mod 2^{d-1}=i}e^{\mathrm{i}k\cdot l}\psi_l$ (i labels sublattices) and write a vector $\psi_k$ that collects these $\psi_{k,i}$ components. The staggered phase minimizes the fermion doubling problem by half its Brillouin zone in the first $d-1$ directions (i.e., $-\pi/2<k_i<\pi/2$, $\forall i=1,...,d-1$). Now the only doubler is the two Weyl points $(...,0,0)$ and $(...,0,\pi)$. The Hamiltonian in momentum space reads

\begin{equation}
    H=\sum_k \psi_k^\dagger\left(\sum_{i=1}^d\sin k_i\alpha_i\right)\psi_k.\label{Eq.stagbas_D}
\end{equation}
with $\alpha_i$ the staggered matrices of dimension $2^{d-1}$, given by $\alpha_1=\sigma^{10...00}$, $\alpha_2=\sigma^{31...00}$, ..., $\alpha_{d-1}=\sigma^{33...31}$, and $\alpha_d=\sigma^{33...33}$.

Though the staggered fermion model minimizes the fermion doubling problem, it still involves intrinsic doubling. The intrinsic doubling is counted by the quotient of total components (including two Weyl points) $2^d$ by the number of components for the root states written in Clifford algebra basis dim$_{\mathbb{C}}\psi_{\mathcal{C}\ell(d)}$. The intrinsic doubling $\nu_{stag}(d)=2^d/\text{dim}_{\mathbb{C}}\psi_{\mathcal{C}\ell(d)}$ is demonstrated in Tab.~\ref{tab: stag_D}.

\begin{table}[h]
\renewcommand{\arraystretch}{1.5}
\centering
\caption{The minimum complex degree of freedom for free Majorana fermion in $d$ spatial dimension is calculated by dim$_{\mathbb{C}}\psi_{\mathcal{C}\ell(d)}$, and the complex degree of freedom ($2^{d-1}$ sublattices and $2$ Weyl points) for staggered Dirac fermion is $2^d$. The intrinsic doubling for staggered fermion is $\nu_{stag}(d)=2^d/$dim$_{\mathbb{C}}\psi_{\mathcal{C}\ell(d)}$ in different spatial dimensions.}
  \begin{tabular}{c|cc|c}
  \hline
    $d$       & $\mathcal{C}\ell(d)$     & $\text{dim}_{\mathbb{C}}\psi_{\mathcal{C}\ell(d)}$    & $\nu_{stag}(d)$       \\ \hline
    1         & $\mathbb{C}(1)\oplus\mathbb{C}(1)$       & 2                                                     & 1                     \\
    2         & $\mathbb{C}(2)$                          & 2                                                     & 2                     \\
$d$+2     & $\mathbb{C}(2)\otimes\mathcal{C}\ell(d)$  & 2$\times\text{dim}_{\mathbb{C}}\psi_{\mathcal{C}\ell(d)}$ & 2$\times\nu_{stag}(d)$\\
\hline
    \end{tabular}\label{tab: stag_D}
\end{table}

\subsubsection{$\mathcal{C}$-$\mathcal{R}$-$\mathcal{T}$-Internal Symmetry in Lattice Model}

In the higher dimension case, the definition of symmetry operators should involve some additional staggered phases in order to preserve the staggered Hamiltonian. To be concrete, we provide their actions on Dirac operators $\psi$'s as follows:

Translational symmetry on each direction $\mathbb{Z}_{L_i}^{T_i} (i=1,2,...,d)$ is generated by unitary operator $T_i$ acting as:

\begin{equation}
    T_i\psi_l=(-)^{\sum_{\mu>i}l_\mu}\psi_{\tilde{T}_i(l)}T_i,\ T_i\psi^\dagger_l=(-)^{\sum_{\mu>i}l_\mu}\psi^\dagger_{\tilde{T}_i(l)}T_i,\label{Eq.trans_D}
\end{equation}
where $\tilde{T}_i(l_1,...,l_i,...,l_d)=(l_1,...,l_i+1,l_d)$ translate $l$ in the $i$-th direction by one site.

$\mathbb{Z}_4^F$ symmetry is generated by unitary operator $(-)^{F/2}$ acting as:

\begin{equation}
    (-)^{F/2}\psi_l=\mathrm{i}\psi_l(-)^{F/2},\ (-)^{F/2}\psi^\dagger_l=-\mathrm{i}\psi^\dagger_l(-)^{F/2}.
\end{equation}

Charge conjugation symmetry $\mathbb{Z}_2^\mathcal{C}$ is generated by unitary operator $\mathcal{C}$ acting as:

\begin{equation}
    \mathcal{C}\psi_l=\psi^\dagger_l\mathcal{C},\ \mathcal{C}\psi^\dagger_l=\psi_l\mathcal{C}.
\end{equation}

Reflection symmetry on each direction $\mathbb{Z}_2^{\mathcal{R}_i}$ is generated by unitary operator $\mathcal{R}_i$ acting as:

\begin{equation}
    \mathcal{R}_i\psi_l=(-)^{l_i}\psi_{\tilde{\mathcal{R}}_i(l)}\mathcal{R}_i,\ \mathcal{R}_i\psi^\dagger_l=(-)^{l_i}\psi^\dagger_{\tilde{\mathcal{R}}_i(l)}\mathcal{R}_i,
\end{equation}
where $\tilde{\mathcal{R}}_i(l_1,...,l_i,...,l_d)=(l_1,...,-l_i,...,l_d)$ reflect $l$ in the $i$-th direction about reflection plane $l_i=0$.

Finally, we can assign time-reversal symmetry $\mathbb{Z}_2^\mathcal{T}$ generated by anti-unitary operator $\mathcal{T}$ acting as:

\begin{equation}
    \mathcal{T}\psi_l=(-)^{\sum_\mu l_\mu}\psi_l\mathcal{T},\ \mathcal{T}\psi^\dagger_l=(-)^{\sum_\mu l_\mu}\psi^\dagger_l\mathcal{T},\ \mathcal{T}\mathrm{i}=-\mathrm{i}\mathcal{T}.
\end{equation}

By checking the action of these operators on low-energy modes (i.e. $\psi_{k=0,i}$ written in sublattice basis), we can find the correspondence of low-energy symmetries (in Tab.~\ref{tab: CPT-D}) on the lattice, as listed in Tab.~\ref{Tab:cor_D}. The unitary transformations connecting staggered basis $\alpha_{stag}$ defined in Eq.~[\ref{Eq.stagbas_D}] and free basis $\alpha_{free}$ defined in Eq.~[\ref{Eq.freebas_D}] are listed in Appendix.~\ref{Ap. bastrans_D}.~\footnote{Note that the definition of $M_\mathcal{C}$ and $M_\mathcal{T}$ involves complex conjugation of $\alpha_i$, which changes under unitary basis transformation. To assign correspondence between IR and UV operators (especially $\mathcal{C}$ and $\mathcal{T}$), we need to redefine their matrices and preserve the group presentation in Appendix.~\ref{Ap. CRTonU(1)}.}

\begin{table*}
\renewcommand{\arraystretch}{1.5}
\centering
\caption{The lattice correspondence of low-energy fermion parity operator $(-)^F$, fermion chiral operator $(-)^{F_L}$, $\mathbb{Z}_4^F$ symmetry operator $(-)^{F/2}$, charge conjugation operator $\mathcal{C}$, reflection operator $\mathcal{R}_1$, and time-reversal operator $\mathcal{T}$.}
    \begin{tabular}{c|cccccccc}
    \hline
      $d$ & $(-)^F$  & $(-)^{F_L}$ &$(-)^{F/2}$ &$\mathcal{C}$  & $\mathcal{R}_1$   & $\mathcal{T}$         \\ \hline
      1         & $(T_1\mathcal{R}_1)^2$ & $T_1$  &$(-)^{F/2}$         & $T_1\mathcal{C}$                  & $\mathcal{R}_1$                             & $\mathcal{T}$                 \\
      2         & $(T_i\mathcal{R}_i)^2$ &        &$(-)^{F/2}$         & $\mathcal{C}$                  & $\mathcal{R}_1$                 & $T_1\mathcal{T}$                 \\
      3         & $(T_i\mathcal{R}_i)^2$ & $T_2T_1$  &$(-)^{F/2}$         & $\mathcal{C}$                  & $\mathcal{R}_1$                             & $T_1\mathcal{T}$                 \\
      4         & $(T_i\mathcal{R}_i)^2$ &        &$(-)^{F/2}$         & $T_3T_2\mathcal{C}$                  & $T_3T_2\mathcal{R}_1$                 & $T_1\mathcal{T}$                 \\
      5         & $(T_i\mathcal{R}_i)^2$ & $\prod_{i=1}^4T_i$  &$(-)^{F/2}$         & $(\prod_{i=1}^4T_i)\mathcal{C}$                  & $T_2T_1\mathcal{R}_1$                             & $T_4T_3T_1\mathcal{T}$                 \\
      6         & $(T_i\mathcal{R}_i)^2$ &        &$(-)^{F/2}$         & $T_2T_1\mathcal{C}$                  & $(\prod_{i=2}^5T_i)\mathcal{R}_1$                 & $(\prod_{i=3}^5T_i)\mathcal{T}$                 \\
      7         & $(T_i\mathcal{R}_i)^2$ & $\prod_{i=1}^6T_1$  &$(-)^{F/2}$         & $T_2T_1\mathcal{C}$                  & $T_3T_1\mathcal{R}_1$                             & $(\prod_{i=1}^3T_i)\mathcal{T}$                 \\
      8         & $(T_i\mathcal{R}_i)^2$ &        &$(-)^{F/2}$         & $\mathcal{C}$                  & $(\prod_{i=2}^7T_i)\mathcal{R}_1$                 & $(\prod_{i=1}^7T_i)\mathcal{T}$                 \\
      \hline
    \end{tabular}\label{Tab:cor_D}
\end{table*}

In lattice model level, the invariant group is $G$ of order $2^{d+3}\prod_{i=1}^dL_i$ ($G\cong \mathbb{D}_{2L_1}\times...\times\mathbb{D}_{2L_d}\times\mathbb{Z}_2^3$~\cite{GAP4}), with its presentation:

\begin{equation}
  \begin{aligned}
    &T_i^{L_i}=1,\ \mathcal{R}_i^2=1,\ \mathcal{C}^2=1,\ \mathcal{T}^2=1,\ [(-)^{F/2}]^4=1,\\
    &(-)^F=[(-)^{F/2}]^2=(T_i\mathcal{R}_i)^2,\ \mathcal{R}_iT_i=(-)^FT_i^\dagger\mathcal{R}_i,\\
    &\mathcal{T}T_i=(-)^FT_i\mathcal{T},\ \mathcal{C}(-)^{F/2}=[(-)^{F/2}]^\dagger \mathcal{C},\\
    &\mathcal{T}(-)^{F/2}=[(-)^{F/2}]^\dagger \mathcal{T},\ T_i(-)^{F/2}=(-)^{F/2}T_i,\\
    &\mathcal{R}_i(-)^{F/2}=(-)^{F/2}\mathcal{R}_i,\ \mathcal{C}T_i=T_i\mathcal{C},\ \mathcal{R}_i\mathcal{T}=\mathcal{TR}_i,\\
    &\mathcal{C}\mathcal{R}_i=\mathcal{R}_i\mathcal{C},\quad \mathcal{CT}=\mathcal{TC}, \ \forall i=1,2,...,d,\\
    &T_iT_j=(-)^FT_jT_i,\ \mathcal{R}_i\mathcal{R}_j=\mathcal{R}_j\mathcal{R}_i,\\
    &\mathcal{R}_iT_j=T_j\mathcal{R}_i,\ \forall i\neq j.
  \end{aligned}
\end{equation}

We can always find an invariant subgroup $\tilde{G}$ isomorphic to the invariant group of the $d$-1 dimensional system:

\begin{equation}
  \begin{aligned}
    &T_i^{L_i}=1,\ \mathcal{R}_i^2=1,\ \mathcal{C}^2=1,\ \mathcal{T}^2=1,\ [(-)^{F/2}]^4=1,\\
    &(-)^F=[(-)^{F/2}]^2=(T_i\mathcal{R}_i)^2,\ \mathcal{R}_iT_i=(-)^FT_i^\dagger\mathcal{R}_i,\\
    &\mathcal{T}T_i=(-)^FT_i\mathcal{T},\ \mathcal{C}(-)^{F/2}=[(-)^{F/2}]^\dagger \mathcal{C},\\
    &\mathcal{T}(-)^{F/2}=[(-)^{F/2}]^\dagger \mathcal{T},\ T_i(-)^{F/2}=(-)^{F/2}T_i,\\
    &\mathcal{R}_i(-)^{F/2}=(-)^{F/2}\mathcal{R}_i,\ \mathcal{C}T_i=T_i\mathcal{C},\ \mathcal{R}_i\mathcal{T}=\mathcal{TR}_i,\\
    &\mathcal{C}\mathcal{R}_i=\mathcal{R}_i\mathcal{C},\quad \mathcal{CT}=\mathcal{TC}, \ \forall i=2,...,d,\\
    &T_iT_j=(-)^FT_jT_i,\ \mathcal{R}_i\mathcal{R}_j=\mathcal{R}_j\mathcal{R}_i,\\
    &\mathcal{R}_iT_j=T_j\mathcal{R}_i,\ \forall i\neq j\geq 2.
  \end{aligned}
\end{equation}
which encourages us to use dimensional reduction way in the classification of SMG in higher dimensions.

\subsubsection{Dimensional Reduction with $2^{d+1}/\text{dim}_{\mathbb{C}}\psi_{\mathcal{C}\ell(d)}$ Copies: Odd-$L$}\label{Sec.dimred_D}

To get an odd-$L_i$ ($\forall i=1,...,d$) system where $\mathcal{R}_i$ is ill-defined, we can set translational defects~\cite{10.21468/SciPostPhys.16.3.064,2023ScPP...15...51C} $T_i$ on an even-$L_i$ ($\forall i=1,...,d$) lattice. The Hamiltonian is then given by:

\begin{equation}
\begin{aligned}
    H=&\frac{\mathrm{i}}{2}\sum_{l_j\neq l_i}(\sum_{l_i=0}^{L_i-2}(-)^{\sum_{\mu<i}l_\mu}\psi^\dagger_l\psi_{\tilde{T}_i(l)}\\
    &+(-)^{\sum_{\mu\neq i}l_\mu}\psi^\dagger_{l_i=L-1,l_j}\psi_{l_i=0,l_j})+h.c.,
\end{aligned}
\end{equation}
with effective boundary condition $\psi_{L_i+l_i}\sim(-)^{\sum_{\mu>i}l_\mu}\psi_{l_i}$ from the action of translation $T_i$ defined in Eq.~[\ref{Eq.trans_D}].

To construct the enlarged Hilbert space, we need to define the twisted Hamiltonian:

\begin{equation}
\begin{aligned}
    H_{tw}=&\frac{\mathrm{i}}{2}\sum_{l_j\neq l_i}(\sum_{l_i=0}^{L_i-2}(-)^{\sum_{\mu<i}l_\mu}\psi^\dagger_l\psi_{\tilde{T}_i(l)}\\
    &-(-)^{\sum_{\mu\neq i}l_\mu}\psi^\dagger_{l_i=L-1,l_j}\psi_{l_i=0,l_j})+h.c.,
\end{aligned}
\end{equation}
and the enlarged Hamiltonian is given by $\tilde{H}=H\oplus H_{tw}$.

With 2 copies of lattice fermion and all $L_i$'s odd, we can assign the symmetry operators in the $0+1$d invariant subgroup as follows:

$\mathbb{Z}_4^F$ symmetry is generated by unitary operator $(-)^{F/2}$ defined as:

\begin{equation}
\begin{aligned}
    (-)^{F/2}=&\left(e^{-\mathrm{i}\frac{\pi}{2}\sum_{\nu,l}\psi_{\nu,l}^\dagger\psi_{\nu,l}}\right)\otimes \sigma^0\\
    =&\left(\prod_{\nu,l} 1-(1+\mathrm{i})\psi_{\nu,l}^\dagger\psi_{\nu,l}\right)\otimes \sigma^0\\
    =&\left(\prod_{\nu,l}\frac{1-\mathrm{i}}{2}-\frac{1+\mathrm{i}}{2}\mathrm{i}\chi_{\nu,l}^{(1)}\chi_{\nu,l}^{(2)}\right)\otimes \sigma^0.
\end{aligned}
\end{equation}

Charge conjugation symmetry $\mathbb{Z}_2^\mathcal{C}$ is generated by unitary operator $\mathcal{C}$ defined as:

\begin{equation}
    \mathcal{C}=\mathrm{i}\left(\prod_{\nu,l}\psi_{\nu,l}^\dagger-\psi_{\nu,l}\right)\otimes\sigma^0=-\mathrm{i}\left(\prod_{\nu,l}\chi_{\nu,l}^{(2)}\right)\otimes\sigma^0.
\end{equation}

Time-reversal symmetry $\mathbb{Z}_2^\mathcal{T}$ is generated by anti-unitary operator $\mathcal{T}$ defined as:

\begin{equation}
    \mathcal{T}=\left(\mathcal{K}\prod_{i=1}^2\prod_{\nu=1}^2\prod_{\sum_{\mu}l_\mu=odd}\chi_{\nu,l}^{(i)}\right)\otimes\sigma^1.
\end{equation}

As we've seen in the $1+1$d case, the invariant group of 2 copies for higher dimensions is also realized projectively:

\begin{equation}
\begin{aligned}
    &\mathcal{C}^2=1,\ \mathcal{T}^2=1,\ [(-)^{F/2}]^4=1,\ (-)^F=[(-)^{F/2}]^2,\\
    &\mathcal{C}(-)^{F/2}={\color{red}-}[(-)^{F/2}]^\dagger\mathcal{C},\ \mathcal{T}(-)^{F/2}=[(-)^{F/2}]^\dagger\mathcal{T},\\
    &\mathcal{CT}={\color{red}-}\mathcal{TC}.
\end{aligned}
\end{equation}

These anomalies guarantee the gapless regime and form obstructions towards SMG.

\subsubsection{Classification for Symmetric Mass Generation}\label{Sec.class_D}

The $\mathbb{Z}_2$ anomalies above cancels with 4 copies of lattice system (i.e. $4\times 2^d/$dim$_{\mathbb{C}}\psi_{\mathcal{C}\ell(d)}$ copies of root states). Our next task is to check if the whole invariant group with all $L_i$ even faithfully reproduce the original lattice invariant group.

Translational symmetry on each direction $\mathbb{Z}_{L_i}^{T_i}$ is generated by unitary operator $T_i$:

\begin{equation}
    T_i=\prod_{k,\nu,l_j\neq l_i}\chi_{\nu,l_i=0,l_j}^{(k)} \prod_{l_i=0}^{L_i-2} \frac{1-(-)^{\sum_{\mu>i}l_\mu}\chi_{\nu,l}^{(k)}\chi_{\nu,\tilde{T}_i(l)}^{(k)}}{\sqrt{2}}.
\end{equation}

$\mathbb{Z}_4^F$ symmetry generated by unitary operator $(-)^{F/2}$:

\begin{equation}
\begin{aligned}
    (-)^{F/2}=&e^{-\mathrm{i}\frac{\pi}{2}\sum_{\nu,l}\psi_{\nu,l}^\dagger\psi_{\nu,l}}\\
    =&\prod_{\nu,l} \left(1-(1+\mathrm{i})\psi_{\nu,l}^\dagger\psi_{\nu,l}\right)\\
    =&\prod_{\nu,l}\left(\frac{1-\mathrm{i}}{2}-\frac{1+\mathrm{i}}{2}\mathrm{i}\chi_{\nu,l}^{(1)}\chi_{\nu,l}^{(2)}\right).
\end{aligned}
\end{equation}

Charge conjugation symmetry $\mathbb{Z}_2^\mathcal{C}$ is generated by unitary operator $\mathcal{C}$:

\begin{equation}
    \mathcal{C}=\prod_{\nu,l} \chi_{\nu,l}^{(2)}.
\end{equation}

Reflection symmetry on each direction $\mathbb{Z}_2^{\mathcal{R}_i}$ is generated by unitary operator $\mathcal{R}_i$:

\begin{equation}
\begin{aligned}
    \mathcal{R}_i=&\prod_{k,\nu,l_j\neq l_i}[\chi_{\nu,l_i=L_i/2,l_j}^{(k)}]^{L_i/2}\left(\prod_{l=L_i/2+1}^{L_i-1}\chi_{\nu,l}^{(k)}\right)\\
    &\prod_{l_i=1}^{L_i/2-1}\frac{1+(-)^{l_i}\chi_{\nu,l}^{(k)}\chi_{\nu,\tilde{\mathcal{R}}_i(l)}^{(k)}}{\sqrt{2}}.
\end{aligned}
\end{equation}

Time-reversal symmetry $\mathbb{Z}_2^\mathcal{T}$ is generated by anti-unitary operator $\mathcal{T}$:

\begin{equation}
    \mathcal{T}=\mathcal{K}\left(\prod_{i,\nu}\prod_{\sum_{\mu}l_\mu=odd}\chi_{\nu,l}^{(i)}\right).
\end{equation}

Through straightforward calculation, we can prove that these symmetries form exactly the original invariant group without anomalies. With 4 copies of Dirac operators on each site, we can assign on-site interaction term found in Sec.~\ref{Sec.0+1d_D} on a given site, and generate the full interaction Hamiltonian by translations on all directions:

\begin{equation}
    H_{int}=\sum_l \psi_{1,l}\psi_{2,l}\psi_{3,l}\psi_{4,l}+\psi^\dagger_{1,l}\psi^\dagger_{2,l}\psi^\dagger_{3,l}\psi^\dagger_{4,l}.
\end{equation}

In conclusion, we show that SMG happens with $4\times 2^d/$dim$_{\mathbb{C}}\psi_{\mathcal{C}\ell(d)}$ copies of root states or 4 copies of staggered Dirac fermion (2 copies of K\"ahler-Dirac fermion~\cite{2023PhRvB.108k5139G}). The classification is listed in Tab.~\ref{Tab: class_D}.

\begin{table}[h]
\renewcommand{\arraystretch}{1.5}
\centering
\caption{The classification of root state, staggered fermion and K\"ahler-Dirac fermion~\cite{2023PhRvB.108k5139G}. The staggered fermion is intrinsically $\nu_{stag}=2^d/$dim$_{\mathbb{C}}\psi_{\mathcal{C}\ell(d)}$ copies of root states and 1/2 copies of K\"ahler-Dirac fermion. $2\mathbb{C}$ is an abbreviation of $\mathbb{C}\oplus\mathbb{C}$.}
  \begin{tabular}{c|c|ccc}
  \hline
    $d\mod 2$       & $\mathcal{C}\ell(d)$     & root state    & staggered & K\"ahler-Dirac      \\ \hline
    0         & $\mathbb{C}(2^{\frac{d}{2}})$ & $\mathbb{Z}_{2^{\frac{d+4}{2}}}$ & $\mathbb{Z}_4$ & $\mathbb{Z}_2$ \\
    1         & $2\mathbb{C}(2^{\frac{d-1}{2}})$& $\mathbb{Z}_{2^{\frac{d+3}{2}}}$ & $\mathbb{Z}_4$ & $\mathbb{Z}_2$ \\
\hline
    \end{tabular}\label{Tab: class_D}
\end{table}

\section{Conclusions and Discussions}

In this work, we systematically analyzed the $\mathcal{C}$-$\mathcal{R}$-$\mathcal{T}$ symmetry fractionalization, anomaly, and symmetric mass generation (SMG) on the lattice using staggered fermion models. We assigned the corresponding lattice symmetries by analogy to the free model. We demonstrated that by promoting these symmetries to many-body Hilbert space in $1+1$d spacetime, the corresponding invariant group was realized projectively, containing multiple anomalies that form obstructions towards gapping. We further demonstrated that 8 copies of Majorana fermion (or 4 copies of Dirac fermion with $U(1)$ broken into $\mathbb{Z}_4^F$) canceled these anomalies and can be gapped by corresponding on-site interactions. For higher dimensions, we showed that by dimensional reduction process, we can reduce higher dimensional system to 0+1d case, and we demonstrated that for any space dimension $d$, every 8 copies of staggered Majorana fermion (or 4 copies of staggered Dirac fermion with $U(1)$ broken into $\mathbb{Z}_4^F$) can generate a symmetric mass through on-site interaction.

Moreover, our discussion is not confined to the staggered fermion model on a square lattice but is also geometrical. For Majorana (real) fermion, translational symmetries $\mathbb{Z}_{L_i}^{T_i}$ ($\forall i=1,...,d$) and $\mathbb{Z}_2^{\mathcal{T}}$ on the lattice are essential for the SMG. For Dirac (complex) fermion, translational symmetries $\mathbb{Z}_{L_i}^{T_i}$ ($\forall i=1,...,d$) and $\mathbb{D}_8^{F,\mathcal{C}}\times\mathbb{Z}_2^{\mathcal{CT}}$ or its subgroups (e.g. $\mathbb{Z}_4^{\mathcal{C}F}$) on lattice are essential for SMG.

Though we've specified the on-site stabilizer interaction on the lattice, we still need stronger mathematical tools to design more complicated non-on-site SMG interactions.

After we finished our work, we found Refs.~\cite{2023arXiv231109978H,2025arXiv250817115S} which also give detailed discussions on $\mathcal{C}$-$\mathcal{R}$-$\mathcal{T}$ symmetries.

\section{Acknowledgments}

We thank Zheyan Wan and Guanyu Zhu for their helpful discussions.
JW is supported by Harvard University CMSA and 
LIMS fellow fund. YZY is supported by the National Science Foundation Grant No. DMR-2238360.

\appendix

\section{Explicit On-Site Interaction for SMG of Majorana Fermion}\label{Ap.search_M}

In this appendix, we'll show the process and results to explicitly search for possible on-site interaction terms for SMG. For space dimension $d>0$, our search is restricted to the on-site interaction while neglecting those longer-range correlations. After setting our 0+1d interaction on a given site, we can immediately obtain the full interaction Hamiltonian by translations.

After setting the Jordan-Wigner basis (discussed in Sec. \ref{Sec: 0+1d})
we first search for all possible terms preserving fermion parity $\mathbb{Z}_2^F$ generated by $(-)^F=\prod_{\nu=1}^8\chi_\nu=\sigma^{3333}$ and time-reversal symmetry $\mathbb{Z}_2^{\mathcal{T}}$ generated by $\mathcal{T}=\mathcal{K}\prod_{\nu=even}\chi_\nu=-\mathcal{K}\sigma^{1212}$. These results are listed as:

\begin{equation}
\begin{aligned}
    &\chi_1\chi_2\chi_3\chi_4=-\sigma^{3300},&&\chi_1\chi_2\chi_3\chi_5=\sigma^{3210},\\
    &\chi_1\chi_2\chi_4\chi_5=-\sigma^{3110},&&\chi_1\chi_3\chi_4\chi_5=\sigma^{2010},\\
    &\chi_2\chi_3\chi_4\chi_5=-\sigma^{1010},&&\chi_1\chi_2\chi_3\chi_6=\sigma^{3220},\\
    &\chi_1\chi_2\chi_4\chi_6=-\sigma^{3120},&&\chi_1\chi_3\chi_4\chi_6=\sigma^{2020},\\
    &\chi_2\chi_3\chi_4\chi_6=-\sigma^{1020},&&\chi_1\chi_2\chi_5\chi_6=-\sigma^{3030},\\
    &\chi_1\chi_3\chi_5\chi_6=\sigma^{2130},&&\chi_2\chi_3\chi_5\chi_6=-\sigma^{1130},\\
    &\chi_1\chi_4\chi_5\chi_6=\sigma^{2230},&&\chi_2\chi_4\chi_5\chi_6=-\sigma^{1230},\\
    &\chi_3\chi_4\chi_5\chi_6=-\sigma^{0330},&&\chi_1\chi_2\chi_3\chi_7=\sigma^{3231},\\
\end{aligned}
\end{equation}
\begin{equation}
\begin{aligned}
    &\chi_1\chi_2\chi_4\chi_7=-\sigma^{3131},&&\chi_1\chi_3\chi_4\chi_7=\sigma^{2031},\\
    &\chi_2\chi_3\chi_4\chi_7=-\sigma^{1031},&&\chi_1\chi_2\chi_5\chi_7=\sigma^{3021},\\
    &\chi_1\chi_3\chi_5\chi_7=-\sigma^{2121},&&\chi_2\chi_3\chi_5\chi_7=\sigma^{1121},\\
    &\chi_1\chi_4\chi_5\chi_7=-\sigma^{2221},&&\chi_2\chi_4\chi_5\chi_7=\sigma^{1221},\\
    &\chi_3\chi_4\chi_5\chi_7=\sigma^{0321},&&\chi_1\chi_2\chi_6\chi_7=-\sigma^{3011},\\
    &\chi_1\chi_3\chi_6\chi_7=\sigma^{2111},&&\chi_2\chi_3\chi_6\chi_7=-\sigma^{1111},\\
    &\chi_1\chi_4\chi_6\chi_7=\sigma^{2211},&&\chi_2\chi_4\chi_6\chi_7=-\sigma^{1211},\\
    &\chi_3\chi_4\chi_6\chi_7=-\sigma^{0311},&&\chi_1\chi_5\chi_6\chi_7=\sigma^{2301},\\
\end{aligned}
\end{equation}
\begin{equation}
\begin{aligned}
    &\chi_2\chi_5\chi_6\chi_7=-\sigma^{1301},&&\chi_3\chi_5\chi_6\chi_7=\sigma^{0201},\\
    &\chi_4\chi_5\chi_6\chi_7=-\sigma^{0101},&&\chi_1\chi_2\chi_3\chi_8=\sigma^{3232},\\
    &\chi_1\chi_2\chi_4\chi_8=-\sigma^{3132},&&\chi_1\chi_3\chi_4\chi_8=\sigma^{2032},\\
    &\chi_2\chi_3\chi_4\chi_8=-\sigma^{1032},&&\chi_1\chi_2\chi_5\chi_8=\sigma^{3022},\\
    &\chi_1\chi_3\chi_5\chi_8=-\sigma^{2122},&&\chi_2\chi_3\chi_5\chi_8=\sigma^{1122},\\
    &\chi_1\chi_4\chi_5\chi_8=-\sigma^{2222},&&\chi_2\chi_4\chi_5\chi_8=\sigma^{1222},\\
    &\chi_3\chi_4\chi_5\chi_8=\sigma^{0322},&&\chi_1\chi_2\chi_6\chi_8=-\sigma^{3012},\\
    &\chi_1\chi_3\chi_6\chi_8=\sigma^{2112},&&\chi_2\chi_3\chi_6\chi_8=-\sigma^{1112},\\
\end{aligned}
\end{equation}
\begin{equation}
\begin{aligned}
    &\chi_1\chi_4\chi_6\chi_8=\sigma^{2212},&&\chi_2\chi_4\chi_6\chi_8=-\sigma^{1212},\\
    &\chi_3\chi_4\chi_6\chi_8=-\sigma^{0312},&&\chi_1\chi_5\chi_6\chi_8=\sigma^{2302},\\
    &\chi_2\chi_5\chi_6\chi_8=-\sigma^{1302},&&\chi_3\chi_5\chi_6\chi_8=\sigma^{0202},\\
    &\chi_4\chi_5\chi_6\chi_8=-\sigma^{0102},&&\chi_1\chi_2\chi_7\chi_8=-\sigma^{3003},\\
    &\chi_1\chi_3\chi_7\chi_8=\sigma^{2103},&&\chi_2\chi_3\chi_7\chi_8=-\sigma^{1103},\\
    &\chi_1\chi_4\chi_7\chi_8=\sigma^{2203},&&\chi_2\chi_4\chi_7\chi_8=-\sigma^{1203},\\
    &\chi_3\chi_4\chi_7\chi_8=-\sigma^{0303},&&\chi_1\chi_5\chi_7\chi_8=\sigma^{2313},\\
    &\chi_2\chi_5\chi_7\chi_8=-\sigma^{1313},&&\chi_3\chi_5\chi_7\chi_8=\sigma^{0213},\\
\end{aligned}
\end{equation}
\begin{equation}
\begin{aligned}
  &\begin{aligned}
    &\chi_4\chi_5\chi_7\chi_8=-\sigma^{0113},&&\chi_1\chi_6\chi_7\chi_8=\sigma^{2323},\\
    &\chi_2\chi_6\chi_7\chi_8=-\sigma^{1323},&&\chi_3\chi_6\chi_7\chi_8=\sigma^{0223},\\
    &\chi_4\chi_6\chi_7\chi_8=-\sigma^{0123},&&\chi_5\chi_6\chi_7\chi_8=-\sigma^{0033},
  \end{aligned}\\
  &\begin{aligned}
    &\chi_1\chi_2\chi_3\chi_4\chi_5\chi_6\chi_7\chi_8=\sigma^{3333}.
  \end{aligned}  
\end{aligned}
\end{equation}

From these interaction terms, we can pick out four independent ones that commute with each other (i.e., four stabilizers). For convenience, we'll choose $\chi_1\chi_2\chi_3\chi_4=-\sigma^{3300},$ $\chi_1\chi_2\chi_5\chi_6=-\sigma^{3030}$, $\chi_1\chi_3\chi_5\chi_7=-\sigma^{2121}$, $\chi_2\chi_3\chi_5\chi_8=\sigma^{1122}$ as our stabilizers to form an interaction term
\begin{equation}
H_{int}=\chi_1\chi_2\chi_3\chi_4+\chi_1\chi_2\chi_5\chi_6+\chi_1\chi_3\chi_5\chi_7+\chi_2\chi_3\chi_5\chi_8.
\end{equation}

Written in the qubit representation, these four stabilizers enjoy the same properties of $Z_1$, $Z_2$, $Z_3$, and $Z_4$, so we can easily obtain the degeneracy of the eigenstates is 1, 4, 6, 4, 1 and the ground state is unique.

\section{Explicit On-Site Interaction for SMG of Dirac Fermion}\label{Ap.search_D}

\subsection{$U(1)$ Breaks to $\mathbb{Z}_2^F$}

For Dirac fermions, it's convenient for us to write their operators in terms of Majorana fermion operators (i.e. $\psi_{\nu}=(\chi_{\nu}^{(1)}+\mathrm{i}\chi_{\nu}^{(2)})/2$), and $\chi_{2\nu-1}=\chi_\nu^{(1)},\chi_{2\nu}=\chi_\nu^{(2)}$ can be set in Jordan-Wigner basis (discussed in Sec.~\ref{Sec.0+1d_D}). For brevity, we denote $\chi_{\nu}^{(1)}$ and $\chi_{\nu}^{(2)}$ as $\chi_{2\nu-1}$ and $\chi_{2\nu}$. We first search for all possible terms preserving fermion parity $\mathbb{Z}_2^F$ generated by $(-)^F=\prod_{\nu=1}^8\chi_\nu=\sigma^{3333}$, charge conjugation symmetry $\mathbb{Z}_2^\mathcal{C}$ generated by $\mathcal{C}=\prod_{\nu=even}\chi_{\nu}=-\sigma^{1212}$ and time-reversal symmetry $\mathbb{Z}_2^\mathcal{T}$ generated by $\mathcal{T}=\mathcal{K}$. These results are listed as:

\begin{equation}
\begin{aligned}
    &\chi_1\chi_2\chi_3\chi_4=-\sigma^{3300},&&\chi_1\chi_2\chi_4\chi_5=-\sigma^{3110},\\
    &\chi_2\chi_3\chi_4\chi_5=-\sigma^{1010},&&\chi_1\chi_2\chi_3\chi_6=\sigma^{3220},\\
    &\chi_1\chi_3\chi_4\chi_6=\sigma^{2020},&&\chi_1\chi_2\chi_5\chi_6=-\sigma^{3030},\\
    &\chi_2\chi_3\chi_5\chi_6=-\sigma^{1130},&&\chi_1\chi_4\chi_5\chi_6=\sigma^{2230},\\
    &\chi_3\chi_4\chi_5\chi_6=-\sigma^{0330},&&\chi_1\chi_2\chi_4\chi_7=-\sigma^{3131},\\
    &\chi_2\chi_3\chi_4\chi_7=-\sigma^{1031},&&\chi_1\chi_3\chi_5\chi_7=-\sigma^{2121},\\
    &\chi_2\chi_4\chi_5\chi_7=\sigma^{1221},&&\chi_1\chi_2\chi_6\chi_7=-\sigma^{3011},\\
    &\chi_2\chi_3\chi_6\chi_7=-\sigma^{1111},&&\chi_1\chi_4\chi_6\chi_7=\sigma^{2211},\\
\end{aligned}
\end{equation}
\begin{equation}
\begin{aligned}
    &\chi_3\chi_4\chi_6\chi_7=-\sigma^{0311},&&\chi_2\chi_5\chi_6\chi_7=-\sigma^{1301},\\
    &\chi_4\chi_5\chi_6\chi_7=-\sigma^{0101},&&\chi_1\chi_2\chi_3\chi_8=\sigma^{3232},\\
    &\chi_1\chi_3\chi_4\chi_8=\sigma^{2032},&&\chi_1\chi_2\chi_5\chi_8=\sigma^{3022},\\
    &\chi_2\chi_3\chi_5\chi_8=\sigma^{1122},&&\chi_1\chi_4\chi_5\chi_8=-\sigma^{2222},\\
    &\chi_3\chi_4\chi_5\chi_8=\sigma^{0322},&&\chi_1\chi_3\chi_6\chi_8=\sigma^{2112},\\
    &\chi_2\chi_4\chi_6\chi_8=-\sigma^{1212},&&\chi_1\chi_5\chi_6\chi_8=\sigma^{2302},\\
    &\chi_3\chi_5\chi_6\chi_8=\sigma^{0202},&&\chi_1\chi_2\chi_7\chi_8=-\sigma^{3003},\\
    &\chi_2\chi_3\chi_7\chi_8=-\sigma^{1103},&&\chi_1\chi_4\chi_7\chi_8=\sigma^{2203},\\
\end{aligned}
\end{equation}
\begin{equation}
\begin{aligned}
  &\begin{aligned}
    &\chi_3\chi_4\chi_7\chi_8=-\sigma^{0303},&&\chi_2\chi_5\chi_7\chi_8=-\sigma^{1313},\\
    &\chi_4\chi_5\chi_7\chi_8=-\sigma^{0113},&&\chi_1\chi_6\chi_7\chi_8=\sigma^{2323},\\
    &\chi_3\chi_6\chi_7\chi_8=\sigma^{0223},&&\chi_5\chi_6\chi_7\chi_8=-\sigma^{0033},\\
  \end{aligned}\\
  &\begin{aligned}
    &\chi_1\chi_2\chi_3\chi_4\chi_5\chi_6\chi_7\chi_8=\sigma^{3333}.
  \end{aligned}  
\end{aligned}
\end{equation}

From these interaction terms, we can pick out four independent ones that commute with each other (i.e. four stabilizers). For convenience, we'll we'll choose $\chi_1\chi_2\chi_3\chi_4=-\sigma^{3300},$ $\chi_1\chi_2\chi_5\chi_6=-\sigma^{3030}$, $\chi_1\chi_3\chi_5\chi_7=-\sigma^{2121}$, $\chi_2\chi_3\chi_5\chi_8=\sigma^{1122}$ as our stabilizers to form an interaction term
\begin{equation}
\begin{aligned}
H_{int}=&\chi_1\chi_2\chi_3\chi_4+\chi_1\chi_2\chi_5\chi_6+\chi_1\chi_3\chi_5\chi_7+\chi_2\chi_3\chi_5\chi_8\\
=&\chi_1^{(1)}\chi_1^{(2)}\chi_2^{(1)}\chi_2^{(2)}+\chi_1^{(1)}\chi_1^{(2)}\chi_3^{(1)}\chi_3^{(2)}\\
    +&\chi_1^{(1)}\chi_2^{(1)}\chi_3^{(1)}\chi_4^{(1)}+\chi_1^{(2)}\chi_2^{(1)}\chi_3^{(1)}\chi_4^{(2)}.
\end{aligned}
\end{equation}

Written in the qubit representation, these four stabilizers enjoy the same properties of $Z_1$, $Z_2$, $Z_3$, and $Z_4$, so we can easily obtain the degeneracy of the eigenstates is 1, 4, 6, 4, 1 and the ground state is unique.

\subsection{$U(1)$ Breaks to $\mathbb{Z}_4^F$}

We first find interaction terms preserving the symmetry $\mathbb{Z}_4^F$ generated by $(-)^{F/2}=\prod_{\nu=1}^8 e^{i\frac{\pi}{2}\psi_{\nu}^\dagger\psi_\nu}=\prod_{\nu=1}^8 (1-(1-i)\psi_\nu^\dagger\psi_\nu)$. The only terms preserving $\mathbb{Z}_4^F$ are:

\begin{equation}
    \begin{aligned}
      &\psi_1\psi_2\psi_3\psi_4+h.c.\\
      = & \frac{1}{8}(\chi_1\chi_3\chi_5\chi_7-\chi_1\chi_3\chi_6\chi_8-\chi_1\chi_4\chi_5\chi_8-\chi_1\chi_4\chi_6\chi_7\\
       & -\chi_2\chi_3\chi_5\chi_8-\chi_2\chi_3\chi_6\chi_7-\chi_2\chi_4\chi_5\chi_7+\chi_2\chi_4\chi_6\chi_8),\\
      &\psi_1\psi_2\psi_3^\dagger\psi_4^\dagger+h.c.\\
      = & \frac{1}{8}(\chi_1\chi_3\chi_5\chi_7-\chi_1\chi_3\chi_6\chi_8+\chi_1\chi_4\chi_5\chi_8+\chi_1\chi_4\chi_6\chi_7\\
       & +\chi_2\chi_3\chi_5\chi_8+\chi_2\chi_3\chi_6\chi_7-\chi_2\chi_4\chi_5\chi_7+\chi_2\chi_4\chi_6\chi_8),\\
      &\psi_1\psi_2^\dagger\psi_3\psi_4^\dagger+h.c.\\
      = & \frac{1}{8}(\chi_1\chi_3\chi_5\chi_7+\chi_1\chi_3\chi_6\chi_8-\chi_1\chi_4\chi_5\chi_8+\chi_1\chi_4\chi_6\chi_7\\
       & +\chi_2\chi_3\chi_5\chi_8-\chi_2\chi_3\chi_6\chi_7+\chi_2\chi_4\chi_5\chi_7+\chi_2\chi_4\chi_6\chi_8),\\
      &\psi_1\psi_2^\dagger\psi_3^\dagger\psi_4+h.c.\\
      = & \frac{1}{8}(\chi_1\chi_3\chi_5\chi_7+\chi_1\chi_3\chi_6\chi_8+\chi_1\chi_4\chi_5\chi_8-\chi_1\chi_4\chi_6\chi_7\\
       & -\chi_2\chi_3\chi_5\chi_8+\chi_2\chi_3\chi_6\chi_7+\chi_2\chi_4\chi_5\chi_7+\chi_2\chi_4\chi_6\chi_8).
    \end{aligned}\label{Eq.cccc}
\end{equation}

We can further check that the eight four-Majorana terms included in Eq.~[\ref{Eq.cccc}]

\begin{equation}
  \begin{aligned}
    &\chi_1\chi_3\chi_5\chi_7=-\sigma^{2121},&&\chi_1\chi_3\chi_6\chi_8=\sigma^{2112},\\
    &\chi_1\chi_4\chi_5\chi_8=-\sigma^{2222},&&\chi_1\chi_4\chi_6\chi_7=\sigma^{2211},\\
    &\chi_2\chi_3\chi_5\chi_8=\sigma^{1122},&&\chi_2\chi_3\chi_6\chi_7=-\sigma^{1111},\\
    &\chi_2\chi_4\chi_5\chi_7=\sigma^{1221},&&\chi_2\chi_4\chi_6\chi_8=-\sigma^{1212}
  \end{aligned}
\end{equation}
preserve fermion parity $(-)^F=\prod_{\nu=1}^8\chi_{\nu}=\sigma^{3333}$, charge conjugation $\mathcal{C}=\prod_{\nu=even}\chi_{\nu}=\sigma^{1212}$ and time-reversal symmetry 
$\mathcal{T}=\mathcal{K}$.

Our next task is to check if these interaction terms can generate a mass, i.e., if there are exactly 4 independent stabilizers that commute with each other, and we find that by setting stabilizers:

\begin{equation}
\begin{aligned}
  &Z_1=\chi_1\chi_3\chi_5\chi_7=-\sigma^{2121},\ Z_2=\chi_1\chi_3\chi_6\chi_8=\sigma^{2112},\\
  &Z_3=\chi_1\chi_4\chi_5\chi_8=-\sigma^{2222},\ Z_4=\chi_2\chi_3\chi_5\chi_8=\sigma^{1122},
\end{aligned}
\end{equation}
we can rewrite the interaction terms above as:

\begin{equation}
  \begin{aligned}
    &\psi_1\psi_2\psi_3\psi_4+h.c.\\
    = & \frac{1}{8}(Z_1-Z_2-Z_3+Z_1Z_2Z_3-Z_4+Z_1Z_2Z_4\\
    & +Z_1Z_3Z_4-Z_2Z_3Z_4),\\
    &\psi_1\psi_2\psi_3^\dagger\psi_4^\dagger+h.c.\\
    = & \frac{1}{8}(Z_1-Z_2+Z_3-Z_1Z_2Z_3+Z_4-Z_1Z_2Z_4\\
    & +Z_1Z_3Z_4-Z_2Z_3Z_4),\\
    &\psi_1\psi_2^\dagger\psi_3\psi_4^\dagger+h.c.\\
    = & \frac{1}{8}(Z_1+Z_2-Z_3-Z_1Z_2Z_3+Z_4+Z_1Z_2Z_4\\
    &-Z_1Z_3Z_4-Z_2Z_3Z_4),\\
    &\psi_1\psi_2^\dagger\psi_3^\dagger\psi_4+h.c.\\
    = & \frac{1}{8}(Z_1+Z_2+Z_3+Z_1Z_2Z_3-Z_4-Z_1Z_2Z_4\\
    &-Z_1Z_3Z_4-Z_2Z_3Z_4),
  \end{aligned}
\end{equation}
they all have unique ground state $(Z_1,Z_2,Z_3,Z_4)=(-1,1,1,1),(-1,1,-1,-1),(-1,-1,1,-1),(-1,-1,-1,1)$ with energy -1, respectively. The degeneracy of these interaction terms is 1, 14, 1.

Equipped with this information, we can set the interaction to be:

\begin{equation}
    H_{int}=\psi_1\psi_2\psi_3\psi_4+\psi_1^\dagger\psi_2^\dagger\psi_3^\dagger\psi_4^\dagger,
\end{equation}
which is exactly the four-fermion condensation interaction in charge-4e superconductors.

\section{Commutation Relations in $1+1$d Majorana Chain}\label{Ap. 1+1dchain}

\subsection{Invariant Group on Lattice}

From the group presentation

\begin{equation}
  \begin{aligned}
    &T^L=1,\ \mathcal{R}^2=1,\ \mathcal{T}^2=1,\ (-)^F=(T\mathcal{R})^2,\\
    &(-)^F(-)^F=1,\ \mathcal{R}T=(-)^FT^\dagger\mathcal{R},\\
    &\mathcal{T}T=(-)^FT\mathcal{T},\ \mathcal{RT}=\mathcal{TR},
  \end{aligned}
\end{equation}
we can also derive the following relations:

Commutation relation between translation $T$ and reflection $\mathcal{R}$:

\begin{equation}
  \mathcal{R}T=\mathcal{R}T(-)^F(-)^F=((-)^FT^\dagger\mathcal{R})^\dagger (-)^F=T^\dagger\mathcal{R}(-)^F,
\end{equation}

\begin{equation}
  T\mathcal{R}=T\mathcal{R}TT^\dagger=T\mathcal{R}T\mathcal{R}^2T^\dagger=(-)^F\mathcal{R}T^\dagger.
\end{equation}

Commutation relation between translation $T$ and time-reversion $\mathcal{T}$:

\begin{equation}
  T\mathcal{T}=(-)^F(-)^FT\mathcal{T}=(-)^F\mathcal{T}T.
\end{equation}

Commutation relation between translation $T$ and fermion parity $(-)^F$:

\begin{equation}
  (-)^FT=T\mathcal{R}T\mathcal{R}T=T\mathcal{R}TT^\dagger\mathcal{R}(-)^F=T(-)^F.
\end{equation}

Commutation relation between reflection $\mathcal{R}$ and fermion parity $(-)^F$:

\begin{equation}
  (-)^F\mathcal{R}=T\mathcal{R}T=TT^\dagger\mathcal{R}(-)^F=\mathcal{R}(-)^F.
\end{equation}

Commutation relation between time-reversion $\mathcal{T}$ and fermion parity $(-)^F$:

\begin{equation}
\begin{aligned}
  (-)^F\mathcal{T}=&(-)^F(-)^F(-)^F\mathcal{T}=(-)^FT\mathcal{P}(-)^FT\mathcal{PT}\\
  =&(-)^FT\mathcal{PT}T\mathcal{P}=\mathcal{T}T\mathcal{P}T\mathcal{P}=\mathcal{T}(-)^F.
\end{aligned}
\end{equation}

\subsection{Projective Invariant Group on Lattice}

In the projective invariant group of 1 copy:

\begin{equation}
  \begin{aligned}
    &T^L=1,\ \mathcal{R}^2=1,\ \mathcal{T}^2=1,\ (-)^F=(T\mathcal{R})^2,\\
    &(-)^F(-)^F={\color{red}-1},\ \mathcal{R}T={\color{red}-}(-)^FT^\dagger\mathcal{R},\\
    &\mathcal{T}T={\color{red}-}(-)^FT\mathcal{T},\ \mathcal{RT}=\mathcal{TR},
  \end{aligned}
\end{equation}
these relations may attach an additional minus sign:

Commutation relation between translation $T$ and reflection $\mathcal{R}$:

\begin{equation}
  \mathcal{R}T={\color{red}-}\mathcal{R}T(-)^F(-)^F=((-)^FT^\dagger\mathcal{R})^\dagger (-)^F={\color{red}-}T^\dagger\mathcal{R}(-)^F,
\end{equation}

\begin{equation}
  T\mathcal{R}=T\mathcal{R}TT^\dagger=T\mathcal{R}T\mathcal{R}^2T^\dagger=(-)^F\mathcal{R}T^\dagger.
\end{equation}

Commutation relation between translation $T$ and time-reversion $\mathcal{T}$:

\begin{equation}
  T\mathcal{T}={\color{red}-}(-)^F(-)^FT\mathcal{T}=(-)^F\mathcal{T}T.
\end{equation}

Commutation relation between translation $T$ and fermion parity $(-)^F$:

\begin{equation}
  (-)^FT=T\mathcal{R}T\mathcal{R}T={\color{red}-}T\mathcal{R}TT^\dagger\mathcal{R}(-)^F={\color{red}-}T(-)^F.
\end{equation}

Commutation relation between reflection $\mathcal{R}$ and fermion parity $(-)^F$:

\begin{equation}
  (-)^F\mathcal{R}=T\mathcal{R}T={\color{red}-}TT^\dagger\mathcal{R}(-)^F={\color{red}-}\mathcal{R}(-)^F.
\end{equation}

Commutation relation between time-reversion $\mathcal{T}$ and fermion parity $(-)^F$:

\begin{equation}
\begin{aligned}
  (-)^F\mathcal{T}=&{\color{red}-}(-)^F(-)^F(-)^F\mathcal{T}={\color{red}-}(-)^FT\mathcal{P}(-)^FT\mathcal{PT}\\
  =&(-)^FT\mathcal{PT}T\mathcal{P}={\color{red}-}\mathcal{T}T\mathcal{P}T\mathcal{P}={\color{red}-}\mathcal{T}(-)^F.
\end{aligned}
\end{equation}

For the projectively realized invariant group of more copies, we can go through the same process above to derive these commutation relations.

\section{Unitary Transformation between Staggered Basis and Free Basis of Majorana Fermion}\label{Ap. bastrans}

The explicit representation of free basis $\alpha_{free}$ and staggered basis $\alpha_{stag}$ is given in Tabs.~\ref{tab: freebas},\ref{tab: stagbas}.

\begin{table}[h]
\renewcommand{\arraystretch}{1.5}
\centering
\caption{Explicit representation of free model basis.}
  \begin{tabular}{c|cccccccc}
  \hline
    $d$       & $\alpha_1$     & $\alpha_2$    & $\alpha_3$  & $\alpha_4$   & $\alpha_5$   & $\alpha_6$   & $\alpha_7$   & $\alpha_8$     \\ \hline
    1         & $\sigma^3$    & & & & & & & \\
    2         & $\sigma^1$    & $\sigma^3$   & & & & & & \\
    3         & $\sigma^{10}$ &$\sigma^{22}$ &$\sigma^{30}$  & & & & &  \\
    4         & $\sigma^{100}$ &$\sigma^{212}$ &$\sigma^{220}$ &$\sigma^{300}$  & & & &  \\
    5         & $\sigma^{3100}$ &$\sigma^{3212}$ &$\sigma^{3220}$ &$\sigma^{3232}$ &$\sigma^{3300}$  & & &  \\
    6         & $\sigma^{1000}$ & $\sigma^{3100}$ &$\sigma^{3212}$ &$\sigma^{3220}$ &$\sigma^{3232}$ &$\sigma^{3300}$  & &  \\
    7         & $\sigma^{1000}$ & $\sigma^{2002}$ & $\sigma^{3100}$ &$\sigma^{3212}$ &$\sigma^{3220}$ &$\sigma^{3232}$ &$\sigma^{3300}$ &    \\
    8         & $\sigma^{1000}$ & $\sigma^{2002}$ & $\sigma^{2021}$ & $\sigma^{3100}$ &$\sigma^{3212}$ &$\sigma^{3220}$ &$\sigma^{3232}$ &$\sigma^{3300}$    \\
\hline
    \end{tabular}\label{tab: freebas}
\end{table}

\begin{table}[h]
\renewcommand{\arraystretch}{1.5}
\centering
\caption{Explicit representation of staggered model basis. (Some $\otimes \sigma^0$s are neglected for the sake of brevity.)}
  \begin{tabular}{c|cccccccc}
  \hline
    $d$       & $\alpha_1$     & $\alpha_2$    & $\alpha_3$  & $\alpha_4$   & $\alpha_5$   & $\alpha_6$   & $\alpha_7$   & $\alpha_8$     \\ \hline
    1         & $\sigma^1$    & & & & & & & \\
    2         & $\sigma^1$    & $\sigma^3$   & & & & & & \\
    3         & $\sigma^{1}$ &$\sigma^{31}$ &$\sigma^{33}$  & & & & &  \\
    4         & $\sigma^{1}$ &$\sigma^{31}$ &$\sigma^{331}$ &$\sigma^{333}$  & & & &  \\
    5         & $\sigma^{1}$ &$\sigma^{31}$ &$\sigma^{331}$ &$\sigma^{3331}$ &$\sigma^{3333}$  & & &  \\
    6         & $\sigma^{1}$ & $\sigma^{31}$ &$\sigma^{331}$ &$\sigma^{3331}$ &$\sigma^{33331}$ &$\sigma^{33333}$  & &  \\
    7         & $\sigma^{1}$ & $\sigma^{31}$ & $\sigma^{331}$ &$\sigma^{3331}$ &$\sigma^{33331}$ &$\sigma^{333331}$ &$\sigma^{333333}$ &    \\
    8         & $\sigma^{1}$ & $\sigma^{31}$ & $\sigma^{331}$ & $\sigma^{3331}$ &$\sigma^{33331}$ &$\sigma^{333331}$ &$\sigma^{3333331}$ &$\sigma^{3333333}$    \\
\hline
    \end{tabular}\label{tab: stagbas}
\end{table}

For $d=1$, the unitary transformation defined by $\alpha_{stag}=U\alpha_{free}U^\dagger$ is given by:

\begin{equation}
    U=e^{\frac{\mathrm{i}\pi}{4}\sigma^2}.
\end{equation}

For $d=2$, the unitary transformation defined by $\alpha_{stag}=U\alpha_{free}U^\dagger$ is given by:

\begin{equation}
    U=\sigma^0.
\end{equation}

For $d=3$, the unitary transformation defined by $\alpha_{stag}=U\alpha_{free}U^\dagger$ is given by:

\begin{equation}
    U=e^{\frac{\mathrm{i}\pi}{4}\sigma^{10}}e^{-\frac{\mathrm{i}\pi}{4}\sigma^{03}}e^{-\frac{\mathrm{i}\pi}{4}\sigma^{13}}.
\end{equation}

For $d=4$, the unitary transformation defined by $\alpha_{stag}=U\alpha_{free}U^\dagger$ is given by:

\begin{equation}
  U=e^{\frac{i\pi}{4}\sigma^{013}}e^{\frac{i\pi}{4}\sigma^{131}}e^{-\frac{i\pi}{4}\sigma^{122}}e^{\frac{i\pi}{4}\sigma^{102}}.
\end{equation}

For $d=5$, the unitary transformation defined by $\alpha_{stag}=U\alpha_{free}U^\dagger$ is given by:

\begin{equation}
\begin{aligned}
  U=&e^{-\frac{i\pi}{4}\sigma^{1213}}e^{\frac{i\pi}{4}\sigma^{0013}} e^{-\frac{i\pi}{4}\sigma^{1330}} e^{\frac{i\pi}{4}\sigma^{0123}}\\
  & e^{\frac{i\pi}{4}\sigma^{1310}} e^{-\frac{i\pi}{4}\sigma^{1220}}e^{-\frac{i\pi}{4}\sigma^{0312}}e^{\frac{i\pi}{4}\sigma^{2100}}.
\end{aligned}
\end{equation}

For $d=6$, the unitary transformation defined by $\alpha_{stag}=U(\alpha_{free}\otimes \sigma^0)U^\dagger$ is given by:

\begin{equation}
\begin{aligned}
  U=&e^{\frac{i\pi}{4}\sigma^{12133}}e^{\frac{i\pi}{4}\sigma^{13211}} e^{\frac{i\pi}{4}\sigma^{00130}} e^{-\frac{i\pi}{4}\sigma^{13310}} \\
  &e^{\frac{i\pi}{4}\sigma^{13220}}e^{\frac{i\pi}{4}\sigma^{01020}}.
\end{aligned}
\end{equation}

For $d=7$, the unitary transformation defined by $\alpha_{stag}=U(\alpha_{free}\otimes \sigma^{00})U^\dagger$ is given by:

\begin{equation}
\begin{aligned}
  U=&e^{\frac{i\pi}{4}\sigma^{132133}}e^{-\frac{i\pi}{4}\sigma^{121331}} e^{\frac{i\pi}{4}\sigma^{120110}} e^{-\frac{i\pi}{4}\sigma^{012300}} \\
  &e^{-\frac{i\pi}{4}\sigma^{131200}}e^{\frac{i\pi}{4}\sigma^{110200}}.
\end{aligned}
\end{equation}

For $d=8$, the unitary transformation defined by $\alpha_{stag}=U(\alpha_{free}\otimes \sigma^{000})U^\dagger$ is given by:

\begin{equation}
\begin{aligned}
  U=&e^{-\frac{i\pi}{4}\sigma^{1321333}}e^{\frac{i\pi}{4}\sigma^{1213331}} e^{-\frac{i\pi}{4}\sigma^{1201310}} e^{\frac{i\pi}{4}\sigma^{0121100}} \\
  &e^{-\frac{i\pi}{4}\sigma^{0002000}}e^{\frac{i\pi}{4}\sigma^{1312000}}e^{\frac{i\pi}{4}\sigma^{0023000}}e^{\frac{i\pi}{4}\sigma^{1102000}}.
\end{aligned}
\end{equation}

\section{Unitary Transformation between Staggered Basis and Free Basis of Dirac Fermion}\label{Ap. bastrans_D}

The explicit representation of free basis $\alpha_{free}$ and staggered basis $\alpha_{stag}$ is given in Tabs.~\ref{tab: freebas_D},\ref{tab: stagbas_D}.

\begin{table}[h]
\renewcommand{\arraystretch}{1.5}
\centering
\caption{Explicit representation of free model basis.}
  \begin{tabular}{c|cccccccc}
  \hline
    $d$       & $\alpha_1$     & $\alpha_2$    & $\alpha_3$  & $\alpha_4$   & $\alpha_5$   & $\alpha_6$   & $\alpha_7$   & $\alpha_8$     \\ \hline
    1         & $\sigma^3$    & & & & & & & \\
    2         & $\sigma^1$    & $\sigma^2$   & & & & & & \\
    3         & $\sigma^{01}$ &$\sigma^{02}$ &$\sigma^{33}$  & & & & &  \\
    4         & $\sigma^{01}$ &$\sigma^{02}$ &$\sigma^{13}$ &$\sigma^{23}$  & & & &  \\
    5         & $\sigma^{001}$ &$\sigma^{002}$ &$\sigma^{013}$ &$\sigma^{023}$ &$\sigma^{333}$  & & &  \\
    6         & $\sigma^{001}$ & $\sigma^{002}$ &$\sigma^{013}$ &$\sigma^{023}$ &$\sigma^{133}$ &$\sigma^{233}$  & &  \\
    7         & $\sigma^{0001}$ & $\sigma^{0002}$ & $\sigma^{0013}$ &$\sigma^{0023}$ &$\sigma^{0133}$ &$\sigma^{0233}$ &$\sigma^{3333}$ &    \\
    8         & $\sigma^{0001}$ & $\sigma^{0002}$ & $\sigma^{0013}$ & $\sigma^{0023}$ &$\sigma^{0133}$ &$\sigma^{0233}$ &$\sigma^{1333}$ &$\sigma^{2333}$    \\
\hline
    \end{tabular}\label{tab: freebas_D}
\end{table}

\begin{table}[h]
\renewcommand{\arraystretch}{1.5}
\centering
\caption{Explicit representation of staggered model basis. (Some $\otimes \sigma^0$s are neglected for the sake of brevity.)}
  \begin{tabular}{c|cccccccc}
  \hline
    $d$       & $\alpha_1$     & $\alpha_2$    & $\alpha_3$  & $\alpha_4$   & $\alpha_5$   & $\alpha_6$   & $\alpha_7$   & $\alpha_8$     \\ \hline
    1         & $\sigma^1$    & & & & & & & \\
    2         & $\sigma^1$    & $\sigma^3$   & & & & & & \\
    3         & $\sigma^{1}$ &$\sigma^{31}$ &$\sigma^{33}$  & & & & &  \\
    4         & $\sigma^{1}$ &$\sigma^{31}$ &$\sigma^{331}$ &$\sigma^{333}$  & & & &  \\
    5         & $\sigma^{1}$ &$\sigma^{31}$ &$\sigma^{331}$ &$\sigma^{3331}$ &$\sigma^{3333}$  & & &  \\
    6         & $\sigma^{1}$ & $\sigma^{31}$ &$\sigma^{331}$ &$\sigma^{3331}$ &$\sigma^{33331}$ &$\sigma^{33333}$  & &  \\
    7         & $\sigma^{1}$ & $\sigma^{31}$ & $\sigma^{331}$ &$\sigma^{3331}$ &$\sigma^{33331}$ &$\sigma^{333331}$ &$\sigma^{333333}$ &    \\
    8         & $\sigma^{1}$ & $\sigma^{31}$ & $\sigma^{331}$ & $\sigma^{3331}$ &$\sigma^{33331}$ &$\sigma^{333331}$ &$\sigma^{3333331}$ &$\sigma^{3333333}$    \\
\hline
    \end{tabular}\label{tab: stagbas_D}
\end{table}

For $d=1$, the unitary transformation defined by $\alpha_{stag}=U\alpha_{free}U^\dagger$ is given by:

\begin{equation}
    U=e^{\frac{\mathrm{i}\pi}{4}\sigma^2}.
\end{equation}

For $d=2$, the unitary transformation defined by $\alpha_{stag}=U\alpha_{free}U^\dagger$ is given by (Note that the unitary transformation will also change the definition of $M_\mathcal{C}$ and $M_\mathcal{T}$, so they need to be redefined with new basis instead of being simply related by unitary transformation $U$):

\begin{equation}
    U=e^{\frac{\mathrm{i}\pi}{4}\sigma^1}.
\end{equation}

For $d=3$, the unitary transformation defined by $\alpha_{stag}=U\alpha_{free}U^\dagger$ is given by:

\begin{equation}
    U=e^{-\frac{\mathrm{i}\pi}{4}\sigma^{33}}e^{-\frac{\mathrm{i}\pi}{4}\sigma^{22}}.
\end{equation}

For $d=4$, the unitary transformation defined by $\alpha_{stag}=U(\alpha_{free}\otimes \sigma^0)U^\dagger$ is given by:

\begin{equation}
    U=e^{\frac{\mathrm{i}\pi}{4}\sigma^{013}}e^{-\frac{\mathrm{i}\pi}{4}\sigma^{131}}e^{-\frac{\mathrm{i}\pi}{4}\sigma^{220}}e^{-\frac{\mathrm{i}\pi}{4}\sigma^{330}}.
\end{equation}

For $d=5$, the unitary transformation defined by $\alpha_{stag}=U(\alpha_{free}\otimes \sigma^0)U^\dagger$ is given by:

\begin{equation}
\begin{aligned}
    U=&e^{\frac{\mathrm{i}\pi}{4}\sigma^{1321}}e^{\frac{\mathrm{i}\pi}{4}\sigma^{0013}}e^{-\frac{\mathrm{i}\pi}{4}\sigma^{1331}}e^{\frac{\mathrm{i}\pi}{4}\sigma^{1210}}\\
    &e^{-\frac{\mathrm{i}\pi}{4}\sigma^{1310}}e^{\frac{\mathrm{i}\pi}{4}\sigma^{1110}}e^{\frac{\mathrm{i}\pi}{4}\sigma^{3020}}e^{-\frac{\mathrm{i}\pi}{4}\sigma^{2030}}.
\end{aligned}
\end{equation}

For $d=6$, the unitary transformation defined by $\alpha_{stag}=U(\alpha_{free}\otimes \sigma^{00})U^\dagger$ is given by:

\begin{equation}
\begin{aligned}
    U=&e^{\frac{\mathrm{i}\pi}{4}\sigma^{00133}}e^{\frac{\mathrm{i}\pi}{4}\sigma^{01331}}e^{-\frac{\mathrm{i}\pi}{4}\sigma^{13310}}e^{\frac{\mathrm{i}\pi}{4}\sigma^{12100}}\\
    &e^{-\frac{\mathrm{i}\pi}{4}\sigma^{13100}}e^{\frac{\mathrm{i}\pi}{4}\sigma^{11100}}e^{\frac{\mathrm{i}\pi}{4}\sigma^{30200}}e^{-\frac{\mathrm{i}\pi}{4}\sigma^{20300}}.
\end{aligned}
\end{equation}

For $d=7$, the unitary transformation defined by $\alpha_{stag}=U(\alpha_{free}\otimes \sigma^{00})U^\dagger$ is given by:

\begin{equation}
\begin{aligned}
    U=&e^{-\frac{\mathrm{i}\pi}{4}\sigma^{133333}}e^{\frac{\mathrm{i}\pi}{4}\sigma^{121321}}e^{\frac{\mathrm{i}\pi}{4}\sigma^{132131}}e^{\frac{\mathrm{i}\pi}{4}\sigma^{012010}}\\
    &e^{-\frac{\mathrm{i}\pi}{4}\sigma^{013310}}e^{\frac{\mathrm{i}\pi}{4}\sigma^{001100}}e^{\frac{\mathrm{i}\pi}{4}\sigma^{123000}}e^{-\frac{\mathrm{i}\pi}{4}\sigma^{003000}}\\
    &e^{\frac{\mathrm{i}\pi}{4}\sigma^{110100}}e^{\frac{\mathrm{i}\pi}{4}\sigma^{300200}}e^{-\frac{\mathrm{i}\pi}{4}\sigma^{200300}}.
\end{aligned}
\end{equation}

For $d=8$, the unitary transformation defined by $\alpha_{stag}=U(\alpha_{free}\otimes \sigma^{000})U^\dagger$ is given by:

\begin{equation}
\begin{aligned}
    U=&e^{-\frac{\mathrm{i}\pi}{4}\sigma^{1333213}}e^{\frac{\mathrm{i}\pi}{4}\sigma^{0133201}}e^{\frac{\mathrm{i}\pi}{4}\sigma^{1332131}}e^{\frac{\mathrm{i}\pi}{4}\sigma^{0132010}}\\
    &e^{\frac{\mathrm{i}\pi}{4}\sigma^{1321310}}e^{\frac{\mathrm{i}\pi}{4}\sigma^{0120100}}e^{-\frac{\mathrm{i}\pi}{4}\sigma^{0133100}}e^{\frac{\mathrm{i}\pi}{4}\sigma^{0011000}}\\
    &e^{\frac{\mathrm{i}\pi}{4}\sigma^{1230000}}e^{\frac{\mathrm{i}\pi}{4}\sigma^{0030000}}e^{\frac{\mathrm{i}\pi}{4}\sigma^{1101000}}e^{\frac{\mathrm{i}\pi}{4}\sigma^{3002000}}\\
    &e^{-\frac{\mathrm{i}\pi}{4}\sigma^{2003000}}.
\end{aligned}
\end{equation}

\section{Presentation of the Invariant Group for Dirac Fermion}\label{Ap. CRTonU(1)}

We assume the canonical conditions $(\mathcal{CR}_1\mathcal{T})^2=1$, $\mathcal{C}(\mathcal{CR}_1\mathcal{T})=(\mathcal{CR}_1\mathcal{T})\mathcal{C}$, and $\mathcal{T}(\mathcal{CR}_1\mathcal{T})=(-)^F(\mathcal{CR}_1\mathcal{T})\mathcal{T}$ for the invariant group.

For $d=0$, the invariant group is given by the presentation:

\begin{equation}
\begin{aligned}
    &\mathcal{C}^2=\mathcal{T}^2=(\mathcal{CT})^2=1,\\
    &\mathcal{C}\mathcal{U}^F(\theta)=\mathcal{U}^F(-\theta)\mathcal{C},\ \mathcal{T}\mathcal{U}^F(\theta)=\mathcal{U}^F(-\theta)\mathcal{T}.
\end{aligned}
\end{equation}

For $d=1$, the invariant group is given by the presentation:

\begin{equation}
\begin{aligned}
    &\mathcal{C}^2=\mathcal{R}_1^2=\mathcal{T}^2=(\mathcal{R}_1\mathcal{T})^2=1,\\
    &(\mathcal{CR}_1)^2=(\mathcal{CT})^2=(-)^F,\\
    &\mathcal{C}\mathcal{U}^F(\theta)=\mathcal{U}^F(-\theta)\mathcal{C},\ \mathcal{R}_1\mathcal{U}^F(\theta)=\mathcal{U}^F(\theta)\mathcal{R}_1,\\
    &\mathcal{T}\mathcal{U}^F(\theta)=\mathcal{U}^F(-\theta)\mathcal{T},\ \mathcal{C}\mathcal{U}^\chi(\theta)=\mathcal{U}^\chi(-\theta)\mathcal{C},\\
    &\mathcal{R}_1\mathcal{U}^\chi(\theta)=\mathcal{U}^\chi(-\theta)\mathcal{R}_1,\ \mathcal{T}\mathcal{U}^\chi(\theta)=\mathcal{U}^\chi(\theta)\mathcal{T} .
\end{aligned}
\end{equation}

For $d=2$, the invariant group is given by the presentation:

\begin{equation}
\begin{aligned}
    &\mathcal{C}^2=\mathcal{R}_1^2=(\mathcal{CR}_1)^2=(\mathcal{R}_1\mathcal{T})^2=1,\\
    &\mathcal{T}^2=(\mathcal{CT})^2=(-)^F,\\
    &\mathcal{C}\mathcal{U}^F(\theta)=\mathcal{U}^F(-\theta)\mathcal{C},\ \mathcal{R}_1\mathcal{U}^F(\theta)=\mathcal{U}^F(\theta)\mathcal{R}_1,\\
    &\mathcal{T}\mathcal{U}^F(\theta)=\mathcal{U}^F(-\theta)\mathcal{T}.
\end{aligned}
\end{equation}

For $d=3$, the invariant group is given by the presentation:

\begin{equation}
\begin{aligned}
    &\mathcal{C}^2=\mathcal{R}_1^2=(\mathcal{CR}_1)^2=(\mathcal{R}_1\mathcal{T})^2=1,\\
    &\mathcal{T}^2=(\mathcal{CT})^2=(-)^F,\\
    &\mathcal{C}\mathcal{U}^F(\theta)=\mathcal{U}^F(-\theta)\mathcal{C},\ \mathcal{R}_1\mathcal{U}^F(\theta)=\mathcal{U}^F(\theta)\mathcal{R}_1,\\
    &\mathcal{T}\mathcal{U}^F(\theta)=\mathcal{U}^F(-\theta)\mathcal{T},\ \mathcal{C}\mathcal{U}^\chi(\theta)=\mathcal{U}^\chi(\theta)\mathcal{C},\\
    &\mathcal{R}_1\mathcal{U}^\chi(\theta)=\mathcal{U}^\chi(-\theta)\mathcal{R}_1,\ \mathcal{T}\mathcal{U}^\chi(\theta)=\mathcal{U}^\chi(-\theta)\mathcal{T} .
\end{aligned}
\end{equation}

For $d=4$, the invariant group is given by the presentation:

\begin{equation}
\begin{aligned}
    &(\mathcal{CR}_1)^2=(\mathcal{CT})^2=1,\\
    &\mathcal{C}^2=\mathcal{R}_1^2=\mathcal{T}^2=(\mathcal{R}_1\mathcal{T})^2=(-)^F,\\
    &\mathcal{C}\mathcal{U}^F(\theta)=\mathcal{U}^F(-\theta)\mathcal{C},\ \mathcal{R}_1\mathcal{U}^F(\theta)=\mathcal{U}^F(\theta)\mathcal{R}_1,\\
    &\mathcal{T}\mathcal{U}^F(\theta)=\mathcal{U}^F(-\theta)\mathcal{T}.
\end{aligned}
\end{equation}

For $d=5$, the invariant group is given by the presentation:

\begin{equation}
\begin{aligned}
    &\mathcal{C}^2=\mathcal{R}_1^2=\mathcal{T}^2=(\mathcal{R}_1\mathcal{T})^2=1,\\
    &(\mathcal{CR}_1)^2=(\mathcal{CT})^2=(-)^F,\\
    &\mathcal{C}\mathcal{U}^F(\theta)=\mathcal{U}^F(-\theta)\mathcal{C},\ \mathcal{R}_1\mathcal{U}^F(\theta)=\mathcal{U}^F(\theta)\mathcal{R}_1,\\
    &\mathcal{T}\mathcal{U}^F(\theta)=\mathcal{U}^F(-\theta)\mathcal{T},\ \mathcal{C}\mathcal{U}^\chi(\theta)=\mathcal{U}^\chi(-\theta)\mathcal{C},\\
    &\mathcal{R}_1\mathcal{U}^\chi(\theta)=\mathcal{U}^\chi(-\theta)\mathcal{R}_1,\ \mathcal{T}\mathcal{U}^\chi(\theta)=\mathcal{U}^\chi(\theta)\mathcal{T} .
\end{aligned}
\end{equation}

For $d=6$, the invariant group is given by the presentation:

\begin{equation}
\begin{aligned}
    &\mathcal{R}_1^2=\mathcal{T}^2=1,\\
    &\mathcal{C}^2=(\mathcal{CR}_1)^2=(\mathcal{R}_1\mathcal{T})^2=(\mathcal{CT})^2=(-)^F,\\
    &\mathcal{C}\mathcal{U}^F(\theta)=\mathcal{U}^F(-\theta)\mathcal{C},\ \mathcal{R}_1\mathcal{U}^F(\theta)=\mathcal{U}^F(\theta)\mathcal{R}_1,\\
    &\mathcal{T}\mathcal{U}^F(\theta)=\mathcal{U}^F(-\theta)\mathcal{T}.
\end{aligned}
\end{equation}

For $d=7$, the invariant group is given by the presentation:

\begin{equation}
\begin{aligned}
    &\mathcal{R}_1^2=\mathcal{T}^2=1,\\
    &\mathcal{C}^2=(\mathcal{CR}_1)^2=(\mathcal{R}_1\mathcal{T})^2=(\mathcal{CT})^2=(-)^F,\\
    &\mathcal{C}\mathcal{U}^F(\theta)=\mathcal{U}^F(-\theta)\mathcal{C},\ \mathcal{R}_1\mathcal{U}^F(\theta)=\mathcal{U}^F(\theta)\mathcal{R}_1,\\
    &\mathcal{T}\mathcal{U}^F(\theta)=\mathcal{U}^F(-\theta)\mathcal{T},\ \mathcal{C}\mathcal{U}^\chi(\theta)=\mathcal{U}^\chi(\theta)\mathcal{C},\\
    &\mathcal{R}_1\mathcal{U}^\chi(\theta)=\mathcal{U}^\chi(-\theta)\mathcal{R}_1,\ \mathcal{T}\mathcal{U}^\chi(\theta)=\mathcal{U}^\chi(-\theta)\mathcal{T} .
\end{aligned}
\end{equation}

For $d=8$, the invariant group is given by the presentation:

\begin{equation}
\begin{aligned}
    &\mathcal{C}^2=\mathcal{T}^2=(\mathcal{R}_1\mathcal{T})^2=(\mathcal{CT})^2=1,\\
    &\mathcal{R}_1^2=(\mathcal{CR}_1)^2=(-)^F,\\
    &\mathcal{C}\mathcal{U}^F(\theta)=\mathcal{U}^F(-\theta)\mathcal{C},\ \mathcal{R}_1\mathcal{U}^F(\theta)=\mathcal{U}^F(\theta)\mathcal{R}_1,\\
    &\mathcal{T}\mathcal{U}^F(\theta)=\mathcal{U}^F(-\theta)\mathcal{T}.
\end{aligned}
\end{equation}

\end{document}